\RequirePackage{lineno}
\documentclass[aps,prd,twocolumn,showpacs,superscriptaddress,groupedaddress]{revtex4}  % for review and submission
\usepackage{graphicx}  % needed for figures
\usepackage{dcolumn}   % needed for some tables
\usepackage{bm}        % for math
\usepackage{amssymb}   % for math
\usepackage{multirow}
\usepackage{epsfig}
\usepackage{amsmath}
\usepackage[integrals]{wasysym}
\usepackage{amsmath,amssymb}

\newcommand{\met}{\mbox{$\rlap{\kern0.15em/}E_T$}}
\newcommand{\wen}{\ensuremath{W\rightarrow e \nu}}

\begin{document}

\widetext 
% the following line is for submission, including submission to the arXiv!!
\hspace{5.2in} \mbox{FERMILAB-PUB-14-514-E}

\title{Erratum: Measurement of the electron charge asymmetry in $\boldsymbol{p\bar{p}\rightarrow W+X \rightarrow e\nu +X}$ decays
in $\boldsymbol{p\bar{p}}$ collisions at $\boldsymbol{\sqrt{s}=1.96}$~TeV}

% remove these 3 lines before journal submittal.
%\centerline{author list dated 18 October 2014}
% end removal before journal submittal
%
\affiliation{LAFEX, Centro Brasileiro de Pesquisas F\'{i}sicas, Rio de Janeiro, Brazil}
\affiliation{Universidade do Estado do Rio de Janeiro, Rio de Janeiro, Brazil}
\affiliation{Universidade Federal do ABC, Santo Andr\'e, Brazil}
\affiliation{University of Science and Technology of China, Hefei, People's Republic of China}
\affiliation{Universidad de los Andes, Bogot\'a, Colombia}
\affiliation{Charles University, Faculty of Mathematics and Physics, Center for Particle Physics, Prague, Czech Republic}
\affiliation{Czech Technical University in Prague, Prague, Czech Republic}
\affiliation{Institute of Physics, Academy of Sciences of the Czech Republic, Prague, Czech Republic}
\affiliation{Universidad San Francisco de Quito, Quito, Ecuador}
\affiliation{LPC, Universit\'e Blaise Pascal, CNRS/IN2P3, Clermont, France}
\affiliation{LPSC, Universit\'e Joseph Fourier Grenoble 1, CNRS/IN2P3, Institut National Polytechnique de Grenoble, Grenoble, France}
\affiliation{CPPM, Aix-Marseille Universit\'e, CNRS/IN2P3, Marseille, France}
\affiliation{LAL, Universit\'e Paris-Sud, CNRS/IN2P3, Orsay, France}
\affiliation{LPNHE, Universit\'es Paris VI and VII, CNRS/IN2P3, Paris, France}
\affiliation{CEA, Irfu, SPP, Saclay, France}
\affiliation{IPHC, Universit\'e de Strasbourg, CNRS/IN2P3, Strasbourg, France}
\affiliation{IPNL, Universit\'e Lyon 1, CNRS/IN2P3, Villeurbanne, France and Universit\'e de Lyon, Lyon, France}
\affiliation{III. Physikalisches Institut A, RWTH Aachen University, Aachen, Germany}
\affiliation{Physikalisches Institut, Universit\"at Freiburg, Freiburg, Germany}
\affiliation{II. Physikalisches Institut, Georg-August-Universit\"at G\"ottingen, G\"ottingen, Germany}
\affiliation{Institut f\"ur Physik, Universit\"at Mainz, Mainz, Germany}
\affiliation{Ludwig-Maximilians-Universit\"at M\"unchen, M\"unchen, Germany}
\affiliation{Panjab University, Chandigarh, India}
\affiliation{Delhi University, Delhi, India}
\affiliation{Tata Institute of Fundamental Research, Mumbai, India}
\affiliation{University College Dublin, Dublin, Ireland}
\affiliation{Korea Detector Laboratory, Korea University, Seoul, Korea}
\affiliation{CINVESTAV, Mexico City, Mexico}
\affiliation{Nikhef, Science Park, Amsterdam, the Netherlands}
\affiliation{Radboud University Nijmegen, Nijmegen, the Netherlands}
\affiliation{Joint Institute for Nuclear Research, Dubna, Russia}
\affiliation{Institute for Theoretical and Experimental Physics, Moscow, Russia}
\affiliation{Moscow State University, Moscow, Russia}
\affiliation{Institute for High Energy Physics, Protvino, Russia}
\affiliation{Petersburg Nuclear Physics Institute, St. Petersburg, Russia}
\affiliation{Instituci\'{o} Catalana de Recerca i Estudis Avan\c{c}ats (ICREA) and Institut de F\'{i}sica d'Altes Energies (IFAE), Barcelona, Spain}
\affiliation{Uppsala University, Uppsala, Sweden}
\affiliation{Taras Shevchenko National University of Kyiv, Kiev, Ukraine}
\affiliation{Lancaster University, Lancaster LA1 4YB, United Kingdom}
\affiliation{Imperial College London, London SW7 2AZ, United Kingdom}
\affiliation{The University of Manchester, Manchester M13 9PL, United Kingdom}
\affiliation{University of Arizona, Tucson, Arizona 85721, USA}
\affiliation{University of California Riverside, Riverside, California 92521, USA}
\affiliation{Florida State University, Tallahassee, Florida 32306, USA}
\affiliation{Fermi National Accelerator Laboratory, Batavia, Illinois 60510, USA}
\affiliation{University of Illinois at Chicago, Chicago, Illinois 60607, USA}
\affiliation{Northern Illinois University, DeKalb, Illinois 60115, USA}
\affiliation{Northwestern University, Evanston, Illinois 60208, USA}
\affiliation{Indiana University, Bloomington, Indiana 47405, USA}
\affiliation{Purdue University Calumet, Hammond, Indiana 46323, USA}
\affiliation{University of Notre Dame, Notre Dame, Indiana 46556, USA}
\affiliation{Iowa State University, Ames, Iowa 50011, USA}
\affiliation{University of Kansas, Lawrence, Kansas 66045, USA}
\affiliation{Louisiana Tech University, Ruston, Louisiana 71272, USA}
\affiliation{Northeastern University, Boston, Massachusetts 02115, USA}
\affiliation{University of Michigan, Ann Arbor, Michigan 48109, USA}
\affiliation{Michigan State University, East Lansing, Michigan 48824, USA}
\affiliation{University of Mississippi, University, Mississippi 38677, USA}
\affiliation{University of Nebraska, Lincoln, Nebraska 68588, USA}
\affiliation{Rutgers University, Piscataway, New Jersey 08855, USA}
\affiliation{Princeton University, Princeton, New Jersey 08544, USA}
\affiliation{State University of New York, Buffalo, New York 14260, USA}
\affiliation{University of Rochester, Rochester, New York 14627, USA}
\affiliation{State University of New York, Stony Brook, New York 11794, USA}
\affiliation{Brookhaven National Laboratory, Upton, New York 11973, USA}
\affiliation{Langston University, Langston, Oklahoma 73050, USA}
\affiliation{University of Oklahoma, Norman, Oklahoma 73019, USA}
\affiliation{Oklahoma State University, Stillwater, Oklahoma 74078, USA}
\affiliation{Brown University, Providence, Rhode Island 02912, USA}
\affiliation{University of Texas, Arlington, Texas 76019, USA}
\affiliation{Southern Methodist University, Dallas, Texas 75275, USA}
\affiliation{Rice University, Houston, Texas 77005, USA}
\affiliation{University of Virginia, Charlottesville, Virginia 22904, USA}
\affiliation{University of Washington, Seattle, Washington 98195, USA}
\author{V.M.~Abazov} \affiliation{Joint Institute for Nuclear Research, Dubna, Russia}
\author{B.~Abbott} \affiliation{University of Oklahoma, Norman, Oklahoma 73019, USA}
\author{B.S.~Acharya} \affiliation{Tata Institute of Fundamental Research, Mumbai, India}
\author{M.~Adams} \affiliation{University of Illinois at Chicago, Chicago, Illinois 60607, USA}
\author{T.~Adams} \affiliation{Florida State University, Tallahassee, Florida 32306, USA}
\author{J.P.~Agnew} \affiliation{The University of Manchester, Manchester M13 9PL, United Kingdom}
\author{G.D.~Alexeev} \affiliation{Joint Institute for Nuclear Research, Dubna, Russia}
\author{G.~Alkhazov} \affiliation{Petersburg Nuclear Physics Institute, St. Petersburg, Russia}
\author{A.~Alton$^{a}$} \affiliation{University of Michigan, Ann Arbor, Michigan 48109, USA}
\author{A.~Askew} \affiliation{Florida State University, Tallahassee, Florida 32306, USA}
% ...
%
% visitor_addresses.tex                       18 October 2014
%  available symbols are:
%  $\ast, \dag, \ddag, \S, \P, $\|$, $\ast\ast$, \dag\dag, \ddag\ddag ,\#
%
\collaboration{The D0 Collaboration\footnote{with visitors from
%{alton}
$^{a}$Augustana College, Sioux Falls, SD, USA,
%{burdin}
$^{b}$The University of Liverpool, Liverpool, UK,
%{grohsjean,deterre}
$^{c}$DESY, Hamburg, Germany,
%{de la cruz-burelo}
$^{d}$Universidad Michoacana de San Nicolas de Hidalgo, Morelia, Mexico
%{partridge}
$^{e}$SLAC, Menlo Park, CA, USA,
%{hesketh}
$^{f}$University College London, London, UK,
%{luna-garcia}
$^{g}$Centro de Investigacion en Computacion - IPN, Mexico City, Mexico,
%{santos}
$^{h}$Universidade Estadual Paulista, S\~ao Paulo, Brazil,
%{meyer}
$^{i}$Karlsruher Institut f\"ur Technologie (KIT) - Steinbuch Centre for Computing (SCC),
D-76128 Karlsruhe, Germany,
%{patwa}
$^{j}$Office of Science, U.S. Department of Energy, Washington, D.C. 20585, USA,
%{cooke}
$^{k}$American Association for the Advancement of Science, Washington, D.C. 20005, USA,
%{borysova}
$^{l}$Kiev Institute for Nuclear Research, Kiev, Ukraine,
%{jabeen}
$^{m}$University of Maryland, College Park, Maryland 20742, USA
and
%{williams}
$^{n}$European Orgnaization for Nuclear Research (CERN), Geneva, Switzerland
%{montgomery}
%$^{?}$Thomas Jefferson National Accelerator Facility, Newport News, VA 23606, USA,
%{falkowski}
%$^{?}$Laboratoire de Physique Theorique, Orsay, FR,
%{hooper,kozminski}
%$^{?}$}Visitor from Lewis University, Romeoville, IL, USA.
%{weber}
%$^{?}$Universit{\"a}t Bern, Bern, Switzerland.
%{deceased}
%{zanabria}
%$^{?}$City Colleges of Chicago, Chicago, IL, USA}
%$^{\ddag}$Deceased.
}} \noaffiliation
\vskip 0.25cm

\date{December 09, 2014}
%General comment: Need consistant tense

\pacs{12.15.Ji, 13.38.Be, 13.85.Qk, 14.60.Cd, 14.70.Fm}
\maketitle
%%%%%%%%%%%%%%%%%%%%%%%%%%%%%%%%

\setcounter{table}{6}

%%%%%%%%%%%%%%%%%%%%%%%%%%%%%%%%

The recent paper~\cite{d0_elec} on the charge asymmetry for electrons from $W$ boson
decay has an error in the Tables VII to XI that show the correlation coefficients
of systematic uncertainties.
The correlation matrix elements shown in the original
publication were the square roots of the calculated values.

The corrected correlation
matrices are shown in Tables~\ref{Tab:Syst_corr_Elec_25_Inf_Neut_25_Inf_n}
to~\ref{Tab:Syst_corr_Elec_35_Inf_Neut_35_Inf_n}.
The table numbers used here correspond directly to those in the paper~\cite{d0_elec}.
The results of the paper~\cite{d0_elec} are unchanged except for these tables.

We thank Stefan Camardo for pointing out the error in the tables.

\begin{table*}
\caption{Correlation matrix of the systematic uncertainties between different $|\eta^e|$ bins for events with $E_T^{e} > 25$~GeV, $\met > 25$~GeV. The ``$|\eta^e|$ bin" represents the indexing of the $\eta^e$ bins used in this analysis.}
\begin{tabular}{c|ccccccccccccc}
\hline
\hline
$|\eta^e|$ bin & 1 & 2 & 3 & 4 & 5 & 6 & 7 & 8 & 9 & 10 & 11 & 12 & 13   \\ \hline
  1 &  1.00 &  0.68 &  0.63 &  0.56 &  0.52 &  0.54 &  0.46 &  0.45 &  0.44 &  0.40 &  0.31 &  0.30 &  0.27 \\
 2 &  &  1.00 &  0.82 &  0.78 &  0.76 &  0.77 &  0.73 &  0.71 &  0.65 &  0.54 &  0.40 &  0.40 &  0.31 \\
 3 &  &  &  1.00 &  0.85 &  0.84 &  0.84 &  0.81 &  0.76 &  0.70 &  0.57 &  0.42 &  0.41 &  0.31 \\
 4 &  &  &  &  1.00 &  0.93 &  0.90 &  0.89 &  0.74 &  0.67 &  0.56 &  0.40 &  0.40 &  0.30 \\
 5 &  &  &  &  &  1.00 &  0.91 &  0.90 &  0.74 &  0.68 &  0.57 &  0.40 &  0.40 &  0.29 \\
 6 &  &  &  &  &  &  1.00 &  0.89 &  0.78 &  0.71 &  0.58 &  0.41 &  0.41 &  0.29 \\
 7 &  &  &  &  &  &  &  1.00 &  0.82 &  0.74 &  0.59 &  0.41 &  0.41 &  0.27 \\
 8 &  &  &  &  &  &  &  &  1.00 &  0.81 &  0.61 &  0.44 &  0.42 &  0.24 \\
 9 &  &  &  &  &  &  &  &  &  1.00 &  0.56 &  0.40 &  0.38 &  0.22 \\
 10 &  &  &  &  &  &  &  &  &  &  1.00 &  0.32 &  0.30 &  0.20 \\
 11 &  &  &  &  &  &  &  &  &  &  &  1.00 &  0.22 &  0.15 \\
 12 &  &  &  &  &  &  &  &  &  &  &  &  1.00 &  0.14 \\
 13 &  &  &  &  &  &  &  &  &  &  &  &  &  1.00 \\
 \hline
\hline
\end{tabular}
\label{Tab:Syst_corr_Elec_25_Inf_Neut_25_Inf_n}
\end{table*}

\begin{table*}
\caption{Correlation matrix of the systematic uncertainties between different $|\eta^e|$ bins for events with $25 < E_T^{e} < 35$~GeV, $\met > 25$~GeV.}
\begin{tabular}{c|ccccccccccccc}
\hline
\hline
$|\eta^e|$ bin& 1 & 2 & 3 & 4 & 5 & 6 & 7 & 8 & 9 & 10 & 11 & 12 & 13   \\ \hline
 1 &  1.00 &  0.29 &  0.27 &  0.27 &  0.28 &  0.29 &  0.29 &  0.28 &  0.22 &  0.20 &  0.18 &  0.17 &  0.16 \\
 2 &  &  1.00 &  0.37 &  0.42 &  0.42 &  0.41 &  0.37 &  0.37 &  0.30 &  0.28 &  0.25 &  0.27 &  0.22 \\
 3 &  &  &  1.00 &  0.47 &  0.47 &  0.44 &  0.40 &  0.40 &  0.35 &  0.32 &  0.28 &  0.29 &  0.23 \\
 4 &  &  &  &  1.00 &  0.58 &  0.55 &  0.51 &  0.49 &  0.41 &  0.39 &  0.33 &  0.38 &  0.28 \\
 5 &  &  &  &  &  1.00 &  0.58 &  0.56 &  0.48 &  0.38 &  0.37 &  0.32 &  0.40 &  0.29 \\
 6 &  &  &  &  &  &  1.00 &  0.57 &  0.49 &  0.39 &  0.37 &  0.32 &  0.39 &  0.29 \\
 7 &  &  &  &  &  &  &  1.00 &  0.52 &  0.41 &  0.38 &  0.31 &  0.40 &  0.29 \\
 8 &  &  &  &  &  &  &  &  1.00 &  0.48 &  0.41 &  0.32 &  0.35 &  0.28 \\
 9 &  &  &  &  &  &  &  &  &  1.00 &  0.40 &  0.29 &  0.29 &  0.24 \\
 10 &  &  &  &  &  &  &  &  &  &  1.00 &  0.26 &  0.27 &  0.22 \\
 11 &  &  &  &  &  &  &  &  &  &  &  1.00 &  0.22 &  0.18 \\
 12 &  &  &  &  &  &  &  &  &  &  &  &  1.00 &  0.21 \\
 13 &  &  &  &  &  &  &  &  &  &  &  &  &  1.00 \\
 \hline
\hline
\end{tabular}
\label{Tab:Syst_corr_Elec_25_35_Neut_25_Inf_n}
\end{table*}

\begin{table*}
\caption{Correlation matrix of the systematic uncertainties between different $|\eta^e|$ bins for events with $25 < E_T^{e} < 35$~GeV, $25< \met < 35$~GeV.}
\begin{tabular}{c|ccccccccccccc}
\hline
\hline
$|\eta^e|$ bin & 1 & 2 & 3 & 4 & 5 & 6 & 7 & 8 & 9 & 10 & 11 & 12 & 13   \\ \hline
 1 &  1.00 &  0.46 &  0.43 &  0.45 &  0.46 &  0.48 &  0.45 &  0.44 &  0.36 &  0.33 &  0.30 &  0.28 &  0.26 \\
 2 &  &  1.00 &  0.51 &  0.56 &  0.57 &  0.56 &  0.54 &  0.55 &  0.49 &  0.45 &  0.39 &  0.40 &  0.34 \\
 3 &  &  &  1.00 &  0.57 &  0.57 &  0.56 &  0.53 &  0.56 &  0.52 &  0.48 &  0.40 &  0.42 &  0.35 \\
 4 &  &  &  &  1.00 &  0.65 &  0.63 &  0.62 &  0.63 &  0.60 &  0.56 &  0.47 &  0.51 &  0.41 \\
 5 &  &  &  &  &  1.00 &  0.66 &  0.66 &  0.65 &  0.61 &  0.57 &  0.48 &  0.54 &  0.41 \\
 6 &  &  &  &  &  &  1.00 &  0.68 &  0.63 &  0.57 &  0.54 &  0.47 &  0.53 &  0.41 \\
 7 &  &  &  &  &  &  &  1.00 &  0.62 &  0.56 &  0.54 &  0.48 &  0.57 &  0.43 \\
 8 &  &  &  &  &  &  &  &  1.00 &  0.59 &  0.55 &  0.46 &  0.50 &  0.39 \\
 9 &  &  &  &  &  &  &  &  &  1.00 &  0.55 &  0.45 &  0.50 &  0.38 \\
 10 &  &  &  &  &  &  &  &  &  &  1.00 &  0.43 &  0.49 &  0.37 \\
 11 &  &  &  &  &  &  &  &  &  &  &  1.00 &  0.41 &  0.31 \\
 12 &  &  &  &  &  &  &  &  &  &  &  &  1.00 &  0.36 \\
 13 &  &  &  &  &  &  &  &  &  &  &  &  &  1.00 \\
\hline
\hline
\end{tabular}
\label{Tab:Syst_corr_Elec_25_35_Neut_25_35_n}
\end{table*}

\begin{table*}
\caption{Correlation matrix of the systematic uncertainties between different $|\eta^e|$ bins for events with $E_T^{e} > 35$~GeV, $\met > 25$~GeV.}
\begin{tabular}{c|ccccccccccccc}
\hline
\hline
$|\eta^e|$ bin & 1 & 2 & 3 & 4 & 5 & 6 & 7 & 8 & 9 & 10 & 11 & 12 & 13   \\ \hline
 1 &  1.00 &  0.75 &  0.66 &  0.59 &  0.56 &  0.57 &  0.47 &  0.47 &  0.44 &  0.42 &  0.34 &  0.28 &  0.25 \\
 2 &  &  1.00 &  0.89 &  0.82 &  0.80 &  0.81 &  0.75 &  0.78 &  0.73 &  0.63 &  0.47 &  0.37 &  0.29 \\
 3 &  &  &  1.00 &  0.90 &  0.88 &  0.89 &  0.85 &  0.89 &  0.81 &  0.69 &  0.51 &  0.38 &  0.29 \\
 4 &  &  &  &  1.00 &  0.95 &  0.93 &  0.92 &  0.89 &  0.75 &  0.69 &  0.53 &  0.37 &  0.27 \\
 5 &  &  &  &  &  1.00 &  0.92 &  0.92 &  0.88 &  0.74 &  0.69 &  0.53 &  0.36 &  0.26 \\
 6 &  &  &  &  &  &  1.00 &  0.90 &  0.89 &  0.77 &  0.69 &  0.52 &  0.36 &  0.27 \\
 7 &  &  &  &  &  &  &  1.00 &  0.89 &  0.75 &  0.68 &  0.51 &  0.35 &  0.24 \\
 8 &  &  &  &  &  &  &  &  1.00 &  0.82 &  0.70 &  0.51 &  0.36 &  0.26 \\
 9 &  &  &  &  &  &  &  &  &  1.00 &  0.62 &  0.44 &  0.33 &  0.24 \\
 10 &  &  &  &  &  &  &  &  &  &  1.00 &  0.40 &  0.28 &  0.21 \\
 11 &  &  &  &  &  &  &  &  &  &  &  1.00 &  0.21 &  0.16 \\
 12 &  &  &  &  &  &  &  &  &  &  &  &  1.00 &  0.12 \\
 13 &  &  &  &  &  &  &  &  &  &  &  &  &  1.00 \\
\hline
\hline
\end{tabular}
\label{Tab:Syst_corr_Elec_35_Inf_Neut_25_Inf_n}
\end{table*}

\begin{table*}
\caption{Correlation matrix of the systematic uncertainties between different $|\eta^e|$ bins for events with $E_T^{e} > 35$~GeV, $\met > 35$~GeV.}
\begin{tabular}{c|ccccccccccccc}
\hline
\hline
$|\eta^e|$ bin & 1 & 2 & 3 & 4 & 5 & 6 & 7 & 8 & 9 & 10 & 11 & 12 & 13   \\ \hline
 1 &  1.00 &  0.82 &  0.72 &  0.63 &  0.55 &  0.58 &  0.39 &  0.39 &  0.39 &  0.33 &  0.26 &  0.24 &  0.24 \\
 2 &  &  1.00 &  0.84 &  0.79 &  0.73 &  0.73 &  0.58 &  0.59 &  0.53 &  0.42 &  0.32 &  0.25 &  0.25 \\
 3 &  &  &  1.00 &  0.87 &  0.84 &  0.81 &  0.71 &  0.73 &  0.62 &  0.47 &  0.35 &  0.24 &  0.23 \\
 4 &  &  &  &  1.00 &  0.88 &  0.84 &  0.78 &  0.80 &  0.65 &  0.49 &  0.36 &  0.24 &  0.22 \\
 5 &  &  &  &  &  1.00 &  0.84 &  0.81 &  0.82 &  0.66 &  0.49 &  0.36 &  0.23 &  0.20 \\
 6 &  &  &  &  &  &  1.00 &  0.75 &  0.76 &  0.62 &  0.46 &  0.35 &  0.23 &  0.20 \\
 7 &  &  &  &  &  &  &  1.00 &  0.77 &  0.60 &  0.44 &  0.33 &  0.21 &  0.16 \\
 8 &  &  &  &  &  &  &  &  1.00 &  0.63 &  0.45 &  0.33 &  0.19 &  0.16 \\
 9 &  &  &  &  &  &  &  &  &  1.00 &  0.37 &  0.28 &  0.18 &  0.15 \\
 10 &  &  &  &  &  &  &  &  &  &  1.00 &  0.22 &  0.16 &  0.13 \\
 11 &  &  &  &  &  &  &  &  &  &  &  1.00 &  0.13 &  0.10 \\
 12 &  &  &  &  &  &  &  &  &  &  &  &  1.00 &  0.09 \\
 13 &  &  &  &  &  &  &  &  &  &  &  &  &  1.00 \\
\hline
\hline
\end{tabular}
\label{Tab:Syst_corr_Elec_35_Inf_Neut_35_Inf_n}
\end{table*}

\clearpage

%%%%%%%%%%%%%%%%%%%%%%%%%%%%%%%%

%\hspace{5.2in} \mbox{FERMILAB-PUB-14-514-E}

\makeatletter
\def\maketitle{%
\par\textbf{\large\bfseries\centering \@title}%
\par}

\makeatother

\title{Measurement of the electron charge asymmetry in $\boldsymbol{p\bar{p}\rightarrow W+X \rightarrow e\nu +X}$ decays
in $\boldsymbol{p\bar{p}}$ collisions at $\boldsymbol{\sqrt{s}=1.96}$~TeV}
\maketitle

\begin{center}
We present a measurement of the electron charge asymmetry 
in $p\bar{p}\rightarrow W+X \rightarrow e\nu +X$ events at a center-of-mass energy of 1.96 TeV,
using data corresponding to 9.7~fb$^{-1}$ of integrated luminosity collected with the D0 detector 
at the Fermilab Tevatron Collider.
The asymmetry is measured as a function of the electron pseudorapidity
and is presented in five kinematic bins based on the electron transverse energy 
and the missing transverse energy in the event.
The measured asymmetry is compared with next-to-leading-order predictions in 
perturbative quantum chromodynamics and provides accurate information 
for the determination of parton distribution functions of the proton. This is the most
precise lepton charge asymmetry measurement to date.

\end{center}

\vspace{.2 in} 
\setcounter{table}{0}

\section{Introduction}
Parton distribution functions (PDFs) are essential elements for cross section calculations at a
hadron collider, and many precision measurements are dominated by the systematic uncertainty 
from PDFs. However, PDFs are not directly calculable within the standard model (SM) and must be 
determined using experimental inputs, including a wide range of scattering processes.
At the Fermilab Tevatron Collider, a proton-antiproton ($p\bar{p}$) collider with a center-of-mass energy 
of 1.96~TeV, the measurement of the electron charge asymmetry in the $p\bar{p}\rightarrow W+X$ process
provides important information for the determination of PDFs, as it is sensitive to the valence 
$u$ and $d$ quark and corresponding antiquark PDF distributions.
In $p\bar{p}$ collisions, $W^{+}$ ($W^{-}$) bosons are produced primarily by the annihilation 
of valence quarks in the proton and antiproton. 
Since $u$ quarks on average carry more momentum than $d$ quarks~\cite{old_pdf1, old_pdf2, old_pdf3}, 
$W^+$ bosons tend to be boosted in the proton direction, while $W^-$ bosons 
tend to be boosted in the antiproton direction. This results in a non-zero $W$ boson production 
charge asymmetry, defined as
\begin{eqnarray}
 A(y_W) = \frac{\frac{d\sigma_{W^+}}{dy_W} - \frac{d\sigma_{W^-}}{dy_W}}
{\frac{d\sigma_{W^+}}{dy_W} + \frac{d\sigma_{W^-}}{dy_W}}~,
\end{eqnarray}
where $d\sigma_{W^{\pm}}/dy_W$ is the differential cross section for $W^{\pm}$ boson
production, and $y_W$ is the $W$ boson rapidity~\cite{d0_coordinate}.

\begin{figure*}
\begin{center}
\epsfig{file=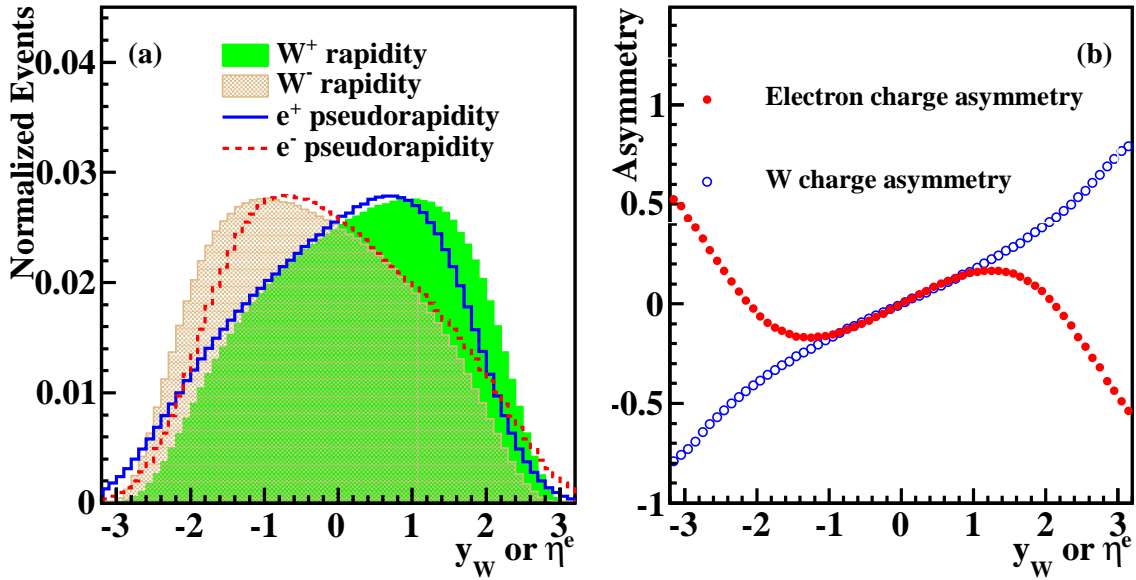, scale=0.8}
\caption{ (color online). (a) The $W$ boson rapidity ($y_W$) and electron pseudorapidity ($\eta^{e}$) 
distributions in $p\bar{p}$ collisions. (b) The charge asymmetry for the $W$ boson and the decay electron.
The electron asymmetry has a ``turn-over" due to 
the convolution of the $W$ boson asymmetry and the $V-A$ structure of the $W$ boson decay.
These predictions were obtained using the MC event generator {\sc resbos}~\cite{resbos} 
with the CTEQ6.6~\cite{cteq66} central PDF set,
using the kinematic cuts $p_T^{e} > 25$~GeV and $p_T^{\nu} > 25$~GeV.} 
\label{fig:gene_asym}
\end{center}
\end{figure*}

The $W$ boson can decay leptonically with a charged lepton and a neutrino
in the final state. The neutrino's presence can be inferred from an imbalance of transverse energy in the
calorimeter, referred to as missing transverse energy ($\met$). 
Reconstruction of the neutrino longitudinal momentum ($p_z^{\nu}$) is not 
feasible due to the unknown longitudinal momentum of the initial state interacting partons.
Without $p_z^{\nu}$, it is impossible to perform a direct measurement of
the $W$ boson charge asymmetry with traditional methods.
Instead we use the lepton pseudorapidity ($\eta$)~\cite{d0_coordinate} distribution which is 
a convolution of the $W$ boson production charge asymmetry and the $V-A$ structure of the $W$
boson decay. With a good understanding of the $V-A$ structure,
the lepton charge asymmetry as a function of lepton pseudorapidity can be used to constrain PDFs.
The comparison between the $W$ boson charge asymmetry and the lepton charge asymmetry
is shown in Fig.~\ref{fig:gene_asym}, using the Monte Carlo (MC) event 
generator {\sc resbos}~\cite{resbos} with the CTEQ6.6~\cite{cteq66} 
central PDF set.

In the $W\rightarrow e\nu$ decay mode used in this analysis, the experimentally measured 
$\wen$ cross section times branching ratio as 
a function of electron pseudorapidity ($\eta^e$) is
 \begin{eqnarray}
  \sigma(\eta^{e}) \times Br(W\rightarrow e\nu) = \frac{N^{e} (\eta^{e})}  {\cal{L}\times {\cal A} \times \epsilon}~,
\end{eqnarray}
where $N^e(\eta^e)$ is the number of events with electron in the $\eta^e$ bin, 
$\cal{A}$ is the acceptance, $\cal{L}$ is the integrated luminosity, and $\epsilon$ is the selection efficiency.
In the simplified case that the acceptances and efficiencies are the same for $W^+$ and $W^-$ bosons,
the electron charge asymmetry, $A$, can be written using the numbers of electrons ($N^{e^-}$) 
and positrons ($N^{e^+}$) in each $\eta^e$ bin as:
 \begin{eqnarray}
 A(\eta^e) = \frac{N^{e^+}(\eta^e) - N^{e^-}(\eta^e)}{N^{e^+}(\eta^e) + N^{e^-}(\eta^e)}.
\end{eqnarray}

The lepton charge asymmetry in $W$ boson decay has been
measured by both the CDF~\cite{CDF_results1, CDF_results2, CDF_results3}
and D0~\cite{d0_results_muon1, d0_results_em_old, d0_muon} Collaborations.
The latest lepton charge asymmetry measurement from the D0
Collaboration was performed in the muon channel using 7.3~fb$^{-1}$ of integrated luminosity~\cite{d0_muon}.
The $W$ boson asymmetry was extracted using missing transverse energy to
estimate the neutrino direction, using 1~fb$^{-1}$
of integrated luminosity by the CDF Collaboration~\cite{cdf_w} and 10~fb$^{-1}$ by the D0 Collaboration~\cite{d0_w}.
The lepton asymmetry has also been measured at the Large Hadron Collider (LHC) in $pp$ 
collisions by the ATLAS~\cite{ATLAS_w} and CMS 
Collaborations~\cite{CMS_w} using integrated luminosities of 35~pb$^{-1}$ and 840~pb$^{-1}$, respectively.
At the LHC, $W$ boson production is dominated by gluons and sea quarks, 
providing different information than the lepton asymmetry measured at the Tevatron. 

In this analysis, we present a new measurement of the electron charge asymmetry 
based on data collected between April 2002 and September 2011
with the D0 detector at $\sqrt{s}=1.96$~TeV, corresponding 
to an integrated luminosity of 9.7~fb$^{-1}$~\cite{d0lumi}.
We measure the electron charge asymmetry in five kinematic bins by selecting 
on the electron transverse energy ($E_T^e$) and event $\met$.
Results from different kinematic bins probe different ranges of $y_W$, and 
thus different ranges of the fraction of proton momentum carried by the parton. 
There are three symmetric bins,
($E_T^e > 25$~GeV, $\met > 25$~GeV), ($25<E_T^e < 35$~GeV, $25<\met < 35$~GeV), 
($E_T^e > 35$~GeV, $\met > 35$~GeV)
and two asymmetric bins, 
($25<E_T^e < 35$~GeV, $\met > 25$~GeV), ($E_T^e > 35$~GeV, $\met > 25$~GeV).
With more data than in previous measurements and more data in the high pseudorapidity region,
we provide information about the PDFs for a broader $x$ range ($0.002 < x < 0.99$ for 
$|\eta^e| < 3.2$) at high $Q^2\approx M^2_W$, where $x$ is the fraction of the 
proton momentum carried by the colliding parton, $Q^2$ is the momentum scale squared, 
and $M_W$ is the $W$ boson mass. 
This analysis improves upon and supersedes the previous D0 electron 
charge asymmetry result~\cite{d0_results_em_old}. 
That result did not include the improved
detector level calibrations discussed in Sec.~\ref{sec:em_energy} and Sec.~\ref{sec:recoil}. 
In addition, it did not include MC modeling of
the difference in efficiency for electrons and positrons for different polarities
of the solenoidal magnet surrounding the tracking region.
This article also provides details of the complementary analysis of Ref.~\cite{d0_w}
where the $W$ boson charge asymmetry is measured
using the same data set.

%%%%%%%%%%%%%%%%%%%%%%%%%%%%
\section{Apparatus}
\begin{figure*}
  \epsfig{file=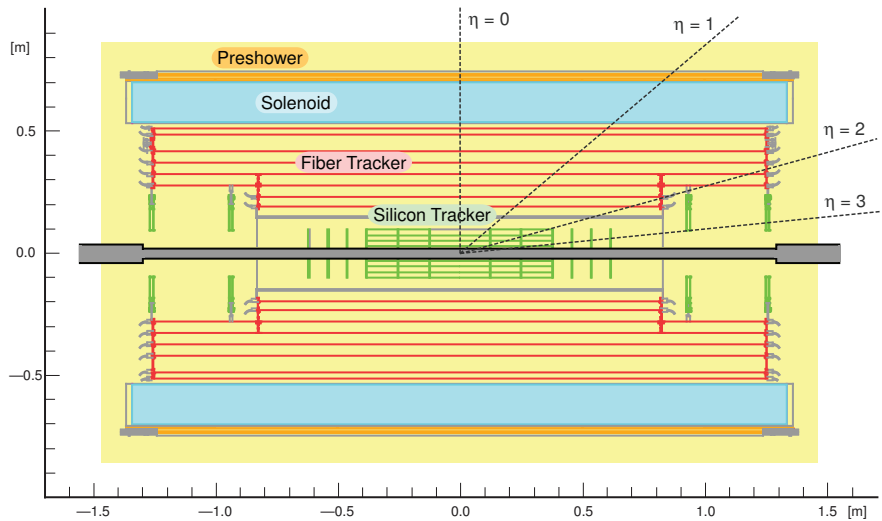,scale=0.75}
  \caption{(color online). Cross sectional view of the D0 central tracking detector in the
$x$-$z$ plane. }
\label{fig:dzero_cft}
\end{figure*}

\begin{figure}
  \epsfig{file=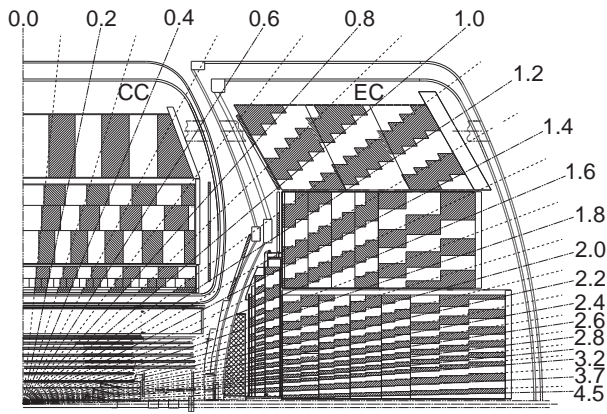,scale=0.6}
  \caption{Schematic view of a portion of the D0 calorimeters showing the transverse 
and longitudinal segmentation patterns. The rays indicate the pseudorapidity measured from the center of the detector ($\eta_{\text{det}}$).}
\label{fig:dzero_cal}
\end{figure}

\indent The D0 detector~\cite{d0det0, d0det} contains central tracking, calorimeter, and 
muon systems. The central tracking system includes a silicon microstrip tracker (SMT) and a central
scintillating fiber tracker (CFT), both located within a 1.9~T superconducting solenoidal magnet,
with designs optimized for tracking and vertexing at pseudorapidity $|\eta_{\text{det}}|<3$ and 
$|\eta_{\text{det}}|<2.5$~\cite{d0_coordinate}, respectively, as shown in Fig.~\ref{fig:dzero_cft}.
Three liquid-argon and uranium calorimeters provide coverage of
$|\eta_{\text{det}}|<3.5$ for electrons.
The central calorimeter (CC) contains the region $|\eta_{\text{det}}| < 1.1$, and two end calorimeters
(EC) extend coverage to $1.5<|\eta_{\text{det}}|<3.5$,
as shown in Fig.~\ref{fig:dzero_cal}.
In the region $1.0 <|\eta_{\text{det}}| < 1.5$, particles 
cross multiple cryostat walls resulting in deterioration of the electron response.
Each calorimeter consists of an inner electromagnetic section (EM)
followed by a hadronic section. The EM calorimeter 
has four longitudinal layers with transverse segmentation of
$\Delta \eta \times \Delta \phi = 0.1\times 0.1$, except for the third layer, where it is $0.05\times 0.05$.
The outer muon system consists of a layer of tracking detectors and scintillation trigger
counters in front of 1.9 T iron toroids, followed by two similar layers after the toroids,
with a coverage of $|\eta_{\text{det}}|<2$. 
The direction of the D0 solenoid and toroid magnetic fields were reversed periodically
during data taking.

The D0 trigger is based on a three-level system. The first level
consists of hardware and firmware components, and the second level combines
information from specific subdetectors to construct a trigger decision based on 
physics quantities.
The software-based third level processes the full event information 
using simplified versions of the offline reconstruction algorithms. 

\section{Event selection}
The $W\rightarrow e\nu$ events for this analysis are selected in several steps. 

\subsection{Trigger selection}
Candidate events must pass at least one of the calorimeter-based single EM triggers.
The trigger towers in the calorimeter are 
$0.2\times 0.2$ in $(\eta, \phi)$ space. At the third trigger level, the EM trigger objects 
must satisfy $E_T^e(\text{trigger})>25$~GeV, or $E_T^e(\text{trigger})>27$~GeV at higher instantaneous luminosity.

\subsection{Lepton transverse energy selection}
We require one EM shower with
transverse energy $E_T^e >25$~GeV measured in the calorimeter,
accompanied by $\met > 25$~GeV. In $W$ boson events, $\met$ is calculated using 
the electron and the vector sum of the transverse components of
the energy deposited in the calorimeter ($u_T$) after subtracting the electron deposit, 
i.e., $\vec{\met}=-(\vec{E_T^e}+\vec{u_T})$.
We also require that the electron has $E_T^e<100$~GeV to ensure 
good charge identification using the momentum of the charged track, described below.

\subsection{Electron selection}
The EM cluster must be in the CC with $|\eta_{\text{det}}| < 1.1$
or in the EC range $1.5<|\eta_{\text{det}}|<3.2$ to allow a precise measurement of its energy.
Electron candidates must be located within the fiducial region of each of the 32 EM calorimeter modules, 
defined as $0.1<\phi_{\text{mod}}<0.9$, where $\phi_{\text{mod}}$ 
is the fractional part of $32\cdot\phi_{\text{trk}}/2\pi$.
The electron energy must be isolated in the calorimeter with
$\left[E_{\text{tot}}(0.4) - E_{\text{EM}}(0.2)\right]/E_{\text{EM}}(0.2)<0.15~(0.10)$ for CC (EC) electrons,
where $E_{\text{tot}}(\cal{R})$ and $E_{\text{EM}}(\cal{R})$ are the total energy and the energy
deposited in the EM section, respectively, within a cone of radius 
$\cal{R}$=$\sqrt{(\Delta \phi)^{2}+(\Delta \eta)^{2}}$ around the electron direction. 
Electron candidates are further required to have at least 90\% of their energy 
deposited in the EM section of the calorimeter
and to have a shower shape (H-matrix)~\cite{hmx, tag_and_probe} consistent with that expected for an electron.

Electron candidates are required to be spatially matched to a reconstructed track
by requiring $|\Delta \eta|<$ 0.05 and $|\Delta \phi|<$ 0.05,
where $|\Delta \eta|$ and $|\Delta \phi|$ are the differences in $\eta$ and $\phi$ between 
the cluster centroid and the extrapolated track.
To reduce the electron charge misidentification probability, the track is further required 
to be of good quality: the track transverse momentum ($p_T^{\text{trk}}$) must be greater than 10~GeV,
the track must pass the central track fitting quality requirement, and the distance of closest approach (dca)
of the track to the beam spot in the plane transverse to the beam direction should be
less than 0.02~cm.
Because the CFT detector does not cover the entire $\eta_{\text{det}}$ region used in the analysis, 
electrons are split into four categories: CC electrons with full CFT coverage, 
EC electrons with full CFT coverage, 
EC electrons with partial CFT coverage, and EC electrons without CFT coverage.
Optimized track quality requirements are employed for the different track categories. 
For tracks with full CFT
coverage, we require that the track must have at least two SMT hits and nine CFT hits.
For tracks with partial CFT coverage, we require that the track must have at least two SMT hits and 
three CFT hits. Finally, for tracks without CFT coverage, we require that the track must have 
at least eight SMT hits and pass a track significance 
($1/p_T^{\text{trk}}\over\sigma (1/p_T^{\text{trk}})$) selection requirement,
where $\sigma (1/p_T^{\text{trk}})$ is the uncertainty on $1/p_T^{\text{trk}}$ due to uncertainties on the tracking
system hit positions. 

\subsection{$W$ boson event selection}
Events are required to have a reconstructed $p\bar{p}$ interaction vertex within
40~cm of the detector center along the $z$ axis
and a reconstructed $W$ boson transverse mass of $50<M_T<130$~GeV,
where $M_T = \sqrt{2E^e_T\met (1-\cos\Delta \phi)}$, and $\Delta \phi$ is
the azimuthal angle between the electron and $\vec{\met}$.
We require $u_T<60$~GeV.
The variable {\it SET} reflects the total activity in the calorimeter and is defined as the scalar 
sum of all of the transverse energy components measured by the calorimeter
except those associated with electron.
Events are required to have 
either ${\it SET} <250$~GeV or ${\it SET} <500$~GeV, where the higher 
{\it SET} threshold is employed for the higher luminosity data-taking periods.

After applying the selection criteria described above, we retain 6,083,198 $W$ boson candidates. 
Of these, 4,466,735 are events with an electron in the CC region, and 1,616,463 
have an electron in the EC region.
The electron charge asymmetry is determined for each of the four electron categories based
on CFT coverage and the results are then combined. Results from different data collection periods are
found to be consistent with each other and are also combined.
We assume charge parity (CP) invariance in the $W$ boson production and decay, and thus report the
folded asymmetry $A(|\eta^e|) = {1\over 2} \left[A\left(\eta^e>0\right) - A\left(\eta^e<0\right)\right]$.
The electron charge asymmetries are measured in thirteen pseudorapidity bins
in the range $|\eta^e| < 3.2$. The bin widths are chosen considering the 
statistics of the sample and the geometry of the detector. 
The selection criteria are identical to those employed in the 
$W$ boson charge asymmetry paper, Ref.~\cite{d0_w},
which also used the entire Run II data set in the electron channel.

%%%%%%%%%%%%%%%%%%%%%%%%%%%%%%%%%%%
\section{Signal and background simulation}

\subsection{Signal}
MC simulations for the $W\rightarrow e\nu$
process are generated using the {\sc pythia}~\cite{pythia} event generator with
CTEQ6.1L PDFs~\cite{cteq},
followed by a detailed {\sc geant}-based simulation~\cite{geant} of the D0 detector response
and overlay of zero-bias events. Zero-bias events are selected from random beam crossings
matching the instantaneous luminosity profile in the data.
This simulation is then improved by correcting for known deficiencies in 
the detector model and for higher-order effects not included in {\sc pythia}.

{\sc pythia} is a leading-order (LO) generator in which the modeling of the $W$ boson $p_T$ is not adequate for 
electroweak (EW) precision measurements. In order to improve the model of the $W$ boson $p_T$, 
we derive a next-to-leading-order (NLO) correction from the ratio of {\sc resbos}~\cite{resbos}
with {\sc photos}~\cite{photos} (to simulate final state radiation, FSR) 
using the CTEQ6.6 central PDF set to {\sc pythia} with the CTEQ6.1L PDF set,
as a function of the $W$ boson $p_T$ and rapidity.

\subsection{MC electron identification efficiency correction}
The MC does not adequately describe the electron identification in the data, and the data and MC 
discrepancies as a function of $\eta^e$ in the forward region are larger than they are in the central region. 

$Z\rightarrow ee$ boson events from data and MC are used to calculate 
electron identification (EMID) corrections using a tag-and-probe
method~\cite{tag_and_probe}. 
In this method, an electron candidate passing tight identification requirements 
is chosen as the tag electron, and then the probe electron is selected
by requiring the invariant mass of the two electrons ($M_{ee}$) to satisfy $70 < M_{ee} < 110$~GeV. 
Probe electrons from this high purity, minimally biased electron sample
are used to tune the MC selection efficiencies.

\begin{center}
\begin {figure*}[htb]
  \epsfig{file=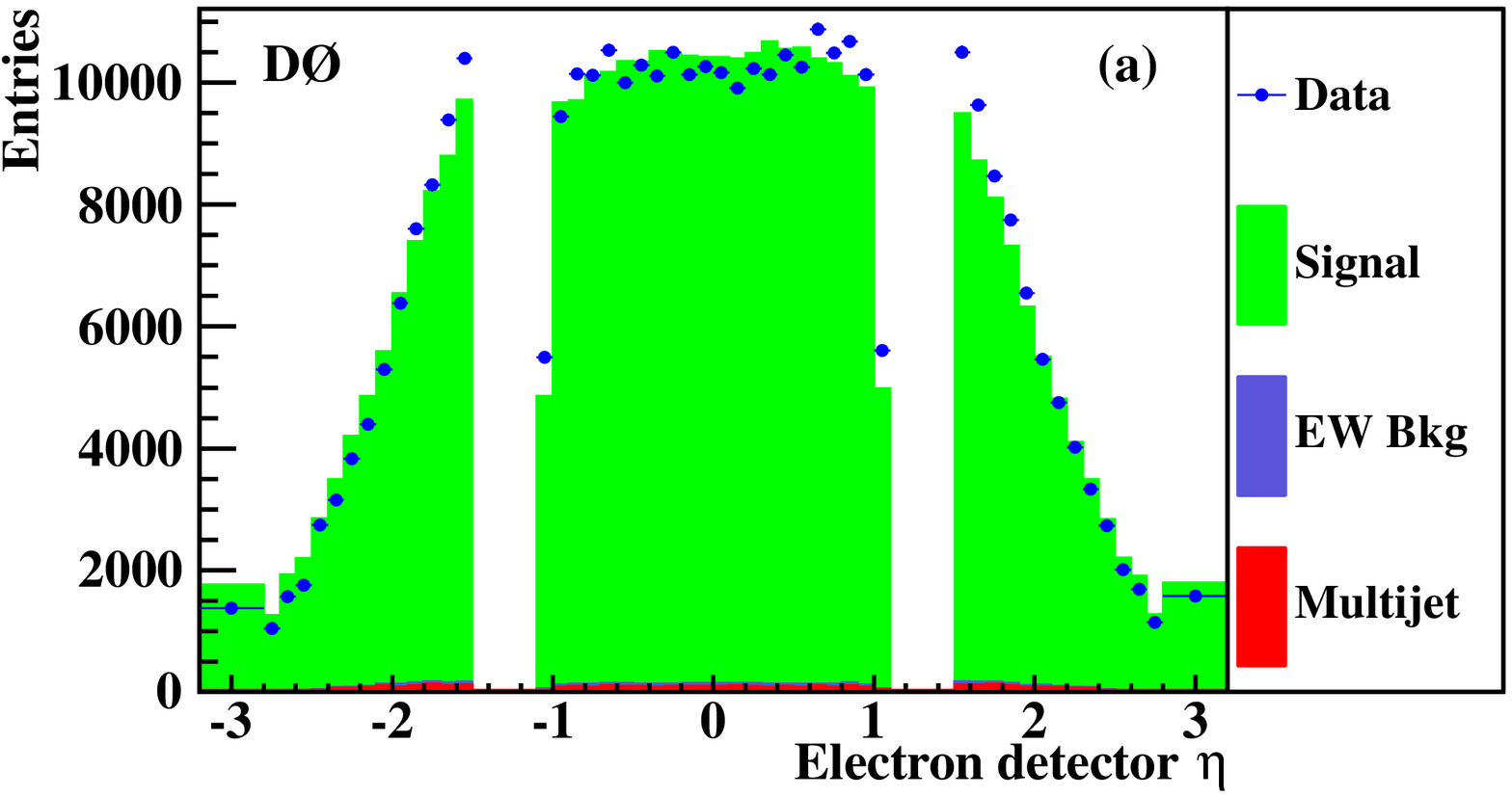,scale=0.44}
  \epsfig{file=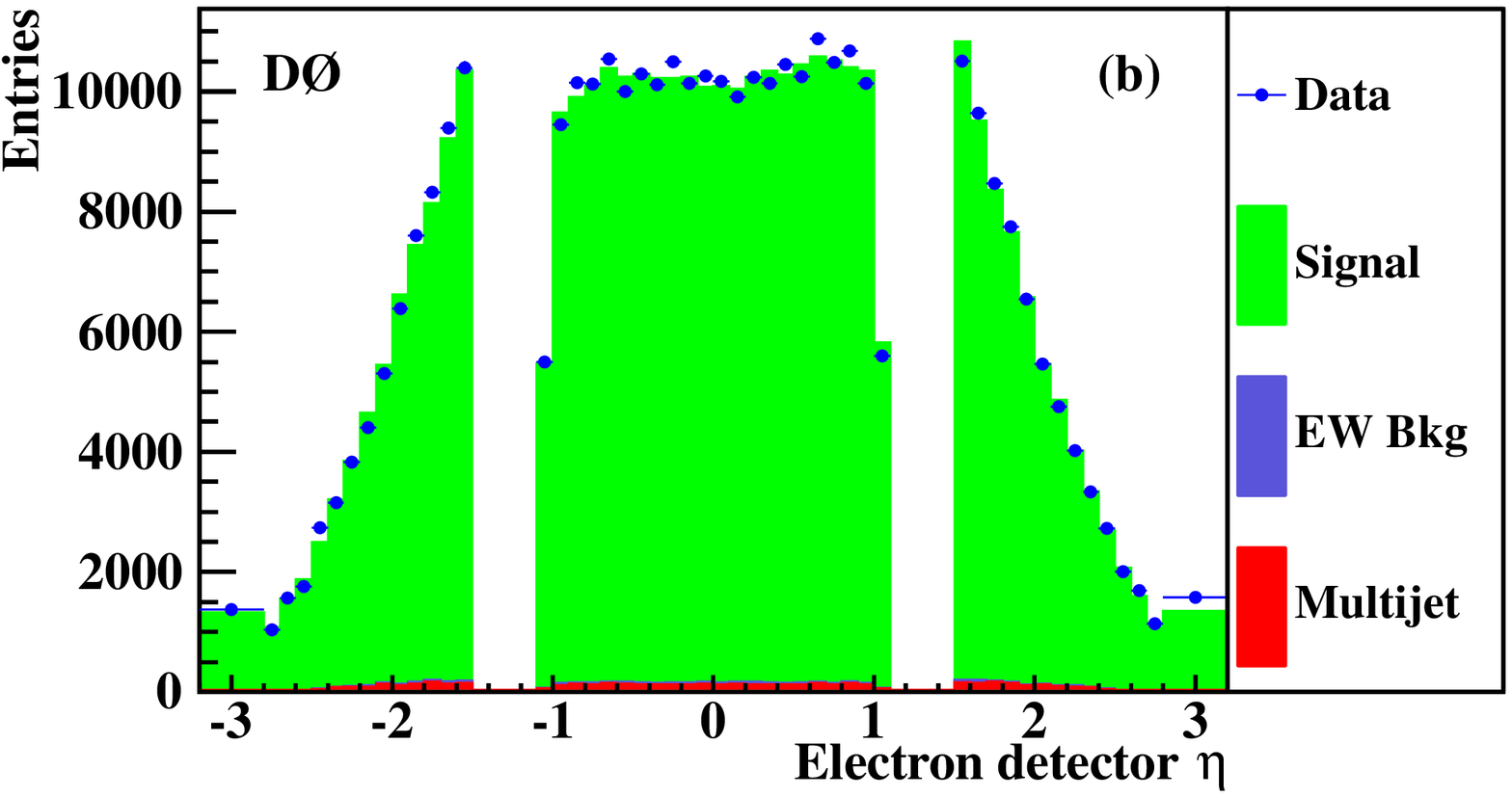,scale=0.44}
  \epsfig{file=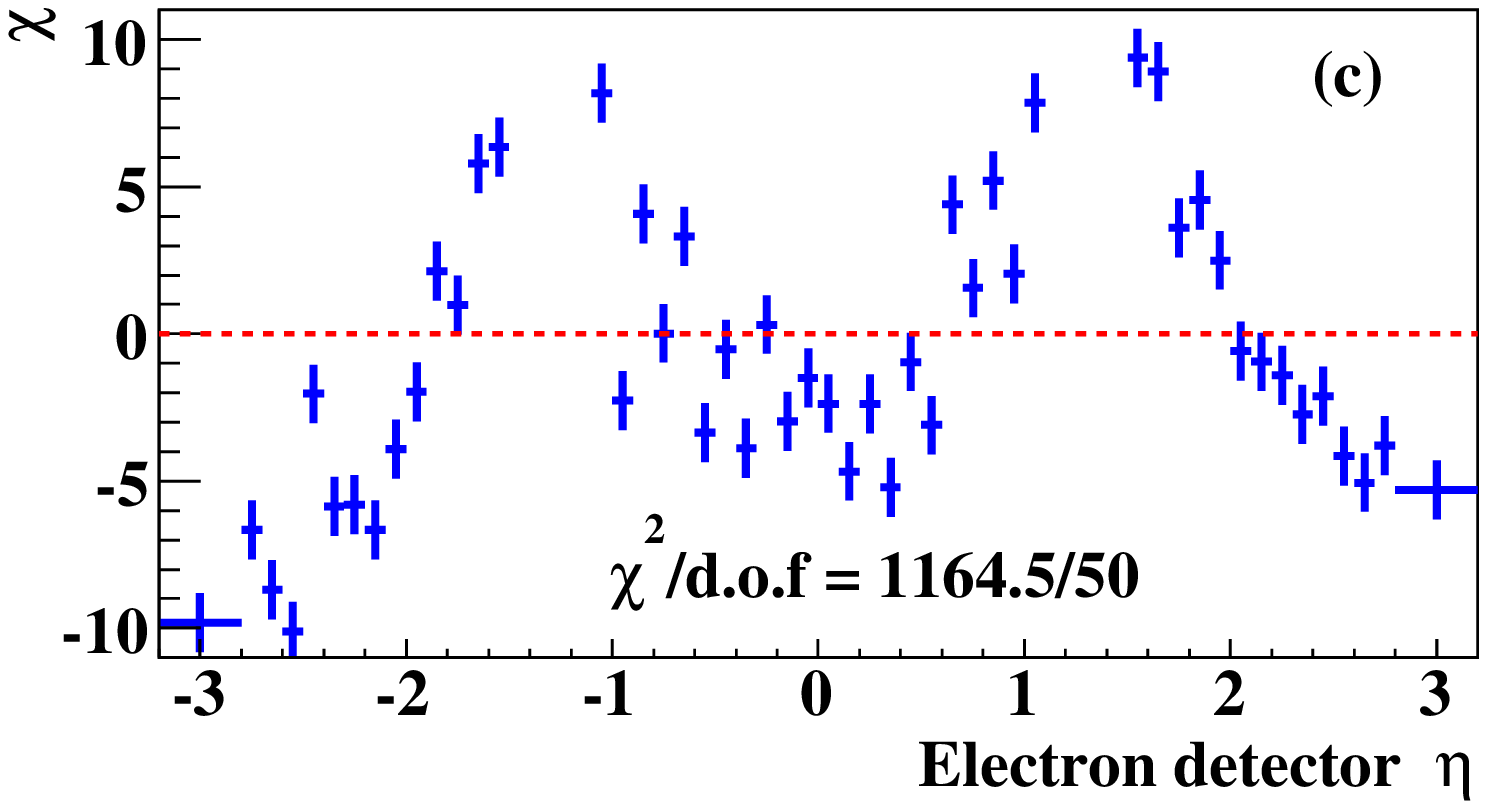,scale=0.44}
  \epsfig{file=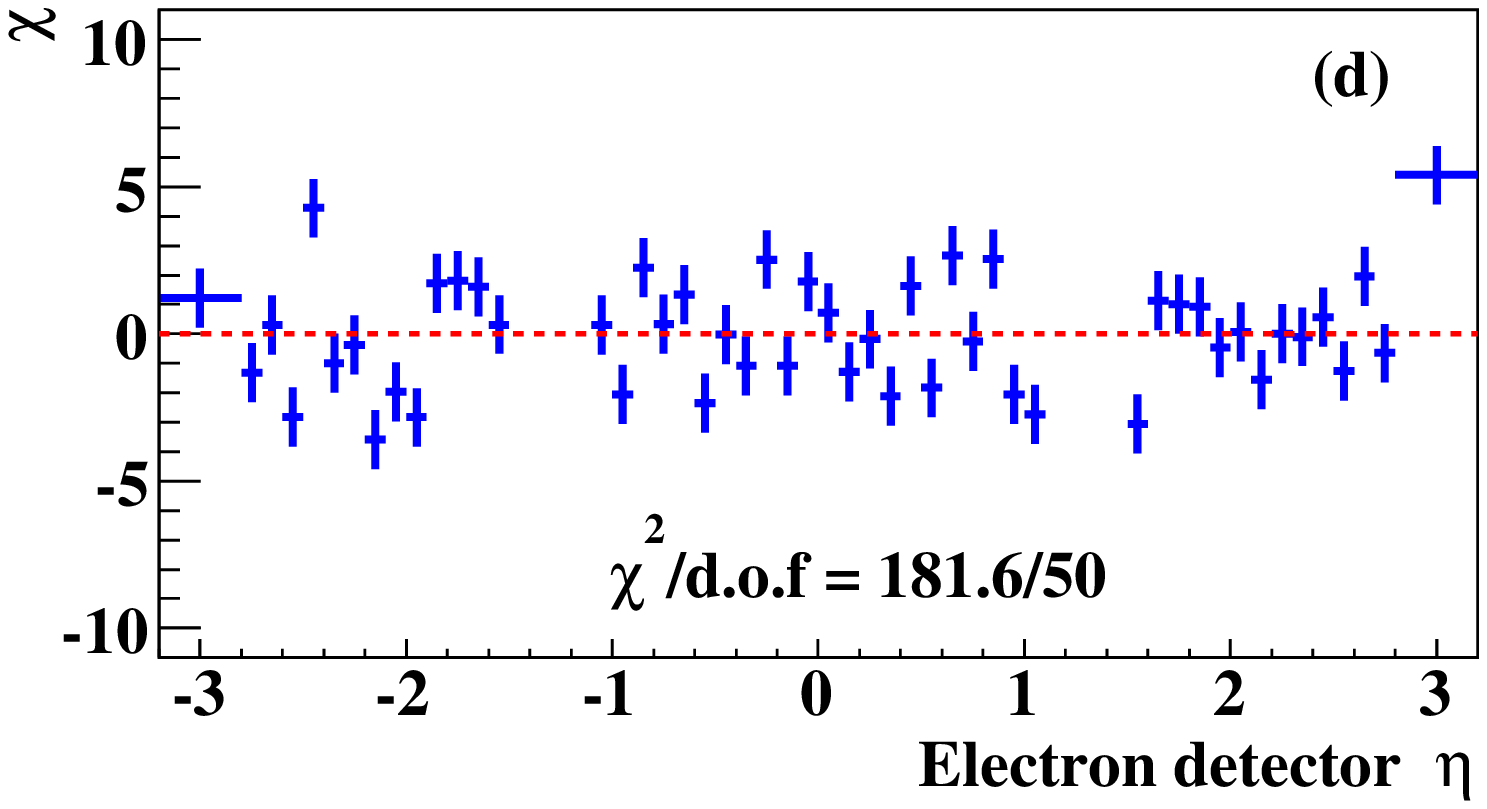,scale=0.44}
  \caption{(color online). Comparisons of the electron $\eta^e_{\text{det}}$ distributions between data and the 
sum of signal and background predictions for selected $Z$ boson events, (a) event distribution and 
(c) value of $\chi_i$ for each bin between data and the MC predictions before
applying the EMID correction, (b) event distribution and (d) value of $\chi_i$ for each bin after
applying the EMID correction.
$\chi_i=\Delta N_i/\sigma_{N_i}$, $\Delta N_i$ is the difference between
the number of data and that of the MC prediction, and $\sigma_{N_i}$ is the statistical uncertainty in each $\eta$ bin.}
\label{fig:zee_comp}
\end{figure*}
\end{center}

To remove the EMID differences between data and MC, 
we apply bin-by-bin efficiency corrections to the MC.
There are multiple dependencies for the corrections, 
particularly for electrons in the forward region.
In the procedure, the corrections are applied as functions of 
electron physical $\eta^e$ (measured with the event vertex), electron detector $\eta$ ($\eta^e_{\text{det}}$),
electron $E_T^e$, electron $\phi$, vertex position in the $z$ direction ($z_{\rm vtx}$), {\it SET}, 
and instantaneous luminosity ($L$) for three selections: the pre-selection 
(Pre-Selection, EM cluster isolation cut), calorimeter-based selection (Cal-ID), 
and track-based selection (Track-Match).
As the number of selected $Z\rightarrow ee$ events is limited, we perform a four-step iterative correction to reduce 
the selection differences between data and MC. As shown in Table~\ref{tab:emid}, 
we first derive a two-dimensional (2-D) correction to remove
the two largest dependencies ($\eta^e_{\text{det}}$ and $E_T^e$). 
Then, using this 2-D correction, we examine the other parameters dependences
and develop a new 2-D correction to remove the largest two remaining dependencies. 
We iterate two more times until all EMID selection data-MC differences are greatly reduced.
The electron $\eta^e_{\text{det}}$ distributions of selected $Z$ boson events before and after applying the EMID 
correction are shown in Fig.~\ref{fig:zee_comp}. 
Reasonable agreement is also observed for $E_T^e$, 
$\eta^e$, $z_{\rm vtx}$, $\phi$, $L$, {\it SET}, and $M_{ee}$
distributions for selected $Z$ boson events after applying EMID corrections.

\begin{table}[htb]
\begin{center}
\caption{Dependencies on the four steps used to determine EMID correction.}
    \begin{tabular}{ccr@{,}lr}
    \hline\hline
      Step & Pre-Selection & \multicolumn{2}{c}{Cal-ID/Track-Match}  \\  \hline
      1    & $\eta^e$    & $\eta^e_{\text{det}}$ & \ $E_T^e$       \\ 
      2    &                                      & $\eta^e$ & \ $z_{\rm vtx}$    \\
      3    &                                      & $\phi$ & \ $L$                          \\ 
      4    &                                      & {\it SET} & \ $E_T^e$                         \\
      \hline\hline
     \end{tabular}
\label{tab:emid}
\end{center}
\end{table}

\subsection{Electron trigger efficiency correction}
We apply the trigger efficiency measured from data
to the MC sample. To estimate the single EM trigger efficiency, 
we use $Z\rightarrow ee$ data and apply the tag-and-probe method. 
The trigger efficiency correction is applied to MC events, as a function of 
$E_T^e$ and $\eta_{\text{det}}^e$, separately for both CC and EC electrons.

 \subsection{Positron/electron efficiency correction} 
The efficiencies for $e^+$ and $e^-$ identification in data and MC
differ, with some difference for the two solenoid polarities also observed. The
effect of different efficiencies for the two magnet polarities is
ameliorated by the fact that the negative and positive solenoid polarity
samples are nearly equal in size. For both data and MC, using a sample
of $Z\rightarrow e^+e^-$ events and a tag-and-probe method, we measure the
identification efficiencies for all four combinations of particle charges ($q$) and
solenoid signs ($p$), and calculate the data and MC efficiency ratio corrections ($K_{\text{eff}}^{q, p}$)
as a function of $\eta^e$ and $E_T$.
For each of these combinations, the MC events are
reweighted to provide agreement with data. 
Figure~\ref{fig:keff} shows the comparison of MC and data after the correction
for positrons with positive solenoid polarity. 

\begingroup
\begin{figure}
\epsfig{file=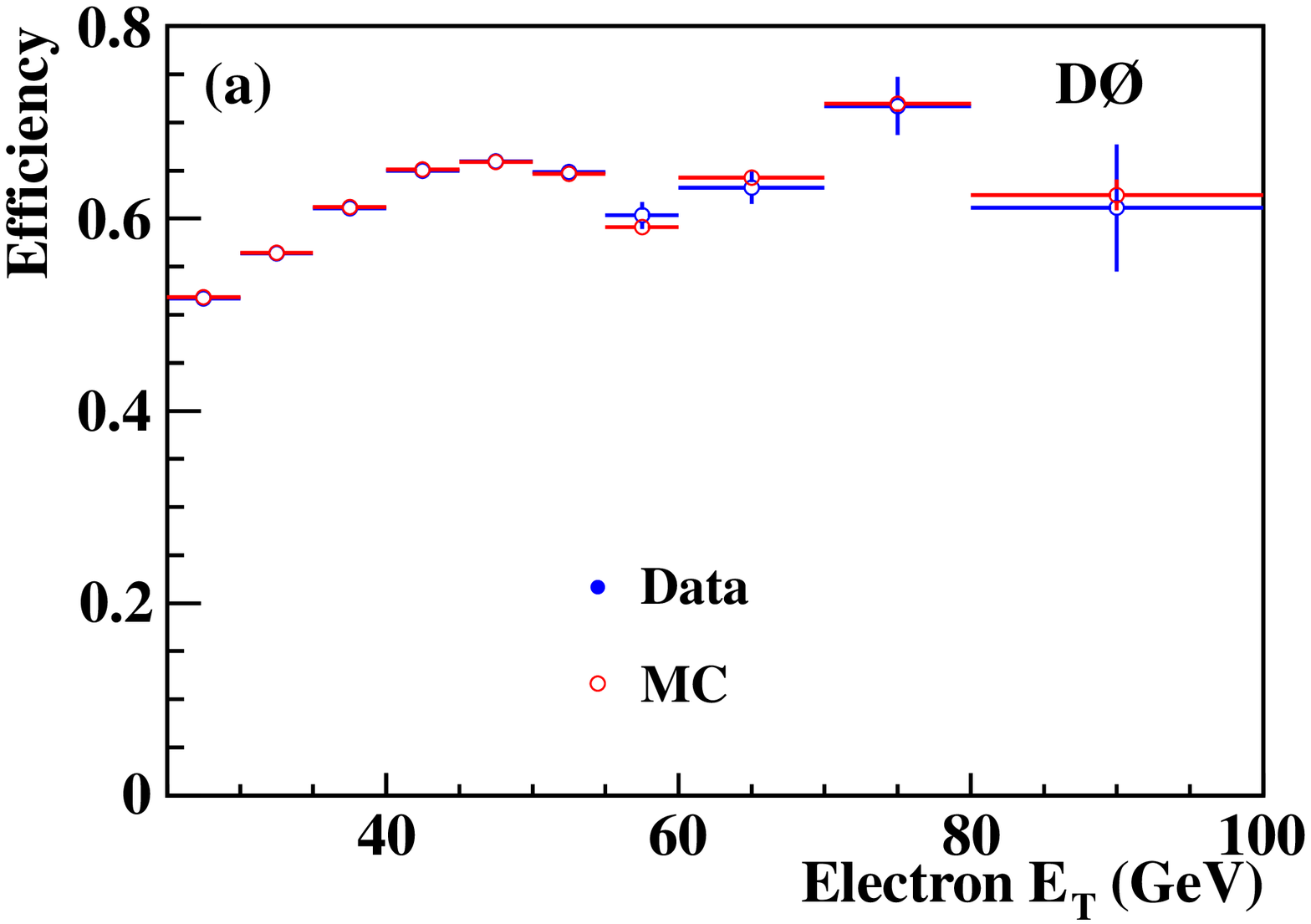,scale =0.45}
\epsfig{file=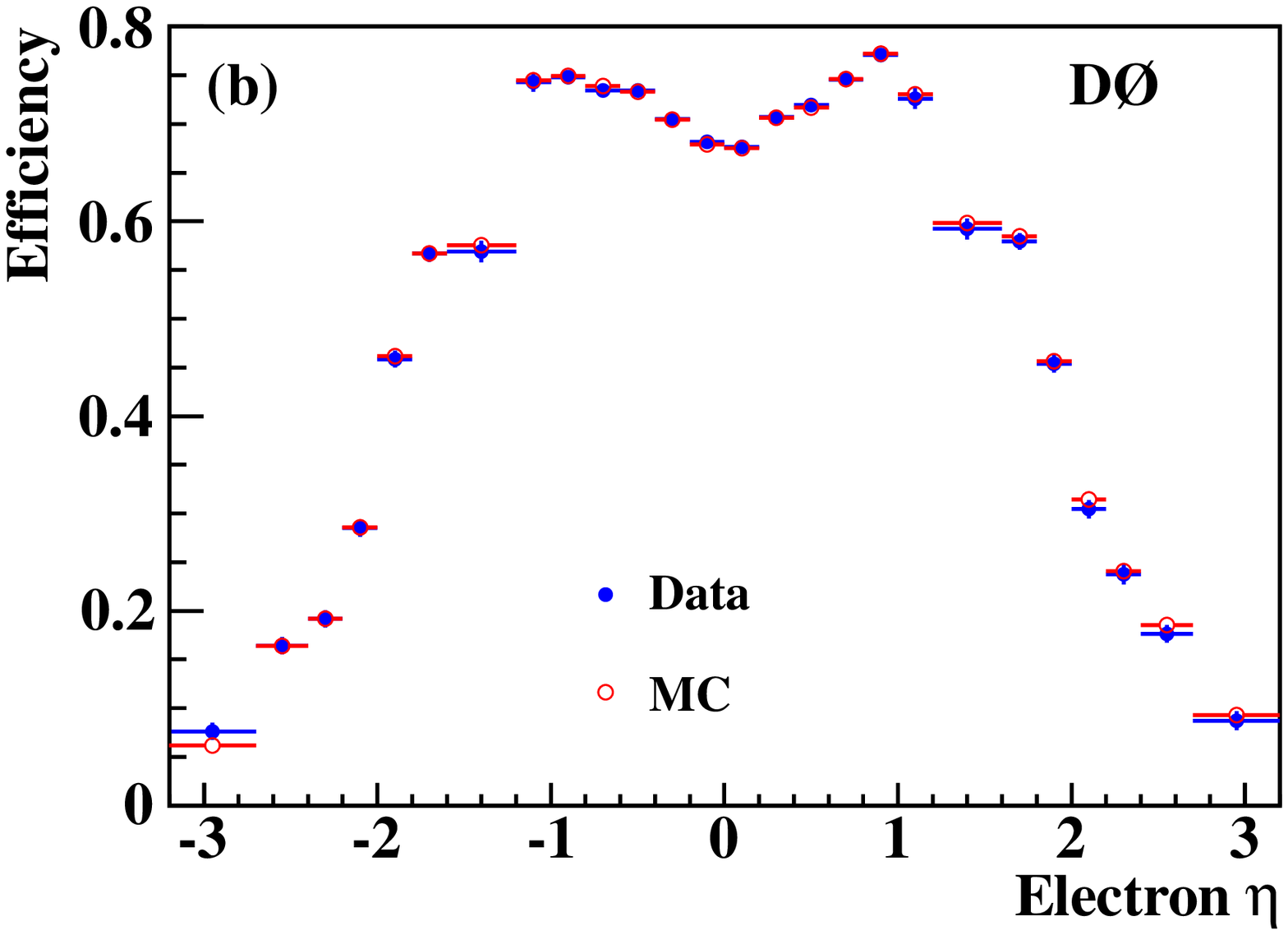,scale =0.45}
\caption{ (color online). Data and MC track matching efficiency comparison after $K^{q, p}_{\text{eff}}$ correction, as a function of (a) electron $E_T$ and (b) electron $\eta$, for $e^+$ with positive solenoid polarity. 
There is an efficiency drop around 55 GeV, which comes from the tag-and-probe method, since in
$Z\rightarrow ee$ events, when the tagged electron has high $E_T$, the other electron in 
the event is soft, resulting in inefficiency in the EMID.}
\label{fig:keff}
\end{figure}
\endgroup

\subsection{Electron energy tuning}
\label{sec:em_energy}
The mismodeling of the passive material in front of the calorimeter results in
energy mismeasurement for electrons. However, there are additional causes of
electron energy mismeasurement. 
The interaction rate of proton and antiproton bunch crossings depends on instantaneous luminosity. 
Events with higher instantaneous luminosity may have more energy deposited in the calorimeter
due to pile-up contributions.
In addition, the {\it SET} will contribute to the electron energy measurement
by adding from a few MeV to a few GeV to the electron energy.
The electron energy reconstruction, especially in the forward region, has strong 
$\eta^e_{\text{det}}$, instantaneous luminosity, and {\it SET} dependences. 
The interplay of these three effects makes a precision measurement of the energy challenging.
To derive a correction, we fit $Z$ boson events in different $\eta^e_{\text{det}}$ bins 
using a {\sc voigt} function~\cite{voigt} combined with an exponential background
to obtain the $Z$ boson mass peak position, and compare the mass peak position with 
the LEP value (91.1876~GeV)~\cite{lep_z}. 
In the mass peak fitting, the multijet background and other SM backgrounds are subtracted.
As shown in Fig.~\ref{fig:zpeak_tune}, there are deviations of more than 2~GeV in the 
value of the $Z$ boson mass peak in the very forward bins before calibration.

An iterative method using {\sc minuit}~\cite{minuit} fitting is employed to reduce 
the electron energy dependences on instantaneous luminosity, {\it SET} and $\eta^e_{\text{det}}$. 
The procedure includes: 

\begin{itemize}
 \item{Instantaneous luminosity tuning: 
 The dependence of the peak position of the $Z$ boson mass on increasing luminosity 
 includes several effects:
 (a) the addition of energy from pile-up and hadronic recoil energy in the electron 
 reconstruction window, (b) the decrease of the energy response due to the high voltage 
 drop on the resistive coating~\cite{d0det0, wmass} on the calorimeter electrodes due to the increased
 ionization at high luminosity, (c) and the decrease of the 
 response due to the over-subtraction of the baseline of
 the signal shape in the calorimeter~\cite{wmass}.
 %The resistitivity of this coating was measured {\it in situ}, at the temperature of liquid argon, to be of
% the order of 180 $M\Omega$ per square, with a large spread from one board to another.
For MC, overlaid zero-bias events contribute to the energy of the electron, 
with high instantaneous luminosity causing a corresponding increase 
of the value of the $Z$ boson peak position.
Thus, different correction factors in sixteen luminosity bins are applied to data and MC, 
according to the instantaneous luminosity of the event.}

 \item{{\it SET} tuning: The {\it SET} affects the electron energy by contributing 
additional energy to the electron shower. The correction factors in thirteen {\it SET} bins are
developed and applied to data and MC, according to the {\it SET} of the event.}

 \item{$\eta^e_{\text{det}}$ tuning: For $Z\rightarrow ee$ events, there are two electrons which will most likely 
be located at different $\eta^e_{\text{det}}$ positions. When tuning the electron energy modeling for
a specific $\eta^e_{\text{det}}$ bin, the tuning is affected by electron energy modeling in 
other $\eta^e_{\text{det}}$ bins,
thus there are strong correlations between bins. 
The procedure employs 44 (CC) or 72 (EC) parameters (22 $\eta^e_{\text{det}}$ bins in the CC region, 
with scale ($\alpha$) and offset ($o$) parameters for each $\eta^e_{\text{det}}$ bin, 
as $E^e_{\text{cor}} = o + \alpha \times E^e$,
where $E^e$ and $E^e_{\text{cor}}$ are the electron energy before and after energy tuning.
There are 24 $\eta^e_{\text{det}}$ bins in the EC region, with scale ($\alpha$), 
offset ($o$), and non-linearity ($\gamma$)
parameters for each $\eta^e_{\text{det}}$ bin, 
as $E^e_{\text{cor}} = o + \alpha \times E^e + \gamma \times (E^e)^2$). 
To take into account substantial differences in statistics between different $\eta^e_{\text{det}}$ bins
and to speed up the procedure, we employ iterative fitting instead of a global fit:

\begin{enumerate}
 \item{Fit the events in the $\eta^e_{\text{det}}$ bin with the largest
statistics (i.e., $0<\eta^e_{\text{det}} <0.1 $ for CC electrons
 and $1.5<\eta^e_{\text{det}}<1.6$ for EC electrons).}
 \item{Fix the parameters for the $\eta^e_{\text{det}}$ bin fit in the previous step, and then fit the events in 
 the next $\eta^e_{\text{det}}$ bin (i.e., $0.1<\eta^e_{\text{det}}<0.2$ for CC electrons
 and $1.6<\eta^e_{\text{det}}<1.7$ for EC electrons).}
 \item{Repeat Step 2 for each $\eta^e_{\text{det}}$ bin.}
 \item{Repeat Steps 1--3 until the fitting results become stable, with a minimum $\chi^2$ value between
 the fitted $Z$ boson mass peak values in each bin and that of the LEP value.}
\end{enumerate}
}
\end{itemize}

The position of the $Z$ boson peak in bins of electron $\eta^e_{\text{det}}$ 
before and after the electron energy tuning 
is shown in Fig.~\ref{fig:zpeak_tune}, demonstrating that good consistency 
is obtained between the LEP measured value~\cite{lep_z} 
and the fitted mass value of the $Z$ boson mass peak after tuning.

After applying the electron energy scale correction, an additional 
energy smearing correction~\cite{tag_and_probe} is applied
to the MC to achieve data-MC agreement for the energy resolution.

\begingroup
\begin{figure}
\epsfig{file=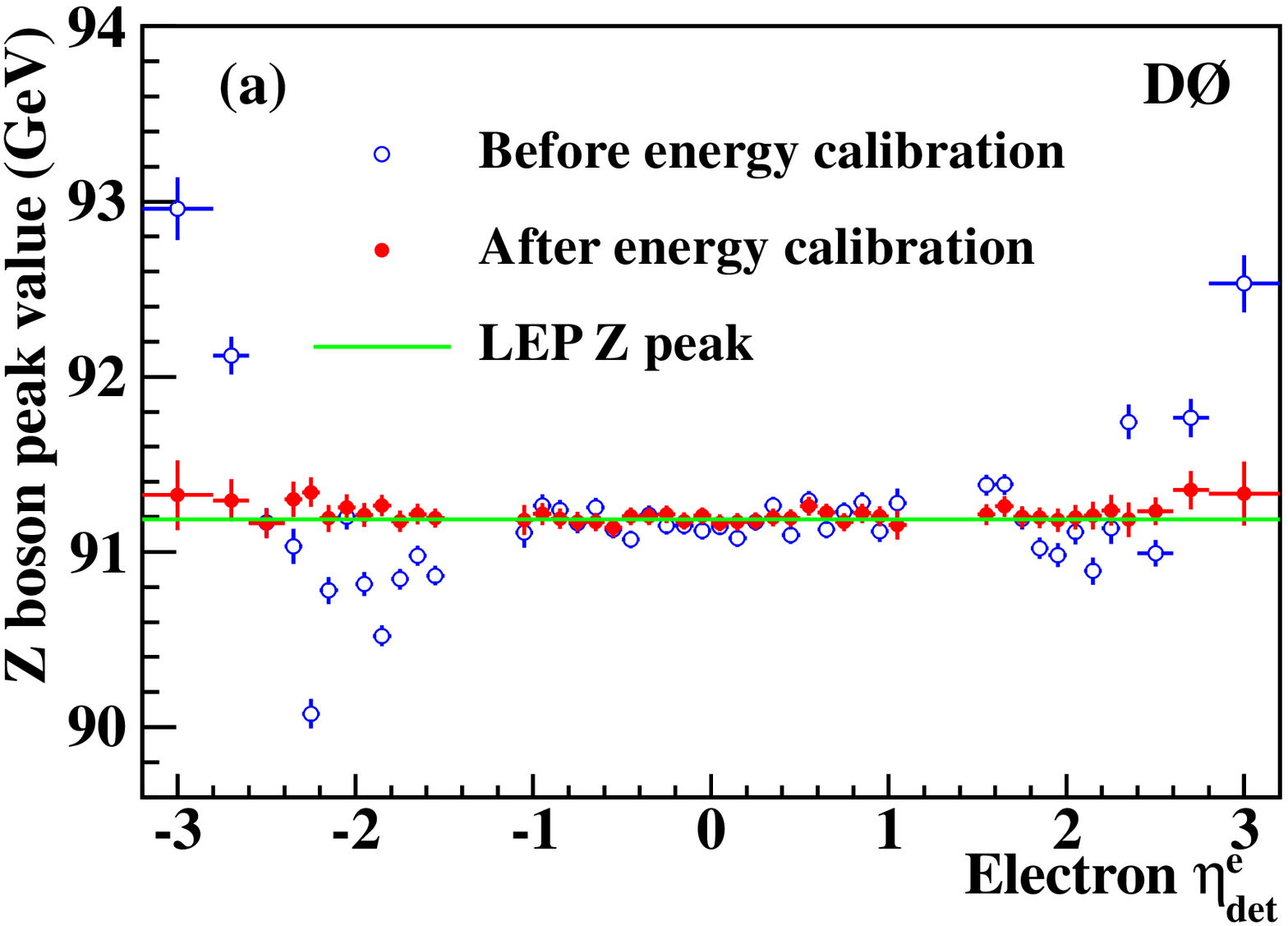, scale=0.4}
\epsfig{file=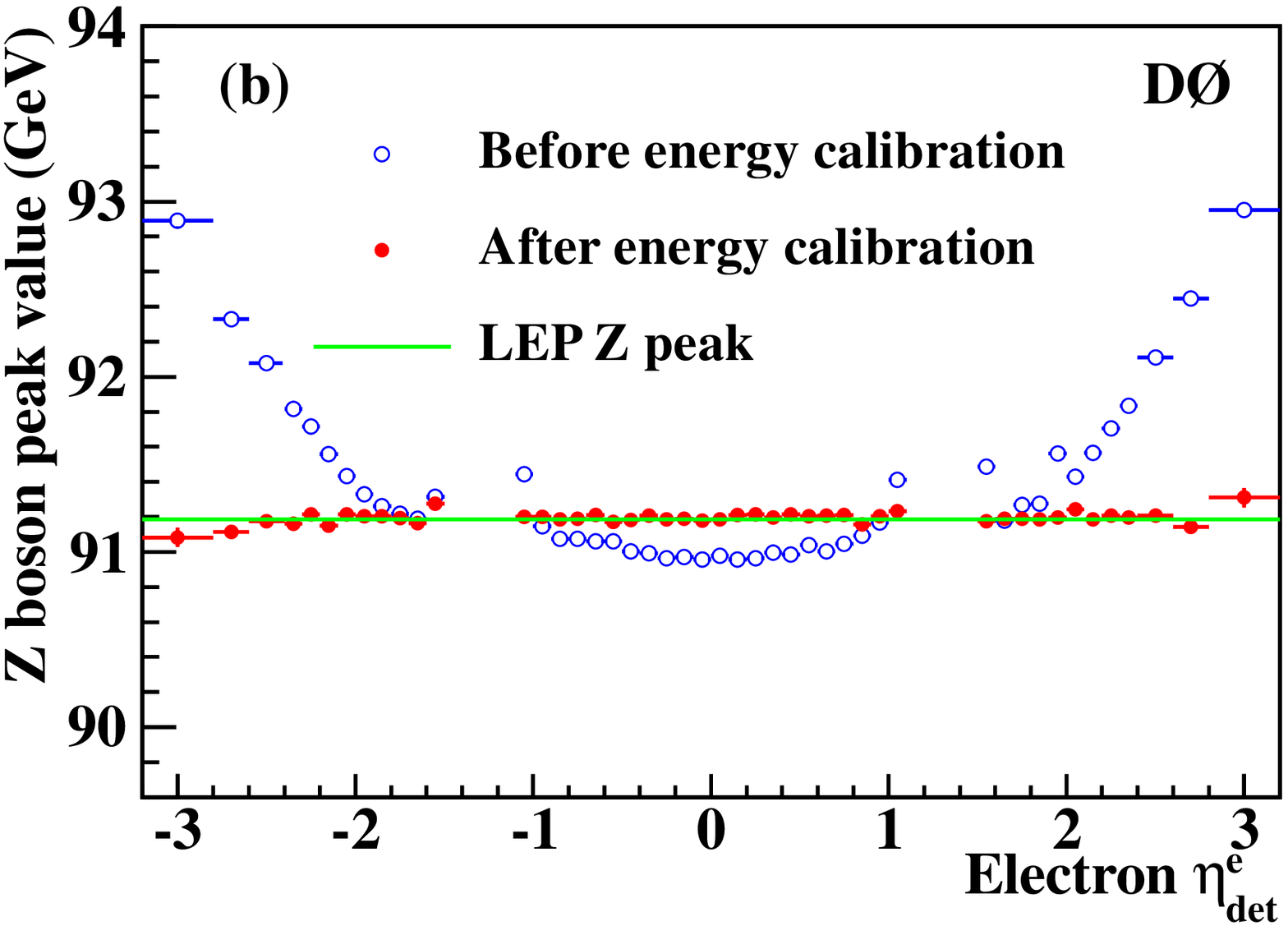, scale=0.4}
\caption{(color online). The fitted mass value of CC-EC events (events with one electron 
in the CC and the other in the EC) in $\eta^e_{\text{det}}$ bins for 
(a) data and (b) MC $Z\rightarrow ee$ events. The open blue points are the $Z$ boson mass peak values
before applying the electron energy calibration, and the solid red points represent the peak
values after applying the electron energy calibration.}
 \label{fig:zpeak_tune}
\end{figure}
\endgroup

\subsection{Recoil system tuning}
\label{sec:recoil}
We also correct the energy response in MC for the hadrons recoiling 
against a $W$ or $Z$ boson.
The recoil system model is needed to determine the $\met$ in $W$ boson events and
is a key component for the electron charge asymmetry measurement.
The response of the calorimeter to the hadronic recoil differs from its response to objects which
shower electromagnetically. This difference occurs because the hadronic calorimeter modules 
differ in construction from the electromagnetic modules and because the process by which hadrons interact in 
material is different from that of electrons and photons.
In principle, if we knew the particle composition of the recoil, it would be possible to simulate the overall recoil
response. However, there is no reliable model to estimate from first principles the particle composition
of the recoil system. Furthermore, many of the recoil particles have low momentum, and 
the energy scale corrections are difficult to calculate for low energy particles.

In this analysis, the hadronic response is directly determined from $Z\rightarrow ee$ data
by comparing the $Z$ boson transverse momentum ($p_T^Z$) measured from the electron pair ($p_T^{ee}$)
to that measured from the recoil system ($u_T$). The particle composition in
the $W$ and $Z$ boson recoil
systems should be very similar, and by averaging over the $Z$ boson sample, we expect to 
derive a hadronic response model that closely approximates that of the $W$ boson sample.

To perform this comparison, a pair of coordinate axes in the transverse plane to the beam is used. 
As shown in Fig.~\ref{fig:eta_xi},
the $\eta$ axis is defined as the inner bisector of the two electron transverse momentum directions,
and the $\xi$ axis is perpendicular to the $\eta$ axis in the transverse plane. 
The $\eta$ direction is defined using electron angle information only, therefore, 
the recoil projection in the $\eta$ direction is minimally sensitive to the 
electron energy resolution.
The projections onto the $\eta$ and $\xi$ axes are denoted $p_{T\eta}^{ee}$,
$p_{T\xi}^{ee}$, $u_{T\eta}$, and $u_{T\xi}$.
The projection of any transverse momentum
$\vec{p_T}$ onto these axes is 
\begin{eqnarray}
 \vec{p_T} = p_{\eta}\hat{\eta}+p_{\xi}\hat{\xi}.
\end{eqnarray}

\noindent The $\eta$ and $\xi$ projections 
enable a good understanding of
the hadronic response by comparing $p_{T\eta}^{ee}$ with $u_{T\eta}$. The momentum 
vectors of the dielectron and hadron systems should be equal and opposite due to momentum conservation.

\begingroup
 \begin{figure}
 \epsfig{file=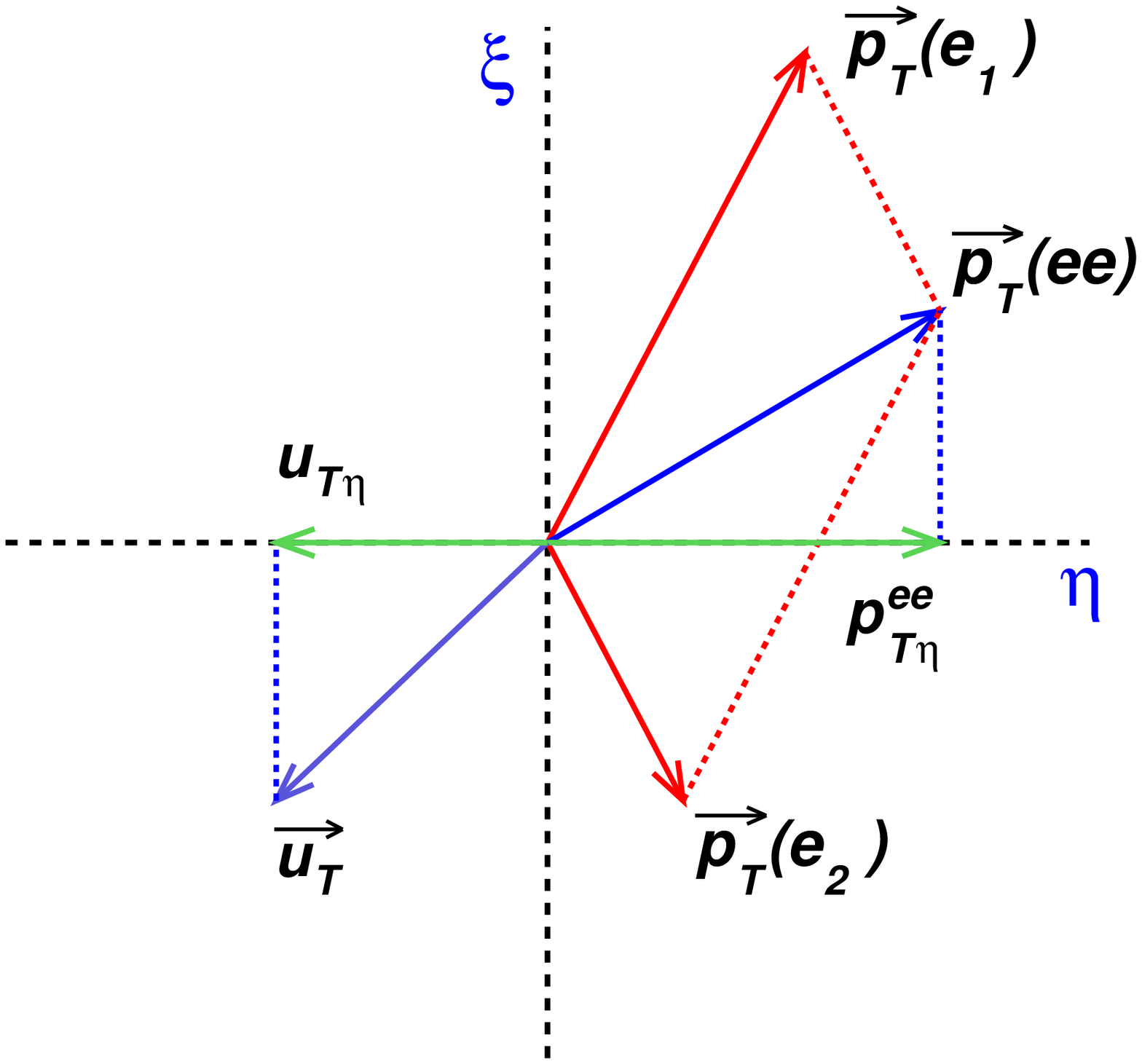,scale=0.45}
  \caption{(color online). Definitions of the $\eta$ and $\xi$ axes in $Z\rightarrow ee$ events, and the dielectron
 and the hadronic recoil system projections in these axes. The $\eta$-$\xi$ plane is transverse to the beam.}
 \label{fig:eta_xi}
 \end{figure}
 \endgroup
 
To improve the recoil modeling in the MC, we determine the 
hadronic scaling, smearing, and offset factors ($\alpha$, $\beta$, and $o$) to MC samples
using {\sc minuit} fitting, as
\begin{eqnarray}
X_{\eta}^{\text{new}} &=& \alpha\times[X_{\eta}^{\text{gen}} + (X_{\eta}^{\text{raw}} - X_{\eta}^{\text{gen}})\times\beta] + o, \nonumber \\
 X_{\xi}^{\text{new}} &=& \alpha\times[X_{\xi}^{\text{gen}} + (X_{\xi}^{\text{raw}} - X_{\xi}^{\text{gen}})\times\beta] + o.
\end{eqnarray}
In these equations, $X$ represents the recoil momentum, 
$X_{\eta}^{\text{new}}$ and $X_{\xi}^{\text{new}}$ are the new recoil system projections in the $\eta$ and
$\xi$ directions after recoil tuning, $X_{\eta}^{\text{raw}}$ and $X_{\xi}^{\text{raw}}$ are the recoil system 
projections in the $\eta$ and $\xi$ directions before recoil tuning, 
and $X_{\eta}^{\text{gen}}$ and $X{\xi}^{\text{gen}}$
are the generator-level recoil system projections in the $\eta$ and $\xi$ directions. 
By varying $\alpha$, $\beta$, and $o$ in the MC,
we achieve good agreement between the MC and data recoil system projections 
in both the $\eta$ and $\xi$ directions 
for each $p_T^Z$ bin. 

We also perform recoil tuning to eliminate {\it SET} dependences. 
An iterative method is used to remove correlations between $p_T^Z$ and {\it SET},
which is done by doing the recoil tuning in each {\it SET} bin, and then, based on the {\it SET} tuning,
performing the tuning for each $p_T^Z$ bin. 
We iterate these two steps until stable and consistent results are obtained.

Additionally, there is a top-bottom asymmetry in the D0 calorimeter coming from
variations in the lengths of calorimeter signal cables. 
We use an additional correction based on the azimuthal angle of the recoil system to reproduce 
this asymmetry in the MC,
and achieve agreement between data and MC.

\subsection{Charge misidentification}
Misidentification of the charge sign of the electron would result in a dilution 
of the measured electron charge asymmetry.
We measure the charge misidentification probability ($Q_{\text{mis}}$) with $Z\rightarrow ee$ events
using the tag-and-probe method. The CC and EC electron charge misidentification probabilities are measured 
using CC-CC events (both electrons in the CC) 
and CC-EC events
separately. In addition to the general electron selection criteria, we use a tighter track significance cut 
to choose tag electrons. This ensures that the track curvature is sufficiently well measured to 
enable a good measurement of the tag track charge. 
We determine the charge misidentification probabilities in the data and MC 
as functions of $\eta^e$ and $E_T^e$. 
The charge misidentification probability in data averaged over $E_T^e$ varies
from 0.2\% at $|\eta^e|=0$ to 8\% at $|\eta^e|=3.0$, as shown in Fig.~\ref{fig:qmisid}.

\begingroup
\begin{figure}
\epsfig{file=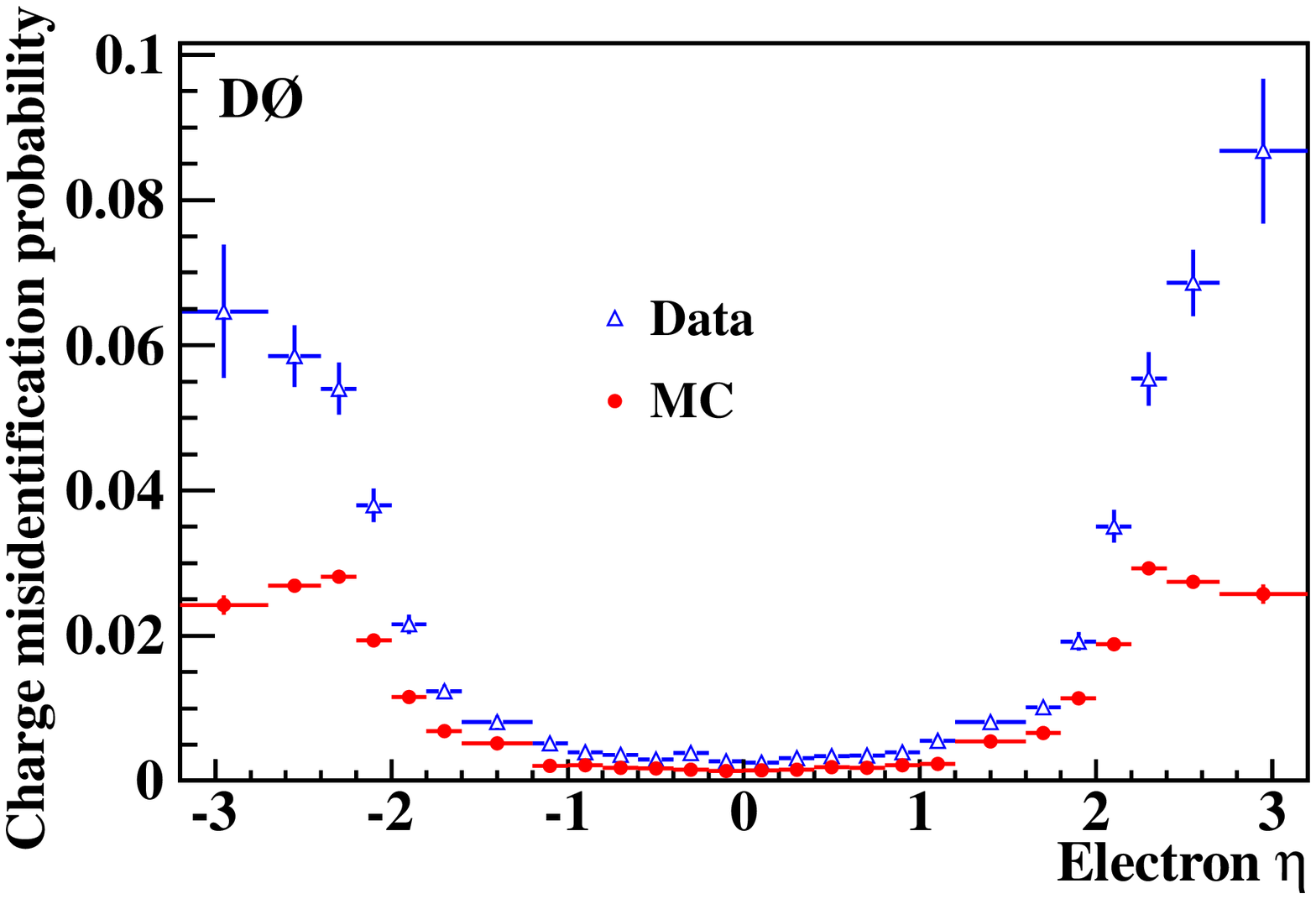, scale=0.44}
\caption{(color online). Charge misidentification probability as a function of $\eta^e$.
The blue triangles represent the measured data charge misidentification probability, 
and the red dots represent that of the MC.}
 \label{fig:qmisid}
\end{figure}
\endgroup

The charge misidentification probability measured in data is roughly 
a factor of three larger than it is in MC,
which is due to MC modeling of the tracking detector, including material modeling deficiencies,
and tracking detector alignment differences between MC and data.
As a function of $\eta^e$ and $E_T^e$, we flip the charge of electrons 
satisfying analysis criteria so that the charge misidentification probability of MC matches that of the data.
This extra electron charge misidentification probability
for each $\eta^e$ and $E_T^e$ bin is applied to the MC used in this analysis. 

\subsection{Backgrounds}
Background contributions, except multijet events, 
are estimated using the {\sc pythia} MC.
This includes $W\rightarrow \tau\nu$ events in which the $\tau$ lepton decays to an electron and a neutrino,
$Z\rightarrow ee$ events in which one of the electrons is not identified, 
and $Z\rightarrow \tau \tau$ events with one tau decaying to an electron and the other not identified.
We normalize these background contributions according to their 
cross sections~\cite{nlo_xsection} and the integrated luminosity.
In the $W\rightarrow \tau\nu$ MC sample, the tau decay
phase space and momentum is not modeled correctly in {\sc pythia} v6,
and we use {\sc tauola}~\cite{tauola}, which applies the correct 
branching fraction for each channel and correctly treats the tau polarization. 

The largest background originates from multijet events
in which one jet is misreconstructed as an electron and there is significant $\met$ in the event.
Even though the probability for a jet to be misidentified as an electron is small due to the track 
requirements, multijet events are the dominant source of background in this analysis
due to the large jet production cross section. 
The multijet background is estimated using collider data by fitting the $W$ boson $M_T$
distribution in the region 50 to 130~GeV (with other SM backgrounds
subtracted) to the sum of the shape predicted by the
$W\rightarrow e\nu$ signal MC and the shape measured from a multijet-enriched sample. 
The multijet-enriched sample is selected by reversing the shower shape (H-matrix) requirement for the
electron candidates~\cite{d0_afb}. 

The background contributions are determined as a function of $\eta^e$, 
and the average contributions in the $M_T$ range of 50~GeV to 130~GeV are
4.0\% from multijet, 2.6\% from $Z\rightarrow ee$, 
2.2\% from $W\rightarrow \tau\nu$, and 0.2\% from $Z\rightarrow \tau \tau$ events.
The $W\rightarrow \tau\nu$ boson background has the same production process as the signal, it
contributes to the raw asymmetry measurement. For the $Z$ boson background, the contribution
is small. The charge of the fake electron in multijet events is random and thus there is no asymmetry in
this background. 

\subsection{Data and MC comparisons}
Comparisons of the $E_T^e$, $\met$, $\eta_{\text{det}}^e$, and $W$ boson $p_T$ of selected 
data events and the sum of the signal and background predictions are shown in Fig.~\ref{fig:data_mc_compare}.
Reasonable agreement between data and prediction is observed for all distributions, 
but there are discrepancies between data and prediction, 
i.e., in the tail region of the $\met$ distribution, so we assign 
systematic uncertainties to account for those discrepancies.

\begingroup
\begin{figure*}
\epsfig{file=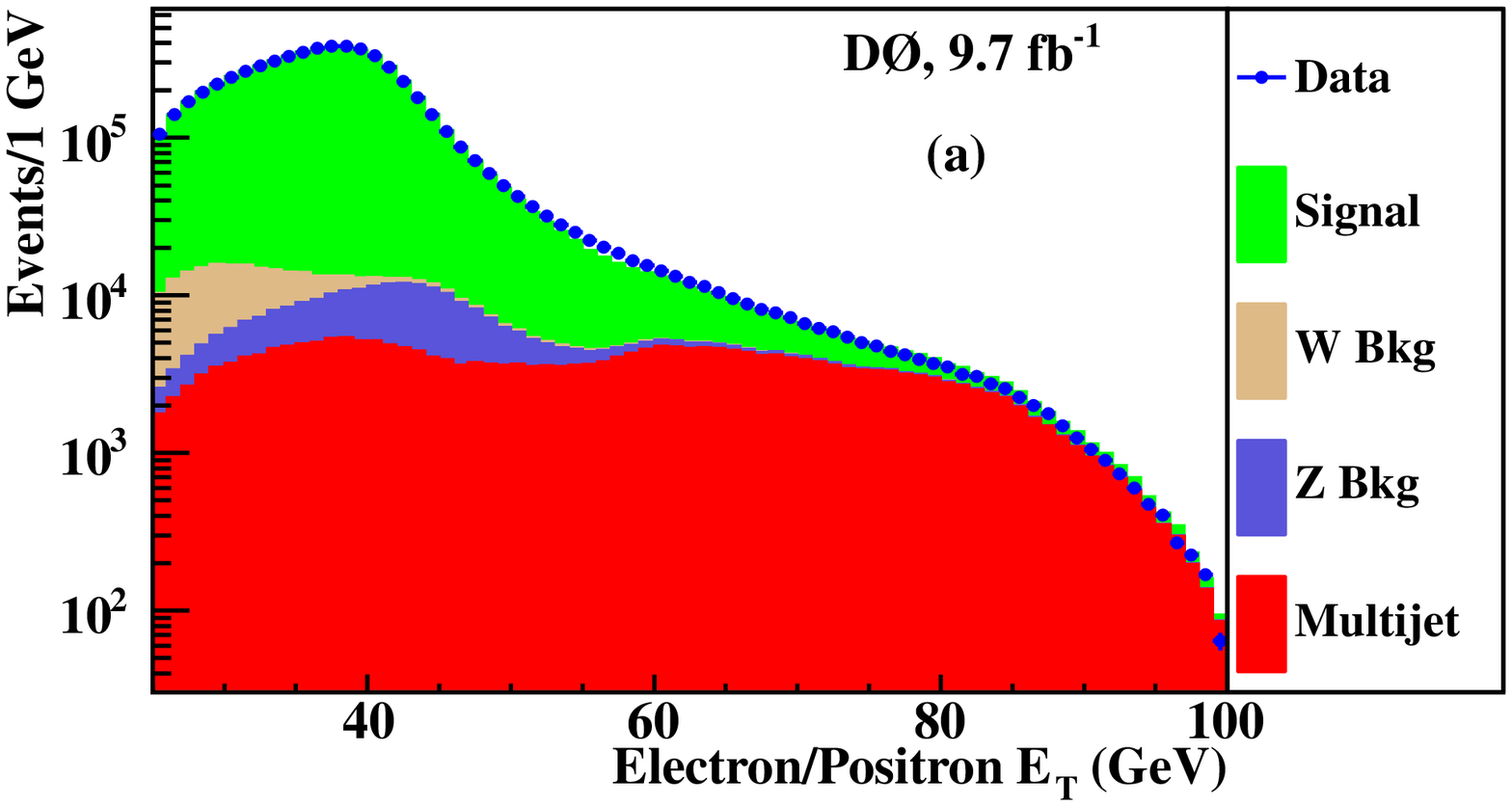, scale=0.44}
\epsfig{file=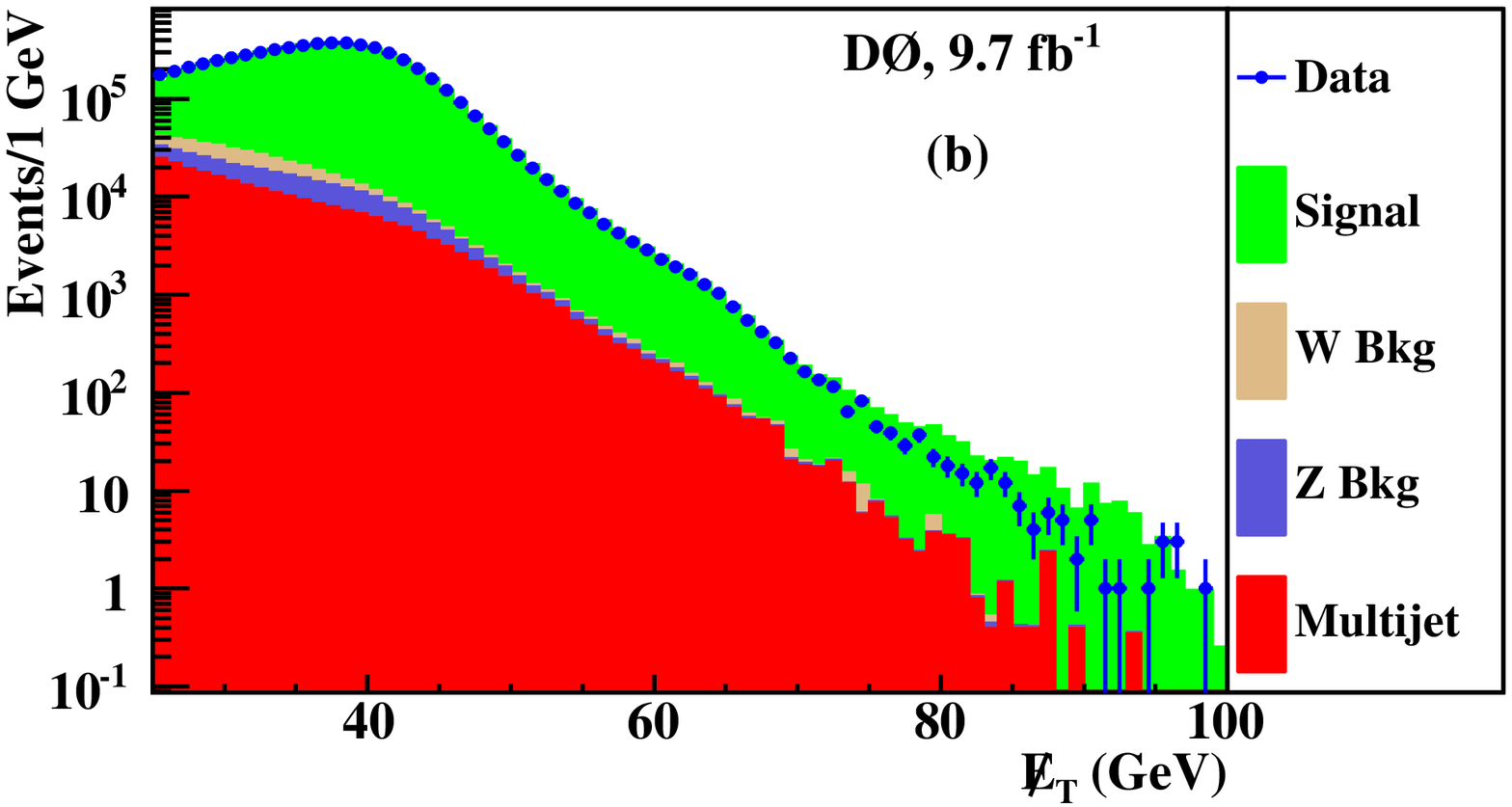, scale=0.44}
\epsfig{file=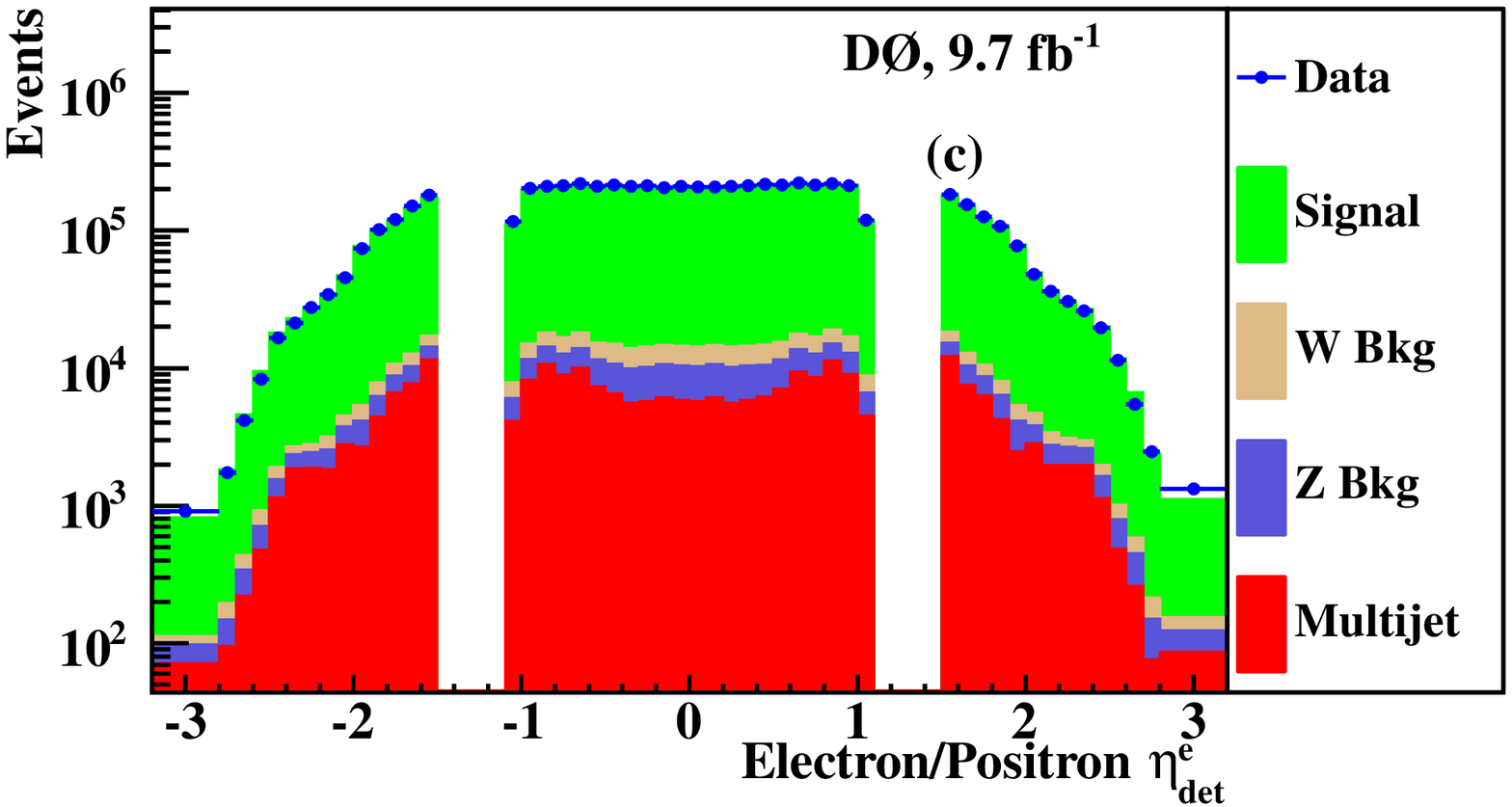, scale=0.44}
\epsfig{file=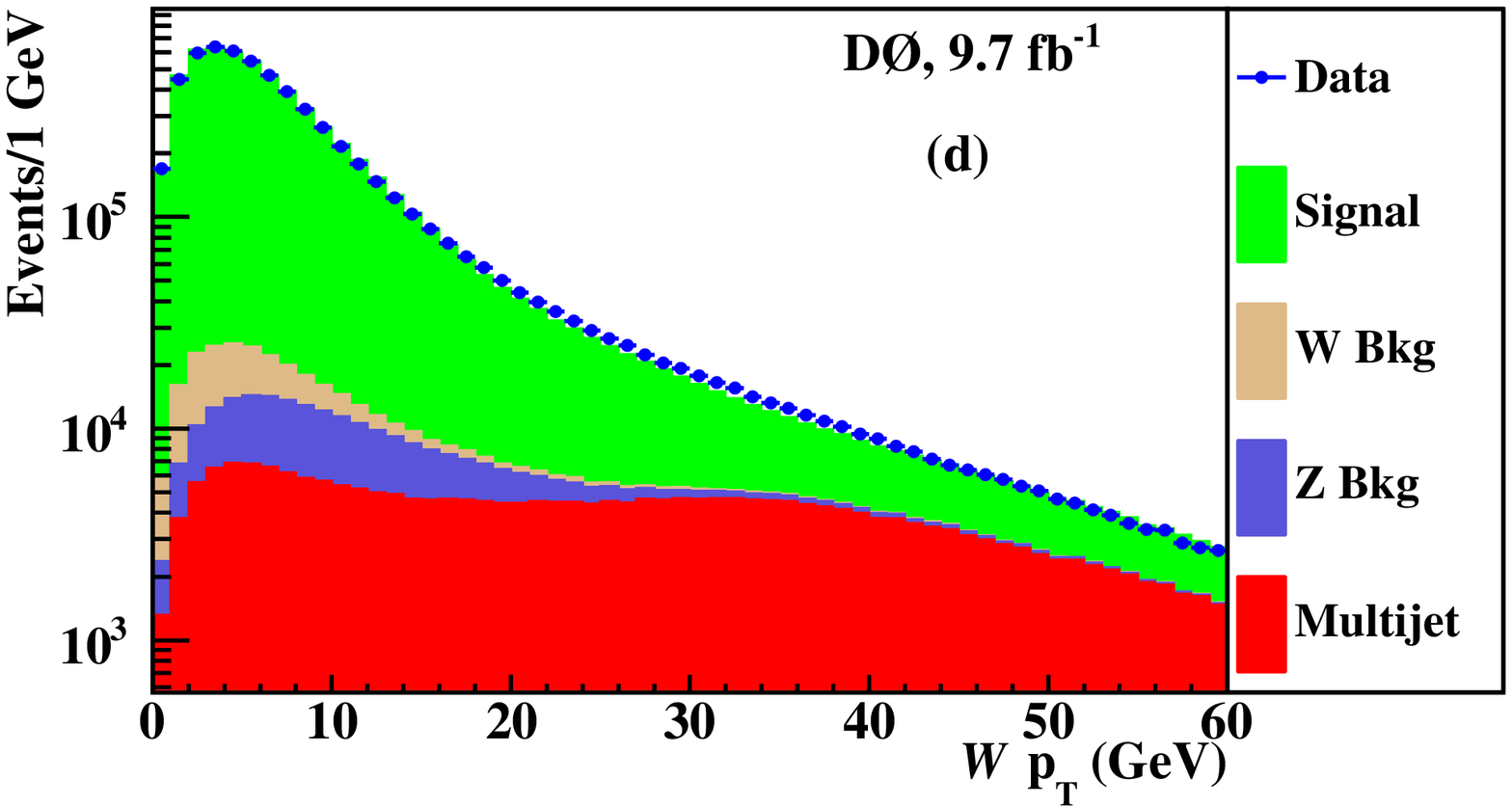, scale=0.44}
\caption{(color online). Comparisons of electron (a) $E_T^e$, (b) $\met$, (c) $\eta^e_{\text{det}}$, 
and (d) $W$ boson $p_T$
between data and the sum of signal and background predictions for selected $W$ boson events. Systematic 
uncertainties are not shown.}
 \label{fig:data_mc_compare}
\end{figure*}
\endgroup

%%%%%%%%%%%%%%%%%%%%%%%%%%%%%%%%%%%%%%%%
\section{Unfolding}
The electron and positron $\eta^e$ distributions after event selection cannot be directly 
compared with generator-level predictions due to detector resolution and acceptance effects.
To correct for the migration of events from one bin to another due to these effects,
an unfolding procedure is performed before comparing the measured asymmetry with predictions.

\subsection{Migration unfolding}
Bin purity is defined as the fraction of events in a bin $i$ for any variable $x$ 
that comes from events that were generated in that bin:
\begin{eqnarray}
 \pi(x, i) = \frac{N_{\text{Reco}}^{\text{Gen}}(x, i)}{N_{\text{Reco}}(x, i)},
\end{eqnarray}
where $N_{\text{Reco}}^{\text{Gen}}(x, i)$ is the number of events in bin $i$ at both the generator and
reconstruction-levels, and $N_{\text{Reco}}(x, i)$ is the number of events in bin $i$ at the reconstruction level.
Our studies show that the migration between $\eta^e$ bins is small, but that the migration 
between the five different kinematic bins in ($E_T^e$, $\met$) is significant, 
with purities varying from 60\% to 90\%. 

The event migration correction uses an unfolding
procedure based on migration matrices determined using the $W$ boson MC.
The migration matrices are derived using an inclusive $W\rightarrow e\nu$ sample
generated using {\sc pythia} with the CTEQ6.1L PDF set.
For each reconstruction-level kinematic bin, 
we construct relevant detector migration matrices for non-overlapping kinematic bins.
These matrices are used to describe events migrating from a given generator-level 
$\eta^e$ and kinematic bin into a different reconstruction-level $\eta^e$ and kinematic bin, as
\begin{eqnarray}
  M_{ij}^{AB} = \frac{N_{\text{Reco, {\it i, B}}}^{\text{Gen, {\it j, A}}}}{N_{\text{Reco, {\it i, B}}}}
\end{eqnarray}
which is the number of events in both the generator-level $\eta^e$ bin $j$
and kinematic bin $A$
and in the reconstruction-level $\eta^e$ bin $i$ and kinematic bin $B$ 
($N_{\text{Reco, {\it i, B}}}^{\text{Gen, {\it j, A}}}$), 
divided by the number of events in the reconstruction-level $\eta^e$ bin $i$ 
and kinematic bin $B$ ($N_{\text{Reco, {\it i, B}}}$).

Using the selected $W$ boson MC events in each kinematic bin and the migration matrices, 
we build the connection between the events selected after reconstruction ($N_{\text{Reco}}(\eta^e, i)$)
and the generator-level events ($N_{\text{Gen({\it j})}}^A$), and use the migration matrices to remove 
the detector resolution effects as
\begin{eqnarray}
  N_{\text{Gen({\it j})}}^A = \sum\limits_{B}\sum\limits_{i} N_{\text{Reco}}(\eta^e, i) \times  M_{ij}^{AB}.
\end{eqnarray}

 \subsection{Acceptance $\times$ efficiency correction} 
After correcting the MC for charge misidentification and migration,
the electron-to-positron ratio at the reconstruction level is still different from that at the generator-level. 
The remaining differences come from 
acceptance times efficiency ($\cal{A}\times \epsilon$) effects.
An $\cal{A}\times \epsilon$ correction is 
performed to account for acceptance and selection criteria effects.
This correction is obtained for each $\eta^e$ bin by accounting for the difference between the 
generator-level and unfolded reconstruction-level asymmetries 
(after charge misidentification and migration corrections).

%%%%%%%%%%%%%%%%%%%%%%%%%%%%%%%%%%%%%%%%
\section{Closure tests}
To verify the validity of the unfolding procedure,
MC closure tests are performed with $W\rightarrow e\nu$ events.
At the generator level, the electron asymmetries for the five kinematic
bins under consideration are obtained using simple kinematic cuts 
(i.e., electron transverse momentum $p_T^e(\text{Gen})>25$~GeV and neutrino 
transverse momentum $p_T^{\nu}(\text{Gen})>25$~GeV). 
At the reconstruction level with the detector simulation included, 
the electron asymmetries are extracted once again. Then, after applying the unfolding procedure 
to the reconstruction-level asymmetries, we expect that the unfolded asymmetries 
will match the generator-level asymmetries.
We perform two closure tests to verify this is the case.

\subsection{Closure test I}
In this closure test, half of the MC events are used to derive the migration matrices
and $\cal{A}\times \epsilon$ correction, and the other half are used as pseudo data. 
This method avoids the bias of applying the corrections to the same sample used to develop the corrections.
Good consistency between the unfolded asymmetry and the generator asymmetry is obtained for each
kinematic bin. An example is shown in Fig.~\ref{fig:closure_fold1} which represents the test
results for the $E_T^e>25$~GeV, $\met>25$~GeV bin after CP folding.

\begingroup
\begin{figure*}
\epsfig{file=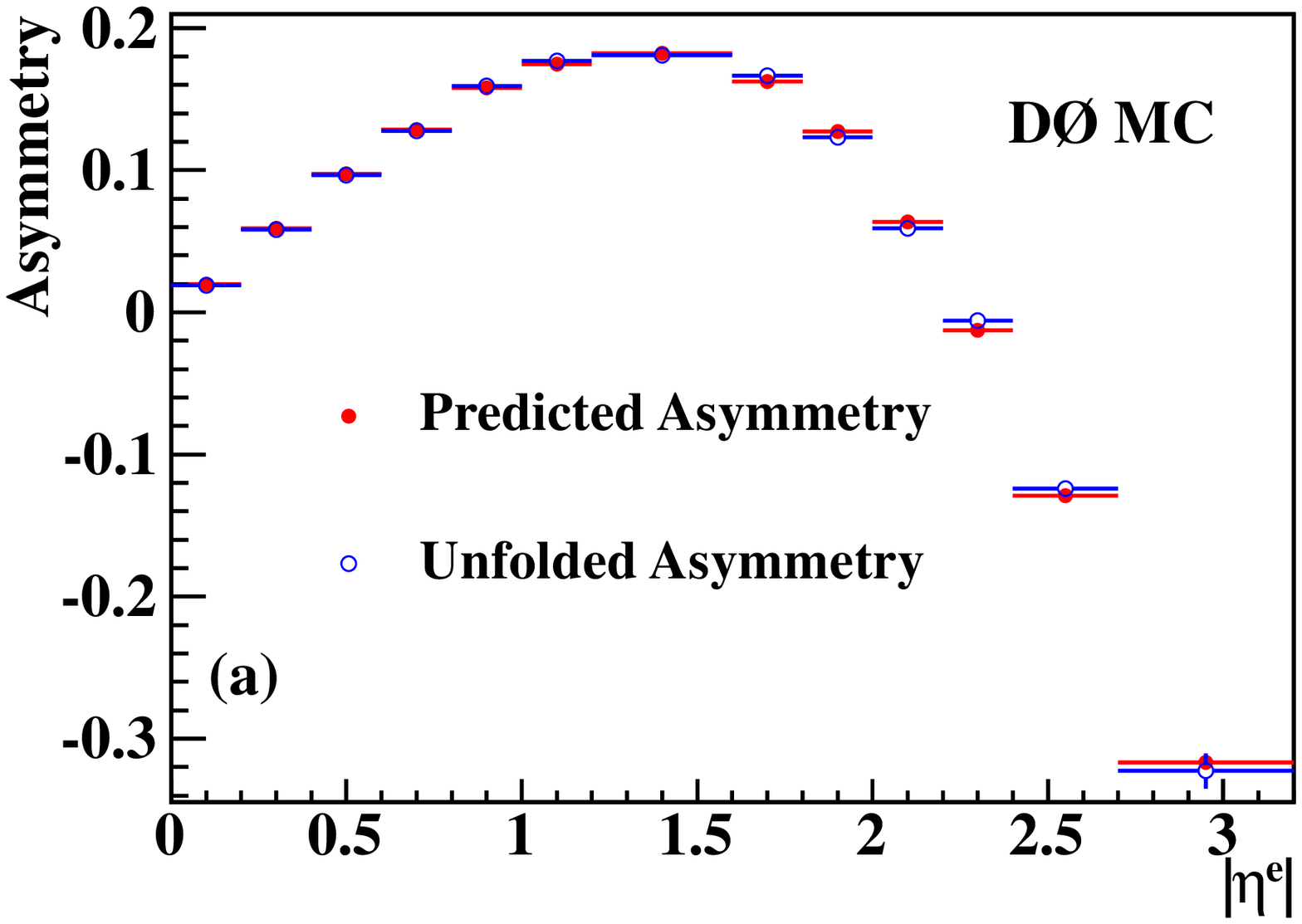,scale =0.44}
\epsfig{file=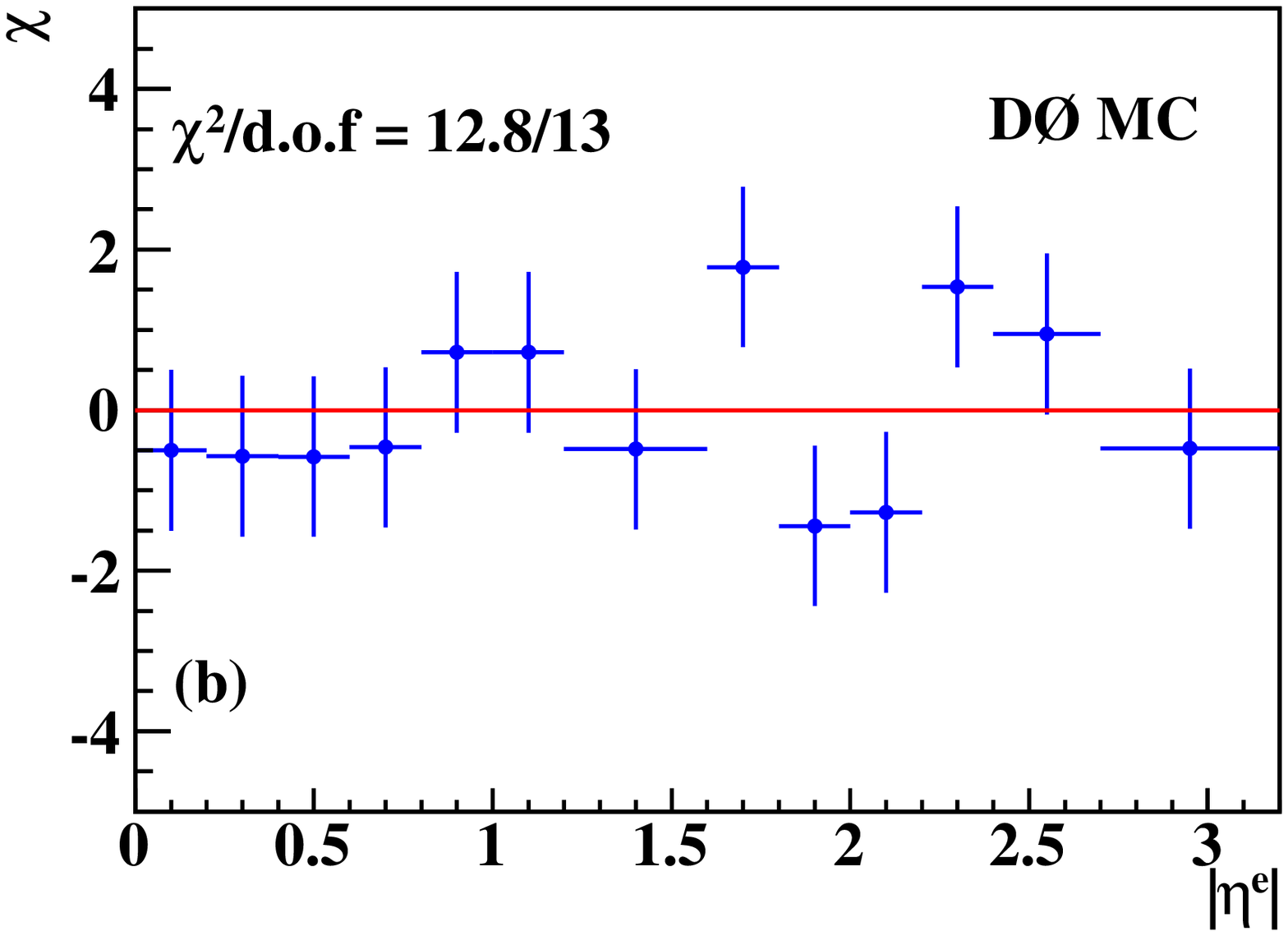,scale =0.44}
\caption{ (color online). (a) Closure test I of the unfolding method for the
kinematic bin $E_T^e>25$~GeV, $\met>25$~GeV, 
using half of the MC sample as input for the unfolding procedure and the other half as pseudo data. 
The solid red points are the {\sc pythia} generator-level electron asymmetries and the open blue points are the unfolded asymmetries. 
The asymmetries are shown after CP-folding.
(b) $\chi$ distribution between predicted asymmetry
and unfolded asymmetry, where $\chi_i=\Delta A_i/\sigma_i$, $\Delta A_i$ is the difference between
the generator-level asymmetry and the unfolded asymmetry, and $\sigma_i$ is the statistical uncertainty in bin $i$.}
\label{fig:closure_fold1}
\end{figure*}
\endgroup

\subsection{Closure test II}
In this closure test, half of the MC events are used to study the migration matrices
and $\cal{A}\times \epsilon$ correction,
and the other half are used as pseudo data, but with the asymmetry distribution modified 
at the generator level (enhanced or suppressed), as shown in Fig.~\ref{fig:closure_fold3}.
To modify the generator-level asymmetry, a reweighting factor based on $\eta^e$ ($f=1~\pm~0.05\times \eta^e$) 
is applied to the number of electrons only, while leaving the number of positrons unchanged. 

\begingroup
\begin{figure*}
\epsfig{file=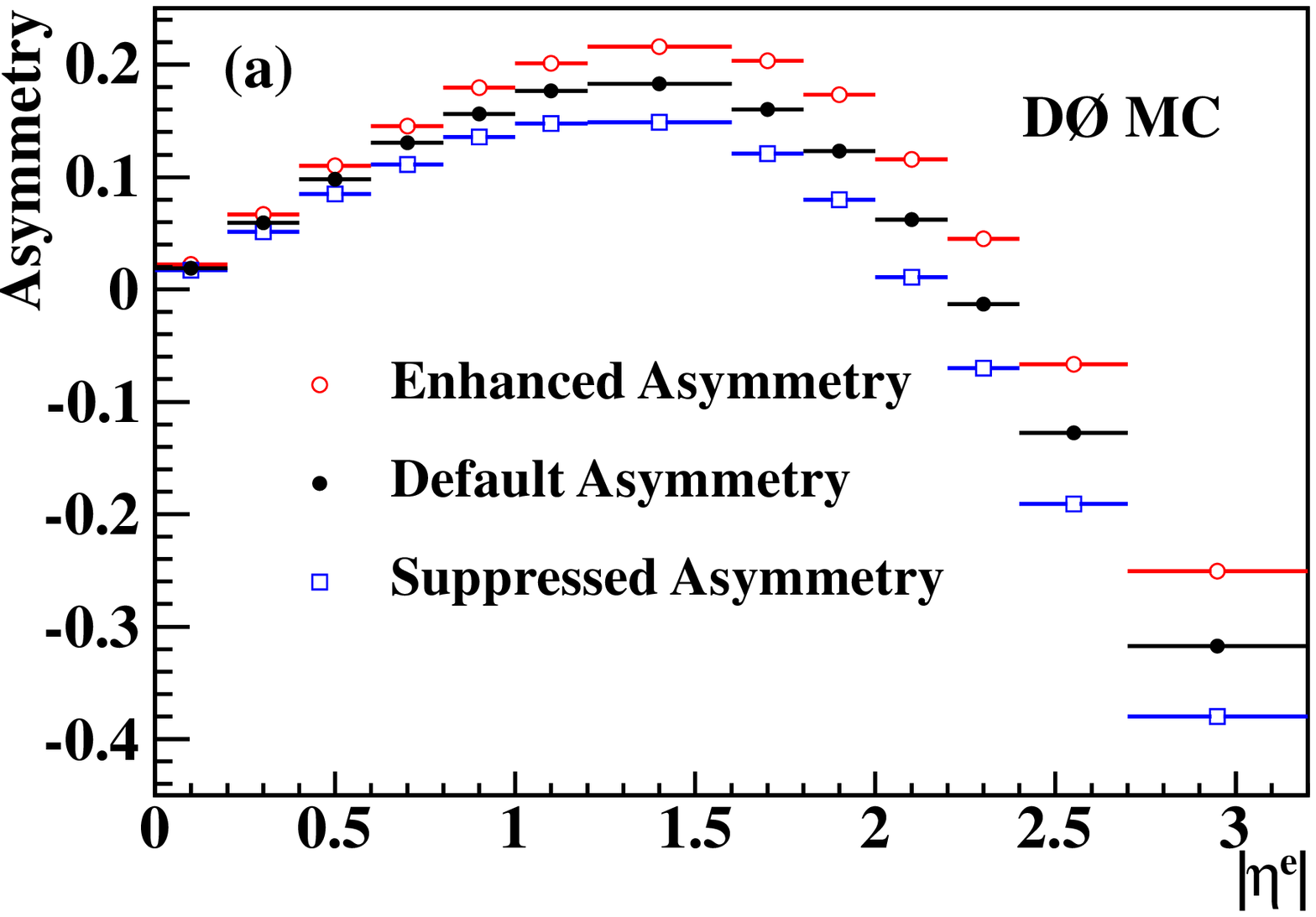,scale=0.44}\\
\epsfig{file=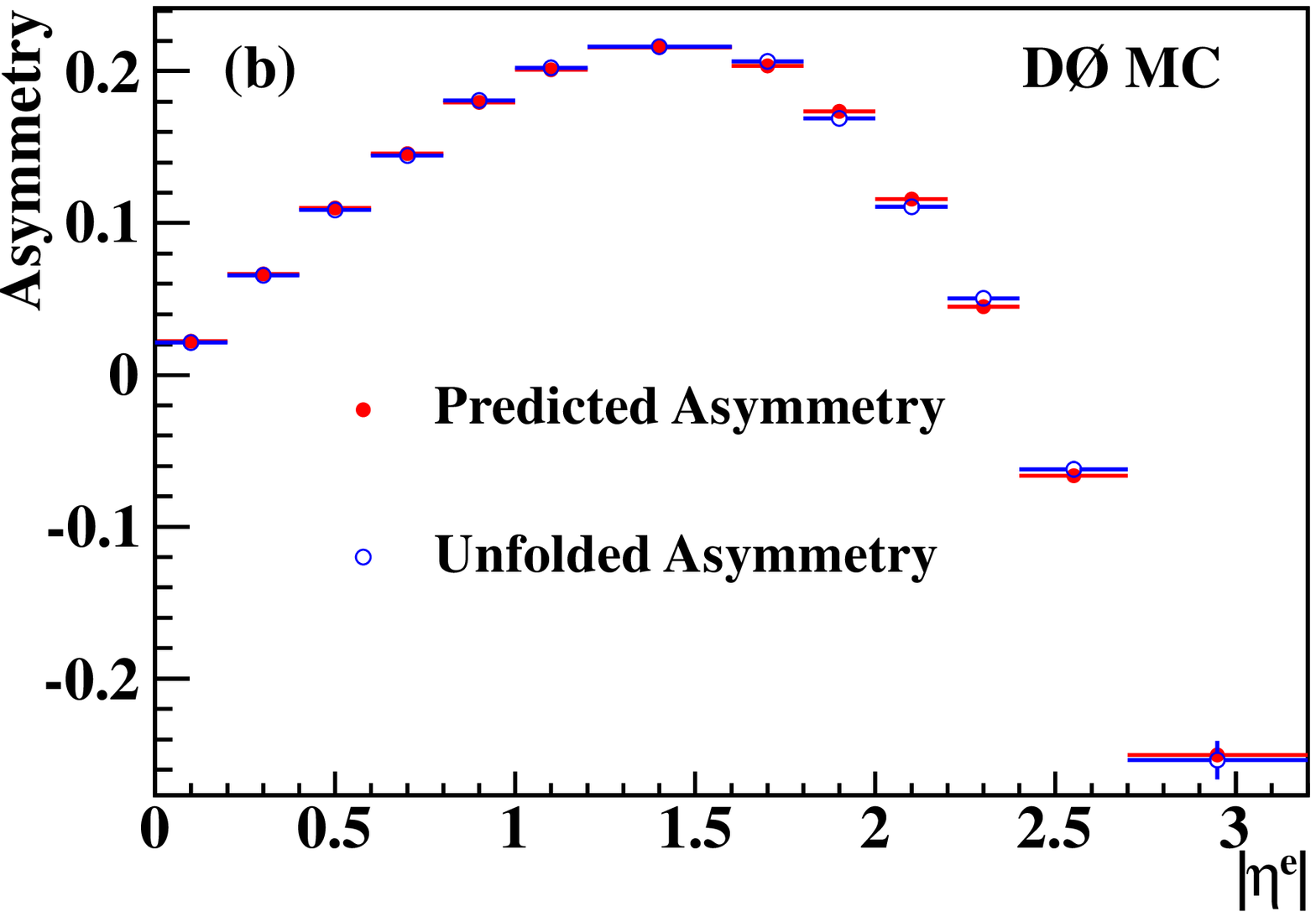,scale=0.44}
\epsfig{file=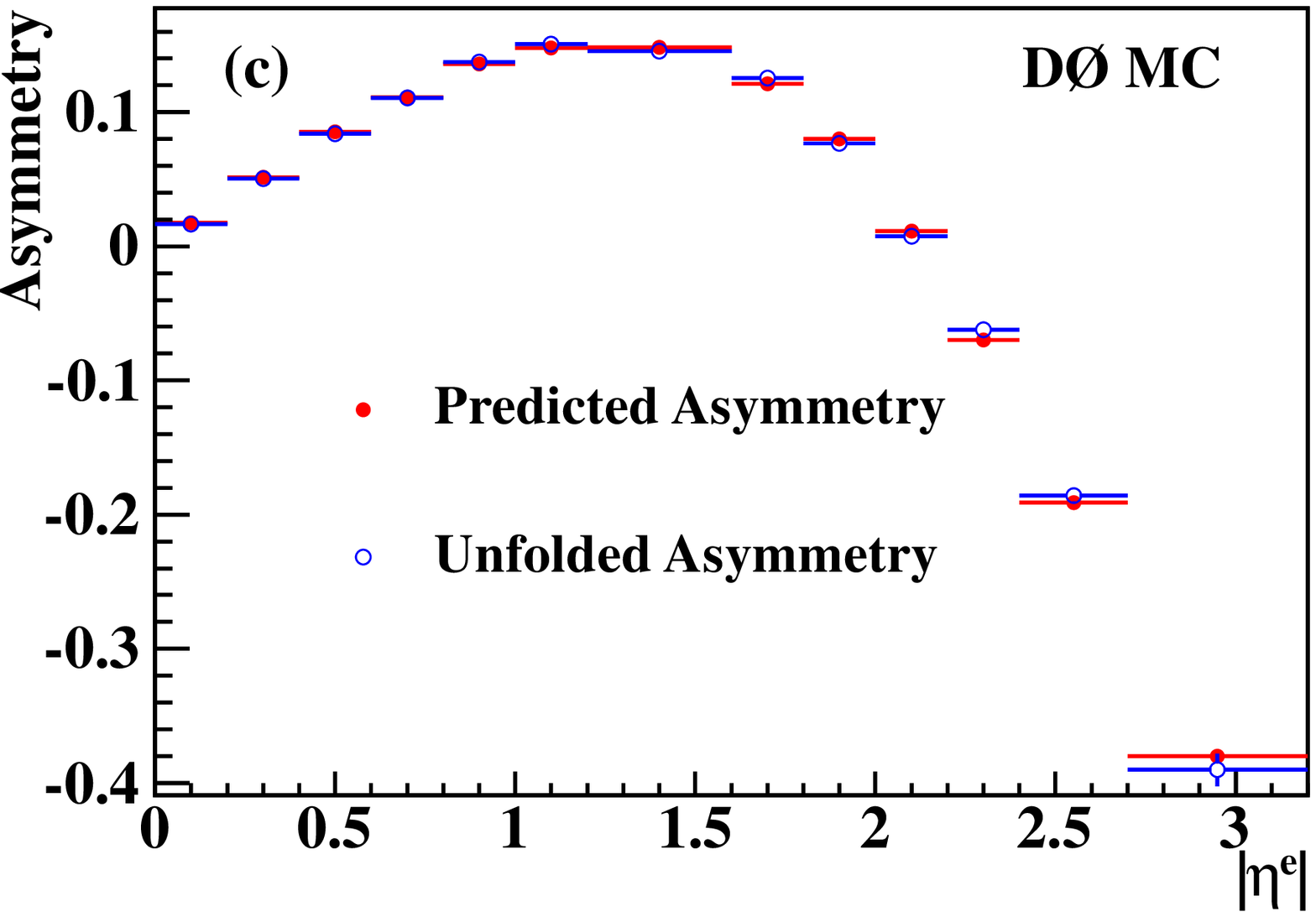,scale=0.44}
\caption{(color online). (a) shows different input asymmetries used in the second closure
test for the kinematic bin $E_T^e>25$~GeV, $\met>25$~GeV. 
The solid black points represent the default asymmetry distribution 
generated using the {\sc pythia} generator with CTEQ6.1L PDFs, the open red points represent
the enhanced input asymmetry, and the open blue squares point represent the
suppressed input asymmetry.
(b) and (c) show closure test II of the unfolding method
using half of the MC sample as input for the unfolding procedure and the other 
half of the MC as pseudo data, where the generator-level asymmetries are
enhanced or suppressed.
A reweighting factor ($f=1-0.05\times\eta^e$ for (b), and 
$f=1+0.05\times\eta^e$ for (c)) has been applied to the number of
electrons to ensure the generator-level asymmetries are far from the default values.
The solid red points are generator-level electron asymmetries, the open blue points are the unfolded asymmetries. 
The asymmetries are shown after CP-folding.}
\label{fig:closure_fold3}
\end{figure*}
\endgroup

Good agreement between the unfolded asymmetry and generator-modified 
asymmetry is obtained for each kinematic bin. 
The plots shown in Fig.~\ref{fig:closure_fold3} correspond to the test results for
the $E_T^e>25$~GeV, $\met>25$~GeV bin with CP-folding.
This test confirms that the migration matrix and $\cal{A}\times \epsilon$ corrections derived
from the predicted asymmetry can be applied without bias for other asymmetries. 
%%%%%%%%%%%%%%%%%%%%%%%%%%%%%%%%%%%%%%%%%%%%%%%%%
\section{Systematic uncertainties}
The systematic uncertainties of the electron charge asymmetry
measurement in the five kinematic bins are summarized in 
Tables~\ref{Tab:Combined_uncer_lept_fold_Elec_25_Inf_Neut_25_Inf}--\ref{Tab:Combined_uncer_lept_fold_Elec_35_Inf_Neut_35_Inf}.
Because this is an asymmetry measurement, some of the uncertainties cancel 
in the measured ratio, i.e., uncertainties from electron identification, energy calibration,
recoil tuning, and background modeling are small compared with the uncertainties
in the individual $e^+$ and $e^-$ distributions. The various sources of uncertainties
that are considered are described below.

\begin{table*} 
\begin{center} 
\caption{Summary of absolute systematic uncertainties for the CP-folded
electron charge asymmetry for kinematic bin $E_T^{e} > 25$~GeV,
$\met > 25$~GeV. The calorimeter has a gap in the range of $1.1<\eta_{\text{det}}^e<1.5$,
so some systematic uncertainties in the $\eta^e$ bin $1.2-1.6$ are large compared to 
those of the neighboring $\eta^e$ bins. 
The uncertainties are multiplied by 1000.} 
\begin{tabular}{c|ccccccccc|c} 
\hline 
\hline 
$\eta^e$ & Gen & EMID   & $K_{\text{eff}}^{q, p}$  & Energy    & Recoil    & Model  & Bkgs& $Q_{\text{mis}}$ &
Unfolding  & Total \\ \hline 
$0.0-0.2$ & 0.06 & 0.02 & 0.20 & 0.03 & 0.04 & 0.28 & 0.26 & 0.54 & 0.82 & 1.08 \\
$0.2-0.4$ & 0.06 & 0.18 & 0.10 & 0.18 & 0.26 & 0.75 & 0.54 & 0.56 & 0.81 & 1.40 \\
$0.4-0.6$ & 0.12 & 0.24 & 0.27 & 0.25 & 0.35 & 1.05 & 0.87 & 0.59 & 0.80 & 1.79 \\
$0.6-0.8$ & 0.07 & 0.34 & 0.04 & 0.34 & 0.49 & 1.32 & 1.81 & 0.60 & 0.80 & 2.55 \\
$0.8-1.0$ & 0.12 & 0.36 & 0.12 & 0.36 & 0.53 & 1.72 & 2.37 & 0.76 & 0.85 & 3.23 \\
$1.0-1.2$ & 0.09 & 0.37 & 0.47 & 0.37 & 0.55 & 2.42 & 2.71 & 1.20 & 1.17 & 4.10 \\
$1.2-1.6$ & 0.03 & 0.42 & 0.64 & 0.39 & 0.58 & 4.10 & 3.94 & 1.67 & 1.04 & 6.11 \\
$1.6-1.8$ & 0.11 & 0.28 & 0.18 & 0.22 & 0.34 & 4.26 & 1.37 & 1.53 & 0.95 & 4.85 \\
$1.8-2.0$ & 0.34 & 0.36 & 1.07 & 0.05 & 0.10 & 4.21 & 1.43 & 2.46 & 1.13 & 5.34 \\
$2.0-2.2$ & 0.37 & 0.36 & 1.38 & 0.04 & 0.07 & 3.33 & 1.75 & 4.37 & 1.47 & 6.14 \\
$2.2-2.4$ & 0.19 & 0.30 & 2.78 & 0.02 & 0.05 & 3.40 & 1.54 & 7.15 & 1.93 & 8.76 \\
$2.4-2.7$ & 0.21 & 0.43 & 5.54 & 0.29 & 0.48 & 4.24 & 2.16 & 8.65 & 2.36 & 11.6 \\
$2.7-3.2$ & 0.05 & 0.87 & 9.00 & 0.81 & 1.30 & 3.48 & 3.99 & 18.9 & 5.48 & 22.3 \\
\hline 
\hline 
\end{tabular} 
\label{Tab:Combined_uncer_lept_fold_Elec_25_Inf_Neut_25_Inf} 
\end{center} 
\end{table*} 

\begin{table*}
\begin{center} 
\caption{Summary of absolute systematic uncertainties for the CP-folded
electron charge asymmetry for kinematic bin $25 < E_T^{e} < 35$~GeV,
$\met > 25$~GeV. The calorimeter has a gap in the range of $1.1<\eta_{\text{det}}^e<1.5$,
so some systematic uncertainties in the $\eta^e$ bin $1.2-1.6$ are large compared to those of the neighboring $\eta^e$ bins. 
The uncertainties are multiplied by 1000.} 
\begin{tabular}{c|ccccccccc|c} 
\hline 
\hline 
$\eta^e$ & Gen & EMID   & $K_{\text{eff}}^{q, p}$  & Energy    & Recoil    & Model  & Bkgs& $Q_{\text{mis}}$ &
Unfolding  & Total \\ \hline 
$0.0-0.2$ & 0.10 & 0.01 & 0.10 & 0.03 & 0.02 & 0.07 & 0.11 & 1.99 & 1.27 & 2.37 \\
$0.2-0.4$ & 0.19 & 0.25 & 0.11 & 0.54 & 0.39 & 0.11 & 0.39 & 1.93 & 1.25 & 2.45 \\
$0.4-0.6$ & 0.54 & 0.39 & 0.27 & 0.80 & 0.56 & 0.28 & 0.55 & 2.10 & 1.25 & 2.80 \\
$0.6-0.8$ & 0.52 & 0.50 & 0.12 & 1.03 & 0.70 & 0.43 & 1.02 & 1.80 & 1.25 & 2.85 \\
$0.8-1.0$ & 0.56 & 0.50 & 0.17 & 0.90 & 0.75 & 0.22 & 1.47 & 1.79 & 1.33 & 3.03 \\
$1.0-1.2$ & 0.53 & 0.50 & 0.34 & 0.74 & 0.78 & 0.56 & 1.88 & 2.50 & 1.88 & 3.93 \\
$1.2-1.6$ & 0.52 & 0.35 & 0.42 & 0.35 & 0.27 & 0.97 & 2.09 & 2.32 & 1.74 & 3.81 \\
$1.6-1.8$ & 0.41 & 0.20 & 0.06 & 0.65 & 0.35 & 1.43 & 1.00 & 2.35 & 1.59 & 3.44 \\
$1.8-2.0$ & 0.71 & 0.72 & 0.39 & 1.07 & 0.86 & 2.72 & 0.72 & 3.82 & 1.90 & 5.40 \\
$2.0-2.2$ & 0.95 & 0.89 & 0.58 & 1.60 & 1.34 & 2.79 & 1.38 & 6.29 & 2.44 & 7.85 \\
$2.2-2.4$ & 0.77 & 0.83 & 0.86 & 2.05 & 1.28 & 1.97 & 1.68 & 9.20 & 3.06 & 10.4 \\
$2.4-2.7$ & 0.47 & 1.25 & 2.64 & 2.25 & 1.68 & 2.08 & 4.66 & 10.3 & 3.40 & 12.7 \\
$2.7-3.2$ & 0.38 & 1.91 & 14.2 & 3.18 & 2.97 & 4.46 & 5.75 & 20.8 & 7.14 & 27.5 \\
\hline 
\hline 
\end{tabular} 
\label{Tab:Combined_uncer_lept_fold_Elec_25_35_Neut_25_Inf} 
\end{center} 
\end{table*} 

\begin{table*} 
\begin{center} 
\caption{Summary of absolute systematic uncertainties for the CP-folded
electron charge asymmetry for kinematic bin $25 < E_T^{e} < 35$~GeV,
$25< \met < 35$~GeV. The calorimeter has a gap in the range of $1.1<\eta_{\text{det}}^e<1.5$,
so some systematic uncertainties in the $\eta^e$ bin $1.2-1.6$ are large compared to those of the neighboring $\eta^e$ bins. 
The uncertainties are multiplied by 1000.} 
\begin{tabular}{c|ccccccccc|c} 
\hline 
\hline 
$\eta^e$ & Gen & EMID   & $K_{\text{eff}}^{q, p}$  & Energy    & Recoil    & Model  & Bkgs& $Q_{\text{mis}}$ &
Unfolding  & Total \\ \hline 
$0.0-0.2$ & 0.06 & 0.08 & 0.10 & 0.13 & 0.17 & 0.34 & 0.08 & 1.99 & 1.77 & 2.70 \\
$0.2-0.4$ & 0.37 & 0.27 & 0.10 & 0.39 & 0.53 & 0.66 & 0.34 & 1.93 & 1.74 & 2.82 \\
$0.4-0.6$ & 0.52 & 0.47 & 0.33 & 0.61 & 0.76 & 0.86 & 0.38 & 2.10 & 1.73 & 3.14 \\
$0.6-0.8$ & 0.39 & 0.51 & 0.13 & 0.68 & 0.91 & 1.23 & 0.60 & 1.80 & 1.74 & 3.14 \\
$0.8-1.0$ & 0.44 & 0.48 & 0.25 & 0.62 & 0.95 & 1.10 & 1.03 & 1.79 & 1.87 & 3.28 \\
$1.0-1.2$ & 0.31 & 0.37 & 0.27 & 0.47 & 0.89 & 1.60 & 1.38 & 2.50 & 2.70 & 4.40 \\
$1.2-1.6$ & 0.11 & 0.32 & 0.37 & 0.36 & 0.59 & 1.95 & 1.97 & 2.32 & 2.52 & 4.48 \\
$1.6-1.8$ & 0.68 & 0.84 & 0.10 & 0.87 & 1.14 & 1.28 & 1.07 & 2.35 & 2.24 & 4.07 \\
$1.8-2.0$ & 0.97 & 1.68 & 0.54 & 1.78 & 2.38 & 2.59 & 1.33 & 3.82 & 2.69 & 6.57 \\
$2.0-2.2$ & 1.23 & 1.88 & 0.61 & 2.04 & 3.34 & 3.77 & 2.34 & 6.29 & 3.42 & 9.58 \\
$2.2-2.4$ & 0.39 & 1.35 & 0.92 & 1.96 & 2.84 & 3.36 & 2.83 & 9.20 & 4.16 & 11.7 \\
$2.4-2.7$ & 0.27 & 1.63 & 1.99 & 2.39 & 3.43 & 6.38 & 6.13 & 10.3 & 4.43 & 15.1 \\
$2.7-3.2$ & 0.35 & 2.19 & 13.7 & 3.34 & 4.81 & 9.76 & 5.37 & 20.8 & 8.98 & 29.4 \\
\hline 
\hline 
\end{tabular} 
\label{Tab:Combined_uncer_lept_fold_Elec_25_35_Neut_25_35} 
\end{center} 
\end{table*} 

\begin{table*} 
\begin{center} 
\caption{Summary of absolute systematic uncertainties for the CP-folded
electron charge asymmetry for kinematic bin $E_T^{e} > 35$~GeV, $\met
> 25$~GeV. The calorimeter has a gap in the range of $1.1<\eta_{\text{det}}^e<1.5$,
so some systematic uncertainties in the $\eta^e$ bin $1.2-1.6$ are large compared to those of the neighboring $\eta^e$ bins. 
The uncertainties are multiplied by 1000.} 
\begin{tabular}{c|ccccccccc|c} 
\hline 
\hline 
$\eta^e$ & Gen & EMID   & $K_{\text{eff}}^{q, p}$  & Energy    & Recoil    & Model  & Bkgs& $Q_{\text{mis}}$ &
Unfolding  & Total \\ \hline 
$0.0-0.2$ & 0.04 & 0.03 & 0.26 & 0.04 & 0.06 & 0.37 & 0.31 & 0.48 & 1.06 & 1.29 \\
$0.2-0.4$ & 0.04 & 0.16 & 0.12 & 0.24 & 0.23 & 1.14 & 0.53 & 0.54 & 1.04 & 1.77 \\
$0.4-0.6$ & 0.20 & 0.18 & 0.25 & 0.30 & 0.26 & 1.89 & 0.95 & 0.58 & 1.03 & 2.48 \\
$0.6-0.8$ & 0.28 & 0.29 & 0.31 & 0.46 & 0.42 & 2.27 & 2.20 & 0.63 & 1.02 & 3.47 \\
$0.8-1.0$ & 0.26 & 0.35 & 0.39 & 0.52 & 0.51 & 2.72 & 2.82 & 0.82 & 1.07 & 4.25 \\
$1.0-1.2$ & 0.56 & 0.39 & 0.77 & 0.52 & 0.56 & 3.76 & 3.11 & 1.34 & 1.47 & 5.42 \\
$1.2-1.6$ & 0.47 & 0.60 & 1.34 & 0.69 & 0.80 & 6.22 & 5.47 & 2.16 & 1.28 & 8.86 \\
$1.6-1.8$ & 0.46 & 0.58 & 0.84 & 0.72 & 0.79 & 6.31 & 3.22 & 1.93 & 1.16 & 7.60 \\
$1.8-2.0$ & 0.65 & 0.86 & 1.64 & 0.70 & 0.76 & 6.29 & 1.65 & 3.15 & 1.36 & 7.68 \\
$2.0-2.2$ & 0.70 & 0.73 & 1.88 & 0.61 & 0.51 & 5.54 & 3.19 & 5.92 & 1.77 & 9.18 \\
$2.2-2.4$ & 0.68 & 0.38 & 3.20 & 0.69 & 0.35 & 5.28 & 4.31 & 10.6 & 2.37 & 13.3 \\
$2.4-2.7$ & 0.46 & 0.71 & 5.23 & 0.90 & 0.72 & 5.32 & 2.47 & 14.6 & 3.05 & 16.9 \\
$2.7-3.2$ & 1.43 & 0.31 & 14.8 & 0.87 & 0.32 & 8.99 & 3.48 & 35.0 & 7.94 & 40.0 \\
\hline 
\hline 
\end{tabular} 
\label{Tab:Combined_uncer_lept_fold_Elec_35_Inf_Neut_25_Inf} 
\end{center} 
\end{table*} 

\begin{table*} 
\begin{center} 
\caption{Summary of absolute systematic uncertainties for the CP-folded
electron charge asymmetry for kinematic bin $E_T^{e} > 35$~GeV, $\met
> 35$~GeV. The calorimeter has a gap in the range of $1.1<\eta_{\text{det}}^e<1.5$,
so some systematic uncertainties in the $\eta^e$ bin $1.2-1.6$ are large compared to those of the neighboring $\eta^e$ bins. 
The uncertainties are multiplied by 1000.} 
\begin{tabular}{c|ccccccccc|c} 
\hline 
\hline 
$\eta^e$ & Gen & EMID   & $K_{\text{eff}}^{q, p}$  & Energy    & Recoil    & Model  & Bkgs& $Q_{\text{mis}}$ &
Unfolding  & Total \\ \hline 
$0.0-0.2$ & 0.06 & 0.00 & 0.25 & 0.03 & 0.08 & 0.18 & 0.10 & 0.48 & 1.30 & 1.42 \\
$0.2-0.4$ & 0.05 & 0.04 & 0.12 & 0.07 & 0.16 & 0.59 & 0.17 & 0.54 & 1.27 & 1.53 \\
$0.4-0.6$ & 0.06 & 0.06 & 0.23 & 0.09 & 0.19 & 1.09 & 0.26 & 0.58 & 1.26 & 1.81 \\
$0.6-0.8$ & 0.03 & 0.14 & 0.32 & 0.22 & 0.33 & 1.52 & 0.50 & 0.63 & 1.25 & 2.19 \\
$0.8-1.0$ & 0.10 & 0.20 & 0.41 & 0.29 & 0.45 & 2.11 & 0.70 & 0.82 & 1.32 & 2.81 \\
$1.0-1.2$ & 0.08 & 0.27 & 0.77 & 0.34 & 0.73 & 2.30 & 0.92 & 1.34 & 1.83 & 3.55 \\
$1.2-1.6$ & 0.24 & 0.40 & 1.32 & 0.44 & 0.73 & 3.88 & 1.98 & 2.16 & 1.59 & 5.37 \\
$1.6-1.8$ & 0.57 & 0.44 & 0.89 & 0.44 & 0.85 & 3.97 & 0.94 & 1.93 & 1.40 & 4.95 \\
$1.8-2.0$ & 0.65 & 0.65 & 1.66 & 0.48 & 0.93 & 3.00 & 0.75 & 3.15 & 1.64 & 5.18 \\
$2.0-2.2$ & 1.01 & 0.61 & 1.87 & 0.30 & 0.85 & 2.76 & 1.22 & 5.92 & 2.13 & 7.37 \\
$2.2-2.4$ & 1.71 & 0.35 & 3.27 & 0.47 & 1.00 & 2.99 & 2.07 & 10.6 & 2.83 & 12.2 \\
$2.4-2.7$ & 2.70 & 0.76 & 5.01 & 0.48 & 1.09 & 1.27 & 3.21 & 14.6 & 3.67 & 16.5 \\
$2.7-3.2$ & 3.18 & 0.41 & 16.8 & 0.62 & 0.71 & 3.17 & 3.12 & 35.0 & 9.90 & 40.4 \\
\hline 
\hline 
\end{tabular} 
\label{Tab:Combined_uncer_lept_fold_Elec_35_Inf_Neut_35_Inf} 
\end{center} 
\end{table*} 

\subsection{Systematic uncertainty from the generator-level prediction}
The modeling of the $W$ boson $p_T$ impacts the asymmetry measurements,
and different generators give different predictions, even those at the 
same order (either LO and NLO).
To estimate the uncertainty from the $p_T^W$ modeling, we weight the
$p_T^W$ spectrum from the {\sc pythia} sample to match those distributions from the 
{\sc resbos}~\cite{resbos} and {\sc powheg}~\cite{powheg} generators, separately. 
Then we take the difference resulting from the two weightings as a systematic uncertainty. 

At the generator level, any FSR electrons and photons within a cone of $\Delta \cal{R}$ $< 0.3$ around
an electron are merged with the electron. To estimate the uncertainty from FSR, 
we weight the events with $|M_{\text{Gen}} - M_{\text{Part}}|>1$~GeV by $\pm$10\%, where 
$M_{\text{Gen}}$ and $M_{\text{Part}}$ are the $W$ boson mass at the generator and particle levels, respectively, 
and take the deviation of the asymmetry as the FSR uncertainty. 

The $p_T^W$ modeling and FSR uncertainties are combined in quadrature to form the overall
generator uncertainty. 

\subsection{Systematic uncertainty from EMID and trigger}
To study the uncertainty from the EMID selection, 
we vary the efficiency correction factors by $\pm$1 standard deviation, extract the 
asymmetries with the varied EMID corrections, and take the larger variation in each bin
as a symmetric systematic uncertainty in that bin. 
As expected, the largest contribution is from the track-match efficiency correction.
Similarly, we obtain the systematic uncertainty from the single EM trigger efficiency modeling,
and combine these two uncertainties in quadrature. 

\subsection{Systematic uncertainty from $K^{q, p}_{\text{eff}}$}
The uncertainty from $K^{q, p}_{\text{eff}}$ correction is determined using the same procedure as for the 
determination of the uncertainty from EMID.

\subsection{Systematic uncertainty from electron energy tuning}
To obtain agreement between the data and MC $Z$ boson invariant mass 
distributions, we first perform the energy calibration for both data and MC, 
and then tune the MC with scale and smearing parameters.
To study the uncertainty from these corrections, we vary each of the energy tuning parameters 
by $\pm$1 standard deviation, extract the asymmetries with the varied parameters, and take the 
larger variation in each bin as a symmetric systematic uncertainty in that bin. 
Finally, we combine the uncertainties of all contributing parameters in quadrature
to arrive at one total electron energy tuning uncertainty.
For this uncertainty study, we consider contributions from the energy scale, smearing, offset, and 
non-linearity terms.

\subsection{Systematic uncertainty from recoil modeling}
The uncertainty due to the recoil modeling is determined using the same procedure as the 
determination of the uncertainty from the electron energy tuning.
We consider contributions from scale, smearing, and offset in recoil tuning,
as well as the recoil $\phi$ tuning parameters.

\subsection{Systematic uncertainty from MC modeling}
\begin{figure*}
\epsfig{file=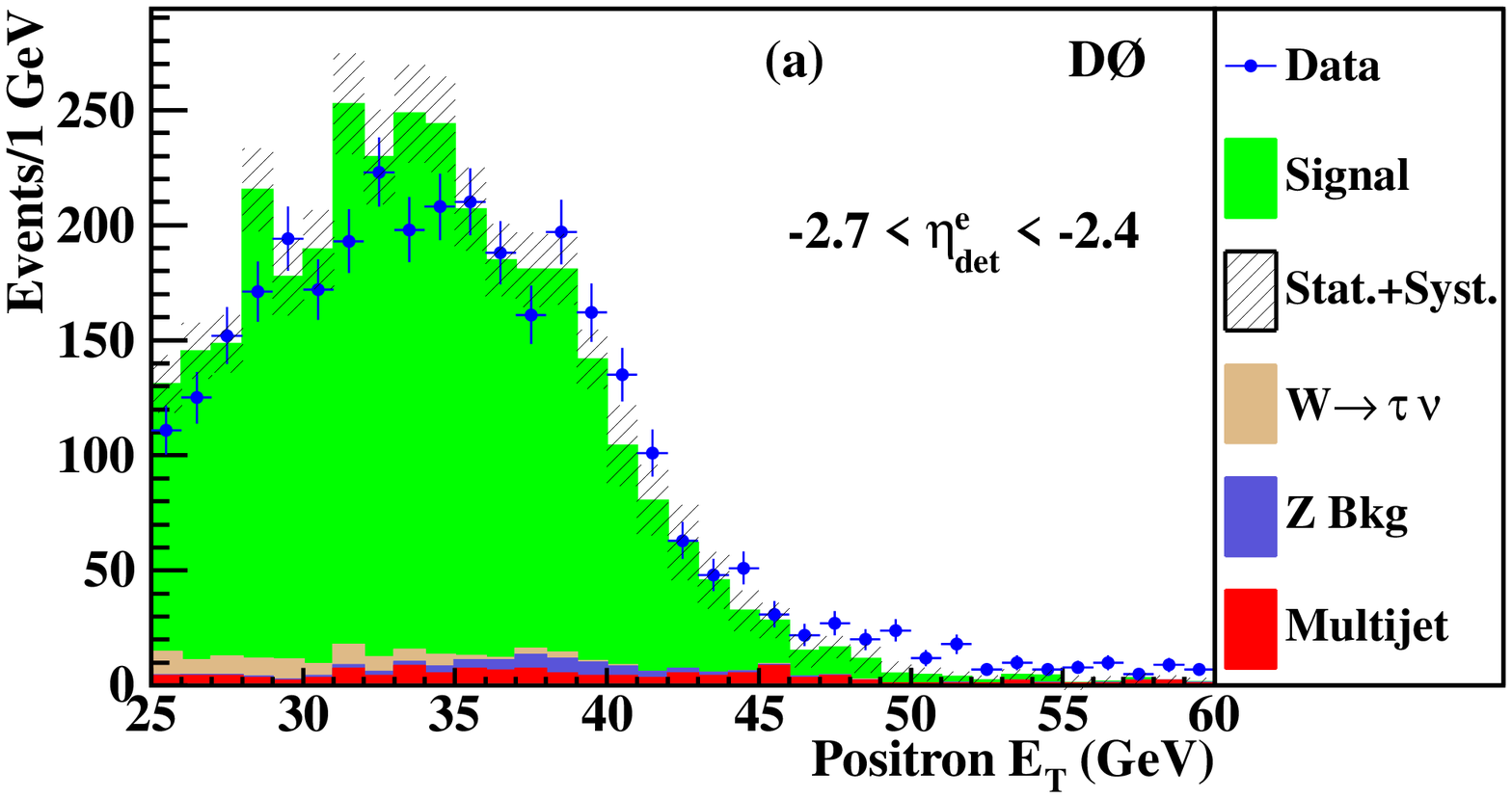, scale=0.44}
\epsfig{file=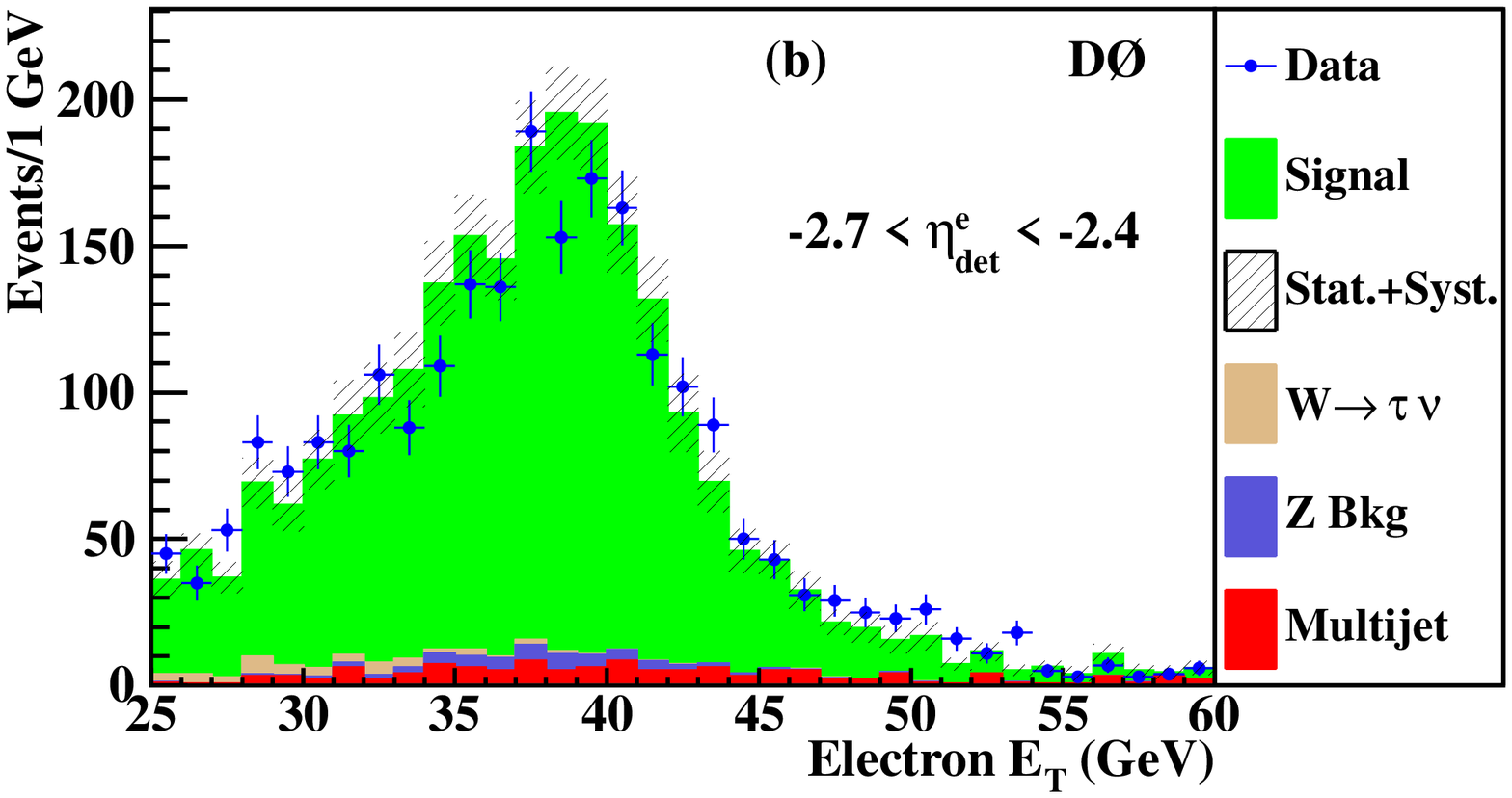, scale=0.44}
\caption{(color online). Comparisons of the electron $E_T$ distribution between data and the
sum of signal and background predictions for selected (a) $W^+$ and (b) $W^-$ 
events in the range $-2.7<\eta^e_{\text{det}}<-2.4$. 
This bin is chosen as the one showing the poorest agreement.
In different rapidity regions, the values of $x$ for $u$ and $d$ quarks are different,
thus the $W^+$ and $W^-$ distributions are different from each other.
The data uncertainty only represents the statistical uncertainty,
and the bands represent the systematic uncertainty on the signal plus backgrounds, without 
any uncertainty from MC modelings.}
 \label{fig:met_compare}
\end{figure*}

The electron charge asymmetry measurement is determined from the numbers of electrons
and positrons in each $\eta^e$ bin. Thus, the differences in the distribution of kinematic quantities between
data and MC will affect the measured asymmetry results. 
In order to minimize the effects from differences between data and MC, the MC sample
is tuned to describe the data, but even after all of the corrections are 
applied, there are discrepancies in the high rapidity region, as shown in Fig.~\ref{fig:met_compare}
for the $E_T^e$ distribution with events in the $-2.7$ to $-2.4$ $\eta^e_{\text{det}}$ range. 
The MC may not be well modeled in some electron $\eta^e$ bins,
and we assign a systematic uncertainty to account for this. 

To estimate the uncertainties from MC sample mismodeling, 
we reweight the $W^+$ and $W^-$ events separately in each electron $\eta^e$ bin,
with the data/MC ratio (obtained separately for $W^+$ and $W^-$ events)
as a function of $M_T$, $E_T^e$, and $\met$.
The larger deviation between the samples with and without the reweighting factors
in each bin is assigned as the symmetric systematic 
uncertainty in that bin. The uncertainties from $M_T$, electron $E_T^e$, and $\met$, 
are combined in quadrature to arrive at a single total MC modeling uncertainty.

\subsection{Systematic uncertainty from background modeling}
The statistical uncertainties in the background MC samples
and the uncertainty in the integrated luminosity measurement contribute 
to the overall asymmetry systematic uncertainty.
For the $Z\rightarrow ee$, $Z\rightarrow \tau\tau$, and $W\rightarrow \tau\nu$ backgrounds 
using NLO cross sections, we vary the integrated luminosity by $\pm$6.1\%~\cite{d0lumi},
extract the asymmetries with the varied integrated luminosity, and take the
larger variation as the systematic uncertainty due to the luminosity.
To study systematic uncertainties from the multijet $M_T$ shape,
we vary the reversed shower shape cuts, 
extract the asymmetries with the different multijet $M_T$ shapes, and take the
larger variation as the systematic uncertainty.
Similarly for the systematic uncertainty in the multijet fraction,
we vary the multijet scale factors in the template fitting by $\pm$1 standard deviation,
extract the asymmetries with the different multijet contributions,
and take the larger variation in each bin as a symmetric systematic uncertainty in that bin.
The uncertainties from luminosity and multijet background are combined in quadrature to arrive
at a single total background modeling uncertainty. 

\subsection{Systematic uncertainty from the electron charge misidentification}
We vary the charge misidentification probability ($Q_{\text{mis}}$) in data by $\pm$1 standard deviation, 
extract the asymmetries with the varied charge misidentification, 
and take the larger variation in each bin as a symmetric systematic uncertainty in that bin.

\subsection{Systematic uncertainty from the unfolding procedure}
To determine the systematic uncertainty due to the limited statistics used
in the calculation of the migration matrices,
we divide the MC sample into ten sub-samples and perform ten pseudo-experiments. 
The root mean squared spread of the ten unfolded asymmetry distributions is divided by $\sqrt{10}$ for each bin,
and this is taken as the systematic uncertainty.
 
An uncertainty on the $\cal{A}\times\epsilon$ corrections arises from the statistics of the $W$ boson
MC samples. This uncertainty is determined when we study the $\cal{A}\times\epsilon$ corrections
($A^{\text{Gen}}-A^{\text{Reco}}$) by varying the $\cal{A}\times\epsilon$ corrections by $\pm$1 standard
deviation and using the larger variation of asymmetry in each bin 
as a symmetric systematic uncertainty in that bin. 

The uncertainties from migration matrices and $\cal{A}\times\epsilon$ are combined in quadrature to arrive
at a single total unfolding procedure uncertainty. 

\subsection{Correlations between systematic uncertainties}
The electron charge asymmetries are measured in different $\eta^e$ bins. 
In the estimation of systematic uncertainties, the migrations 
introduce correlations between different $\eta^e$ bins. 
To estimate the correlations of the systematic uncertainties between different $\eta^e$ bins, 
we study each systematic uncertainty individually, and after determining the correlations as 
explained next, 
we build the correlation matrix in each $\eta^e$ bin for the various systematic uncertainties.

The systematic uncertainties from the generator level, which include $W$ boson 
$p_T$ and FSR modeling, shift the electron charge asymmetry in all of the
$\eta^e$ bins simultaneously. We therefore assume this correlation in the 
asymmetry measurement is 100\%. Similarly, for the electron energy tuning, 
recoil modeling, MC modeling, background modeling, and unfolding procedure,
100\% correlation between each $\eta^e$ bin is assumed when
producing the correlation matrix. The other systematic uncertainties, e.g., EMID
and electron charge misidentification, are obtained using $Z\rightarrow ee$ events
with the same bin size as the electron asymmetry measurement. We therefore assume there is zero 
correlation between $\eta^e$ bins. With the assumptions described above and combined in
quadrature, we build the correlation matrices for each kinematic bin, which are presented 
in Tables~\ref{Tab:Syst_corr_Elec_25_Inf_Neut_25_Inf}--\ref{Tab:Syst_corr_Elec_35_Inf_Neut_35_Inf}.

%%%%%%%%%%%%%%%%%%%%%%%%%%%%%%%%%%%%%%%%%%%%%%%%%%%%%%%%%
\begin{table*}
\caption{Correlation matrix of the systematic uncertainties between different $|\eta^e|$ bins for events with $E_T^{e} > 25$~GeV, $\met > 25$~GeV. The ``$|\eta^e|$ bin" represents the indexing of the $\eta^e$ bins used in this analysis.}
\begin{tabular}{c|ccccccccccccc}
\hline
\hline
$|\eta^e|$ bin & 1 & 2 & 3 & 4 & 5 & 6 & 7 & 8 & 9 & 10 & 11 & 12 & 13   \\ \hline
 1 &  1.00 &  0.83 &  0.79 &  0.74 &  0.72 &  0.73 &  0.68 &  0.67 &  0.66 &  0.63 &  0.56 &  0.55 &  0.52 \\
 2 &  &  1.00 &  0.90 &  0.89 &  0.87 &  0.88 &  0.85 &  0.84 &  0.81 &  0.74 &  0.64 &  0.63 &  0.56 \\
 3 &  &  &  1.00 &  0.92 &  0.92 &  0.92 &  0.90 &  0.87 &  0.84 &  0.76 &  0.65 &  0.64 &  0.55 \\
 4 &  &  &  &  1.00 &  0.96 &  0.95 &  0.94 &  0.86 &  0.82 &  0.75 &  0.63 &  0.63 &  0.55 \\
 5 &  &  &  &  &  1.00 &  0.95 &  0.95 &  0.86 &  0.82 &  0.75 &  0.63 &  0.63 &  0.54 \\
 6 &  &  &  &  &  &  1.00 &  0.94 &  0.88 &  0.84 &  0.76 &  0.64 &  0.64 &  0.54 \\
 7 &  &  &  &  &  &  &  1.00 &  0.90 &  0.86 &  0.77 &  0.64 &  0.64 &  0.52 \\
 8 &  &  &  &  &  &  &  &  1.00 &  0.90 &  0.78 &  0.66 &  0.65 &  0.49 \\
 9 &  &  &  &  &  &  &  &  &  1.00 &  0.75 &  0.63 &  0.62 &  0.47 \\
 10 &  &  &  &  &  &  &  &  &  &  1.00 &  0.56 &  0.55 &  0.44 \\
 11 &  &  &  &  &  &  &  &  &  &  &  1.00 &  0.47 &  0.38 \\
 12 &  &  &  &  &  &  &  &  &  &  &  &  1.00 &  0.38 \\
 13 &  &  &  &  &  &  &  &  &  &  &  &  &  1.00 \\
\hline
\hline
\end{tabular}
\label{Tab:Syst_corr_Elec_25_Inf_Neut_25_Inf}
\end{table*}

\begin{table*}
\caption{Correlation matrix of the systematic uncertainties between different $|\eta^e|$ bins for events with $25 < E_T^{e} < 35$~GeV, $\met > 25$~GeV.}
\begin{tabular}{c|ccccccccccccc}
\hline
\hline
$|\eta^e|$ bin& 1 & 2 & 3 & 4 & 5 & 6 & 7 & 8 & 9 & 10 & 11 & 12 & 13   \\ \hline
 1 &  1.00 &  0.54 &  0.51 &  0.52 &  0.53 &  0.54 &  0.53 &  0.53 &  0.47 &  0.44 &  0.42 &  0.41 &  0.40 \\
 2 &  &  1.00 &  0.61 &  0.65 &  0.65 &  0.64 &  0.61 &  0.61 &  0.55 &  0.53 &  0.50 &  0.52 &  0.46 \\
 3 &  &  &  1.00 &  0.69 &  0.69 &  0.66 &  0.63 &  0.63 &  0.59 &  0.57 &  0.53 &  0.54 &  0.48 \\
 4 &  &  &  &  1.00 &  0.76 &  0.74 &  0.72 &  0.70 &  0.64 &  0.63 &  0.57 &  0.62 &  0.53 \\
 5 &  &  &  &  &  1.00 &  0.76 &  0.75 &  0.69 &  0.62 &  0.61 &  0.57 &  0.63 &  0.54 \\
 6 &  &  &  &  &  &  1.00 &  0.75 &  0.70 &  0.63 &  0.61 &  0.56 &  0.63 &  0.54 \\
 7 &  &  &  &  &  &  &  1.00 &  0.72 &  0.64 &  0.61 &  0.55 &  0.63 &  0.54 \\
 8 &  &  &  &  &  &  &  &  1.00 &  0.69 &  0.64 &  0.56 &  0.59 &  0.53 \\
 9 &  &  &  &  &  &  &  &  &  1.00 &  0.63 &  0.54 &  0.54 &  0.49 \\
 10 &  &  &  &  &  &  &  &  &  &  1.00 &  0.51 &  0.52 &  0.47 \\
 11 &  &  &  &  &  &  &  &  &  &  &  1.00 &  0.47 &  0.42 \\
 12 &  &  &  &  &  &  &  &  &  &  &  &  1.00 &  0.46 \\
 13 &  &  &  &  &  &  &  &  &  &  &  &  &  1.00 \\
\hline
\hline
\end{tabular}
\label{Tab:Syst_corr_Elec_25_35_Neut_25_Inf}
\end{table*}

\begin{table*}
\caption{Correlation matrix of the systematic uncertainties between different $|\eta^e|$ bins for events with $25 < E_T^{e} < 35$~GeV, $25< \met < 35$~GeV.}
\begin{tabular}{c|ccccccccccccc}
\hline
\hline
$|\eta^e|$ bin & 1 & 2 & 3 & 4 & 5 & 6 & 7 & 8 & 9 & 10 & 11 & 12 & 13   \\ \hline
 1 &  1.00 &  0.68 &  0.65 &  0.67 &  0.68 &  0.69 &  0.67 &  0.66 &  0.60 &  0.57 &  0.55 &  0.53 &  0.51 \\
 2 &  &  1.00 &  0.72 &  0.75 &  0.75 &  0.75 &  0.73 &  0.74 &  0.70 &  0.67 &  0.62 &  0.63 &  0.58 \\
 3 &  &  &  1.00 &  0.75 &  0.76 &  0.75 &  0.73 &  0.75 &  0.72 &  0.69 &  0.63 &  0.64 &  0.59 \\
 4 &  &  &  &  1.00 &  0.81 &  0.80 &  0.79 &  0.79 &  0.78 &  0.75 &  0.68 &  0.71 &  0.64 \\
 5 &  &  &  &  &  1.00 &  0.81 &  0.81 &  0.80 &  0.78 &  0.76 &  0.70 &  0.73 &  0.64 \\
 6 &  &  &  &  &  &  1.00 &  0.82 &  0.79 &  0.76 &  0.74 &  0.69 &  0.73 &  0.64 \\
 7 &  &  &  &  &  &  &  1.00 &  0.79 &  0.75 &  0.74 &  0.69 &  0.75 &  0.65 \\
 8 &  &  &  &  &  &  &  &  1.00 &  0.77 &  0.74 &  0.68 &  0.71 &  0.63 \\
 9 &  &  &  &  &  &  &  &  &  1.00 &  0.74 &  0.67 &  0.70 &  0.62 \\
 10 &  &  &  &  &  &  &  &  &  &  1.00 &  0.65 &  0.70 &  0.61 \\
 11 &  &  &  &  &  &  &  &  &  &  &  1.00 &  0.64 &  0.56 \\
 12 &  &  &  &  &  &  &  &  &  &  &  &  1.00 &  0.60 \\
 13 &  &  &  &  &  &  &  &  &  &  &  &  &  1.00 \\
\hline
\hline
\end{tabular}
\label{Tab:Syst_corr_Elec_25_35_Neut_25_35}
\end{table*}

\begin{table*}
\caption{Correlation matrix of the systematic uncertainties between different $|\eta^e|$ bins for events with $E_T^{e} > 35$~GeV, $\met > 25$~GeV.}
\begin{tabular}{c|ccccccccccccc}
\hline
\hline
$|\eta^e|$ bin & 1 & 2 & 3 & 4 & 5 & 6 & 7 & 8 & 9 & 10 & 11 & 12 & 13   \\ \hline
 1 &  1.00 &  0.87 &  0.81 &  0.77 &  0.75 &  0.75 &  0.69 &  0.69 &  0.66 &  0.65 &  0.58 &  0.53 &  0.50 \\
 2 &  &  1.00 &  0.94 &  0.91 &  0.89 &  0.90 &  0.87 &  0.88 &  0.85 &  0.79 &  0.69 &  0.61 &  0.54 \\
 3 &  &  &  1.00 &  0.95 &  0.94 &  0.94 &  0.92 &  0.94 &  0.90 &  0.83 &  0.72 &  0.62 &  0.54 \\
 4 &  &  &  &  1.00 &  0.97 &  0.96 &  0.96 &  0.94 &  0.87 &  0.83 &  0.73 &  0.60 &  0.52 \\
 5 &  &  &  &  &  1.00 &  0.96 &  0.96 &  0.94 &  0.86 &  0.83 &  0.73 &  0.60 &  0.51 \\
 6 &  &  &  &  &  &  1.00 &  0.95 &  0.94 &  0.88 &  0.83 &  0.72 &  0.60 &  0.52 \\
 7 &  &  &  &  &  &  &  1.00 &  0.94 &  0.87 &  0.83 &  0.72 &  0.59 &  0.49 \\
 8 &  &  &  &  &  &  &  &  1.00 &  0.91 &  0.83 &  0.71 &  0.60 &  0.51 \\
 9 &  &  &  &  &  &  &  &  &  1.00 &  0.79 &  0.66 &  0.58 &  0.49 \\
 10 &  &  &  &  &  &  &  &  &  &  1.00 &  0.63 &  0.53 &  0.46 \\
 11 &  &  &  &  &  &  &  &  &  &  &  1.00 &  0.46 &  0.40 \\
 12 &  &  &  &  &  &  &  &  &  &  &  &  1.00 &  0.35 \\
 13 &  &  &  &  &  &  &  &  &  &  &  &  &  1.00 \\
\hline
\hline
\end{tabular}
\label{Tab:Syst_corr_Elec_35_Inf_Neut_25_Inf}
\end{table*}

\begin{table*}
\caption{Correlation matrix of the systematic uncertainties between different $|\eta^e|$ bins for events with $E_T^{e} > 35$~GeV, $\met > 35$~GeV.}
\begin{tabular}{c|ccccccccccccc}
\hline
\hline
$|\eta^e|$ bin & 1 & 2 & 3 & 4 & 5 & 6 & 7 & 8 & 9 & 10 & 11 & 12 & 13   \\ \hline
 1 &  1.00 &  0.91 &  0.85 &  0.80 &  0.74 &  0.76 &  0.63 &  0.62 &  0.62 &  0.58 &  0.51 &  0.49 &  0.49 \\
 2 &  &  1.00 &  0.92 &  0.89 &  0.86 &  0.86 &  0.76 &  0.77 &  0.73 &  0.65 &  0.57 &  0.50 &  0.50 \\
 3 &  &  &  1.00 &  0.93 &  0.92 &  0.90 &  0.84 &  0.86 &  0.79 &  0.68 &  0.59 &  0.49 &  0.48 \\
 4 &  &  &  &  1.00 &  0.94 &  0.92 &  0.88 &  0.89 &  0.81 &  0.70 &  0.60 &  0.49 &  0.47 \\
 5 &  &  &  &  &  1.00 &  0.92 &  0.90 &  0.91 &  0.81 &  0.70 &  0.60 &  0.48 &  0.45 \\
 6 &  &  &  &  &  &  1.00 &  0.87 &  0.87 &  0.79 &  0.68 &  0.59 &  0.48 &  0.45 \\
 7 &  &  &  &  &  &  &  1.00 &  0.88 &  0.78 &  0.66 &  0.57 &  0.46 &  0.41 \\
 8 &  &  &  &  &  &  &  &  1.00 &  0.80 &  0.67 &  0.57 &  0.44 &  0.40 \\
 9 &  &  &  &  &  &  &  &  &  1.00 &  0.61 &  0.52 &  0.42 &  0.39 \\
 10 &  &  &  &  &  &  &  &  &  &  1.00 &  0.47 &  0.40 &  0.36 \\
 11 &  &  &  &  &  &  &  &  &  &  &  1.00 &  0.36 &  0.32 \\
 12 &  &  &  &  &  &  &  &  &  &  &  &  1.00 &  0.30 \\
 13 &  &  &  &  &  &  &  &  &  &  &  &  &  1.00 \\
\hline
\hline
\end{tabular}
\label{Tab:Syst_corr_Elec_35_Inf_Neut_35_Inf}
\end{table*}

%%%%%%%%%%%%%%%%%%%%%%%%%%%
\section{Results}

\begin{center}
\begin{figure*}
\epsfig{file=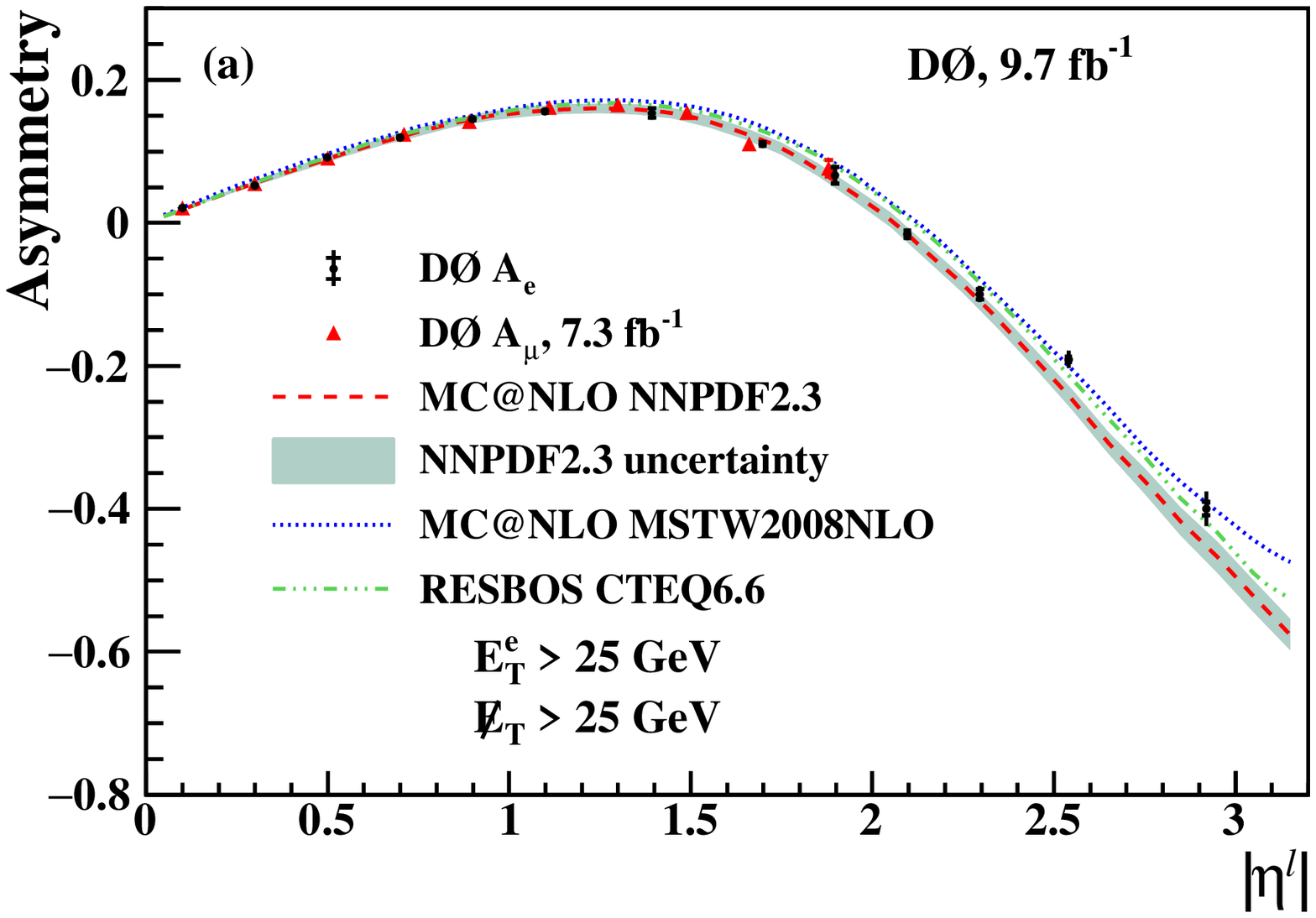, scale=0.43}
\epsfig{file=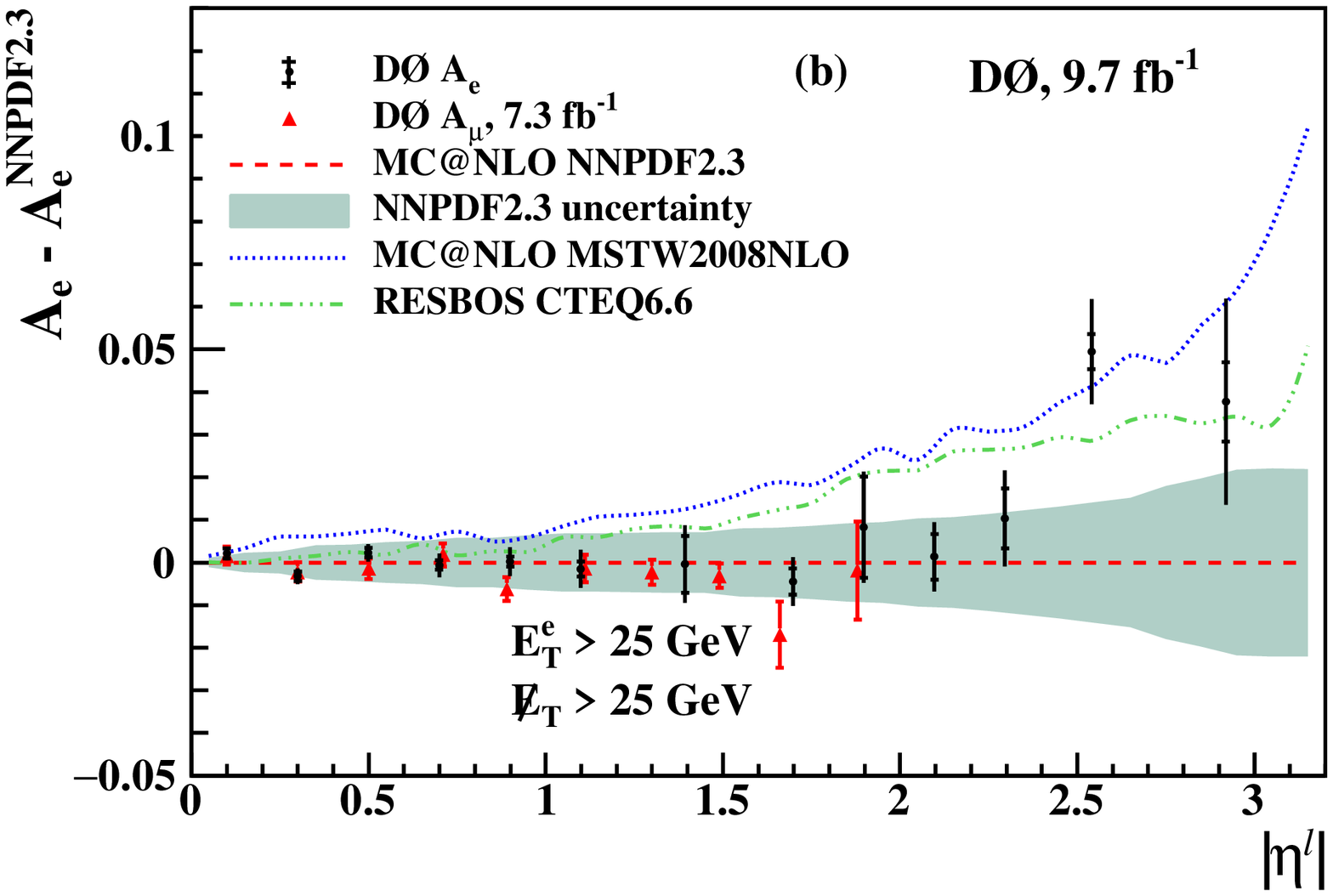, scale=0.43}
\caption{(color online). The lepton charge asymmetry distribution after CP folding 
with symmetric kinematic cuts $E_T^e > 25$~GeV, $\met > 25$~GeV. (a) Comparison between
the measured asymmetry and predictions and (b) the differences between the data and MC predictions and the predicted central value from 
{\sc mc@nlo} using the NNPDF2.3 PDF set.
The black dots show the measured electron charge asymmetry, with the
horizontal bars showing the statistical uncertainty and the vertical lines showing the total uncertainty.
The red triangles show the published D0 muon charge asymmetry~\cite{d0_muon}.
The red dashed lines and cyan bands are the central value and uncertainty band from
{\sc mc@nlo} using the NNPDF2.3 PDF sets. 
The blue dotted lines show the prediction from {\sc mc@nlo} using the MSTW2008NLO central PDF set, 
and the green dot-dashed lines show the prediction from {\sc resbos} using the CTEQ6.6 central PDF set.}
\label{fig:compare_asym1}
\end{figure*}
\end{center}

\begin{center}
\begin{figure*}
\epsfig{file=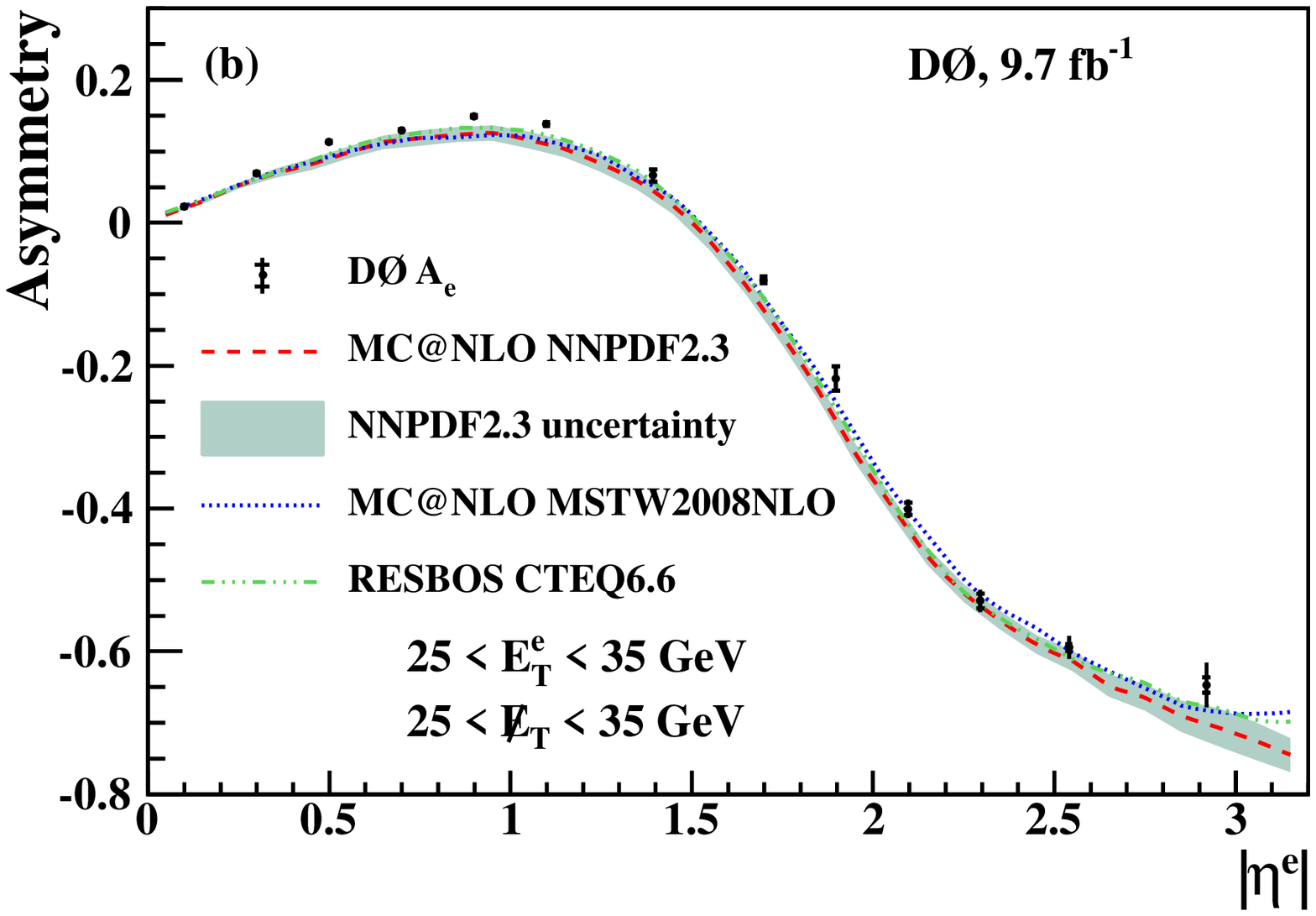, scale=0.43}
\epsfig{file=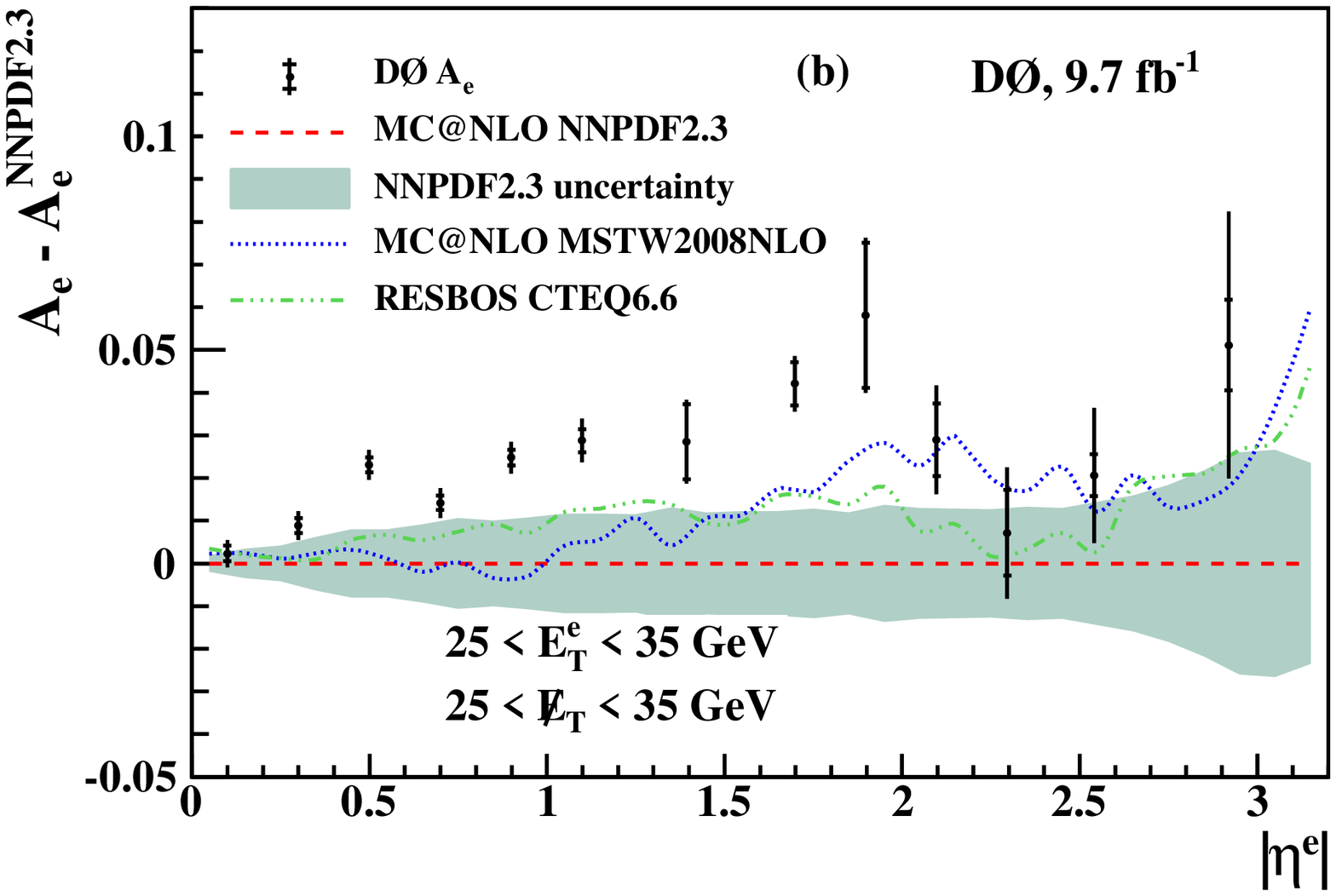, scale=0.43}

\caption{(color online). The electron charge asymmetry distribution after CP folding 
with symmetric kinematic cuts $25<E_T^e < 35$~GeV, $25<\met < 35$~GeV. (a) Comparison between
the measured asymmetry and predictions and (b) the differences between the data and MC predictions and the predicted central value from 
{\sc mc@nlo} using the NNPDF2.3 PDF set.
The black dots show the measured electron charge asymmetry, with the
horizontal bars showing statistical uncertainty and the vertical lines showing the total uncertainty.
The red dashed lines and cyan bands are the central value and uncertainty band from
{\sc mc@nlo} using the NNPDF2.3 PDF sets. 
The blue dotted lines show the prediction from {\sc mc@nlo} using the MSTW2008NLO central PDF set, 
and the green dot-dashed lines show the prediction from {\sc resbos} using the CTEQ6.6 central PDF set.} 
\label{fig:compare_asym2}
\end{figure*}
\end{center}

\begin{center}
\begin{figure*}
\epsfig{file=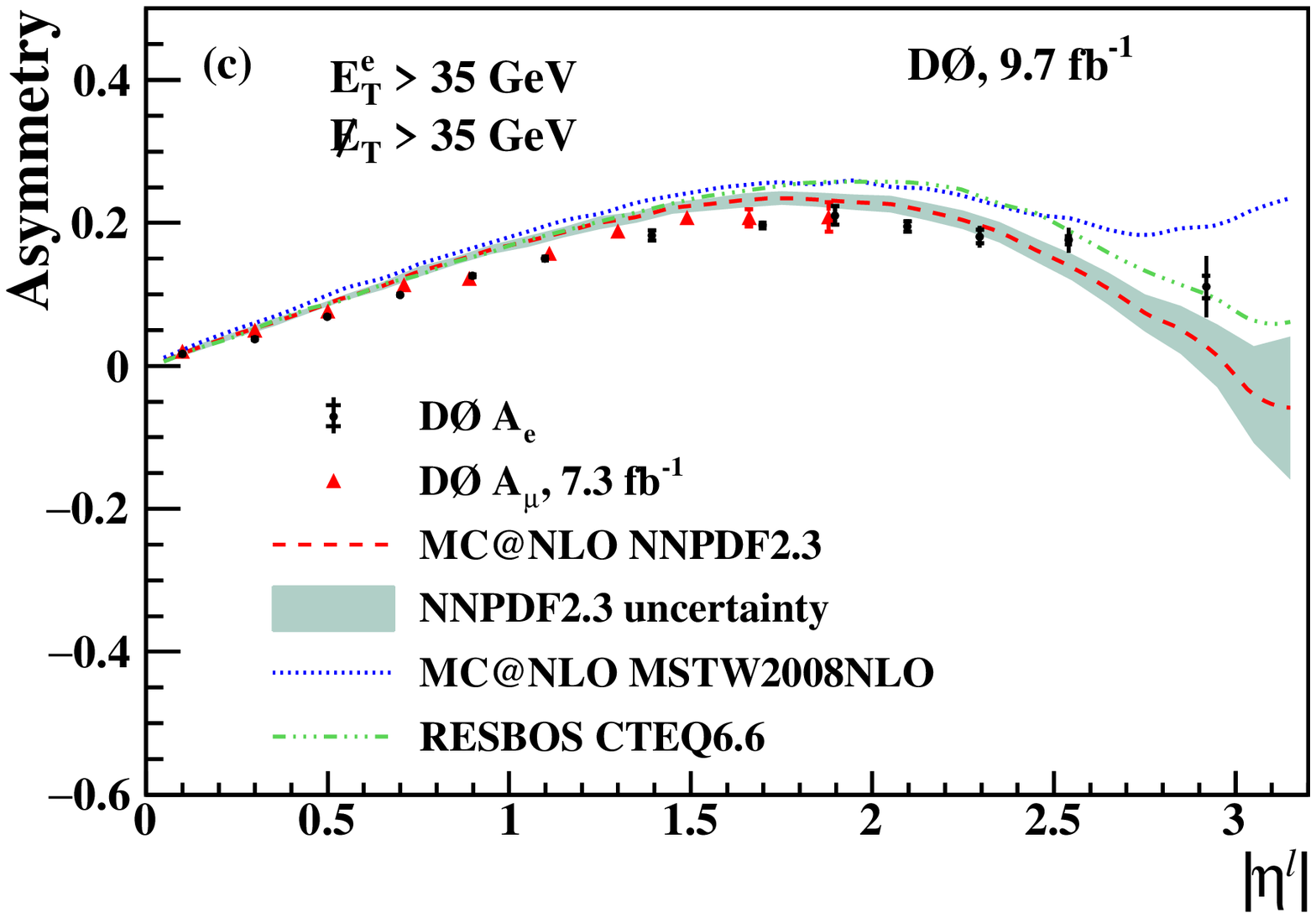, scale=0.43}
\epsfig{file=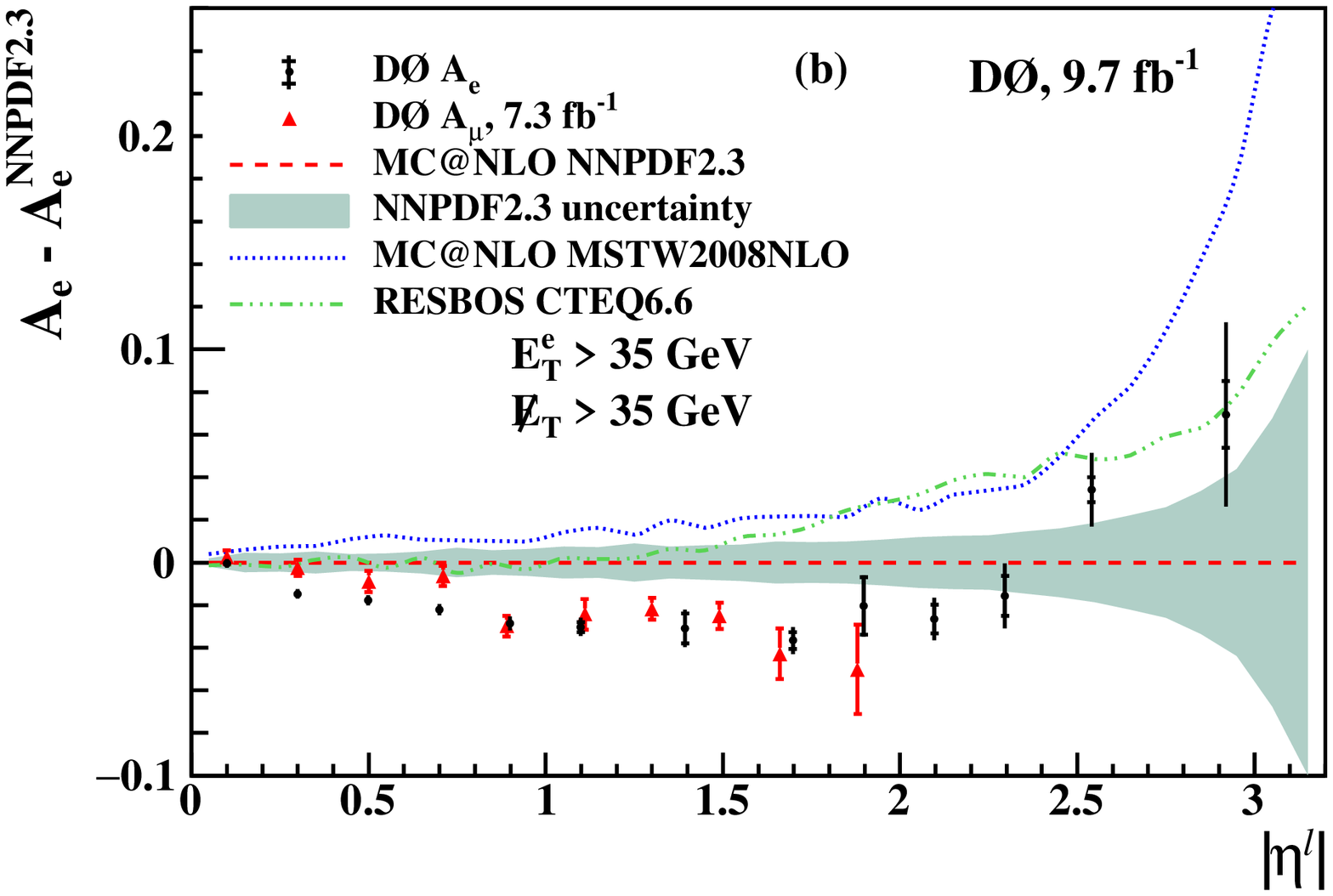, scale=0.43}
\caption{(color online). The lepton charge asymmetry distribution after CP folding 
with symmetric kinematic cuts $E_T^e > 35$~GeV, $\met > 35$~GeV. (a) Comparison between
the measured asymmetry and predictions and (b) the differences between the data and MC predictions and the predicted central value from 
{\sc mc@nlo} using the NNPDF2.3 PDF set.
The black dots show the measured electron charge asymmetry, with the
horizontal bars showing statistical uncertainty and the vertical lines showing the total uncertainty.
The red triangles show the published D0 muon charge asymmetry~\cite{d0_muon}.
The red dashed lines and cyan bands are the central value and uncertainty band from
{\sc mc@nlo} using the NNPDF2.3 PDF sets. 
The blue dotted lines show the prediction from {\sc mc@nlo} using the MSTW2008NLO central PDF set, 
and the green dot-dashed lines show the prediction from {\sc resbos} using the CTEQ6.6 central PDF set.} 
\label{fig:compare_asym3}
\end{figure*}
\end{center}

\begin{center}
\begin{figure*}
\epsfig{file=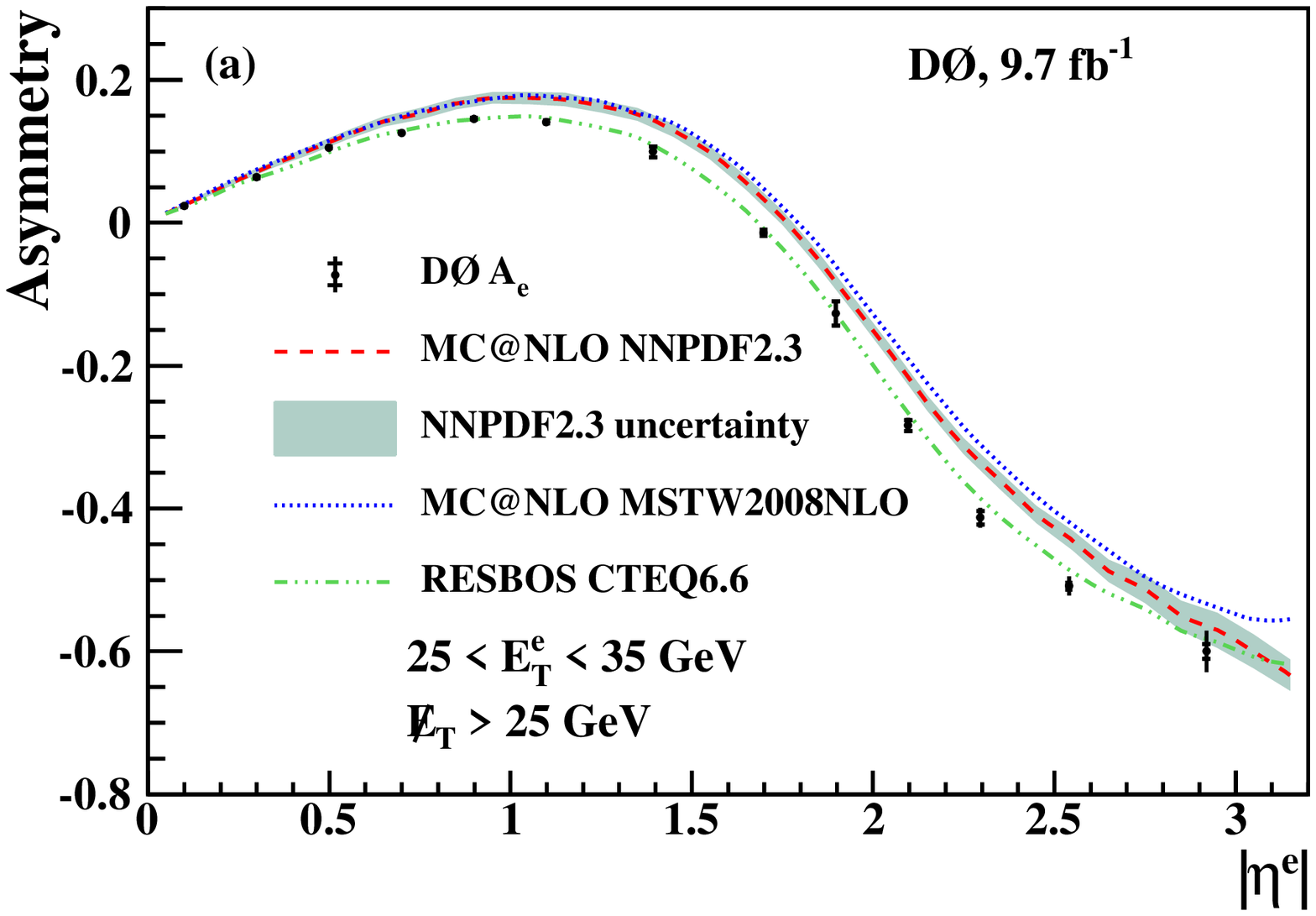, scale=0.43}
\epsfig{file=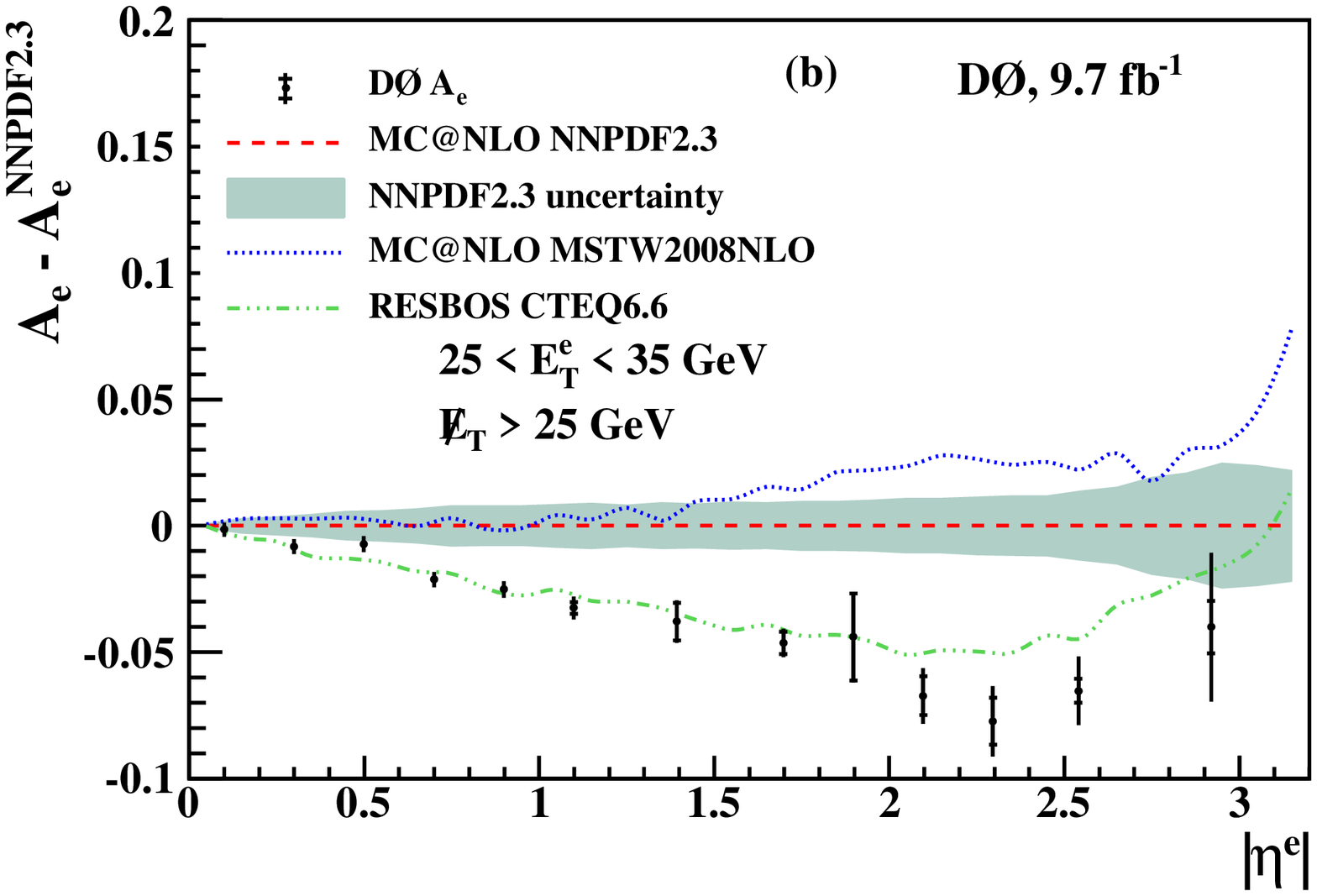, scale=0.43}
\caption{(color online). The electron charge asymmetry distribution after CP folding 
with asymmetric kinematic cuts $25<E_T^e < 35$~GeV, $\met > 25$~GeV. (a) Comparison between
the measured asymmetry and predictions and (b) the differences between the data and MC predictions and the predicted central value from 
{\sc mc@nlo} using the NNPDF2.3 PDF set.
The black dots show the measured electron charge asymmetry, with the
horizontal bars showing statistical uncertainty and the vertical lines showing the total uncertainty.
The red dashed lines and cyan bands are the central value and uncertainty band from
{\sc mc@nlo} using the NNPDF2.3 PDF sets. 
The blue dotted lines show the prediction from {\sc mc@nlo} using the MSTW2008NLO central PDF set, 
and the green dot-dashed lines show the prediction from {\sc resbos} using the CTEQ6.6 central PDF set.}
\label{fig:compare_asym4}
\end{figure*}
\end{center}

\begin{center}
\begin{figure*}
\epsfig{file=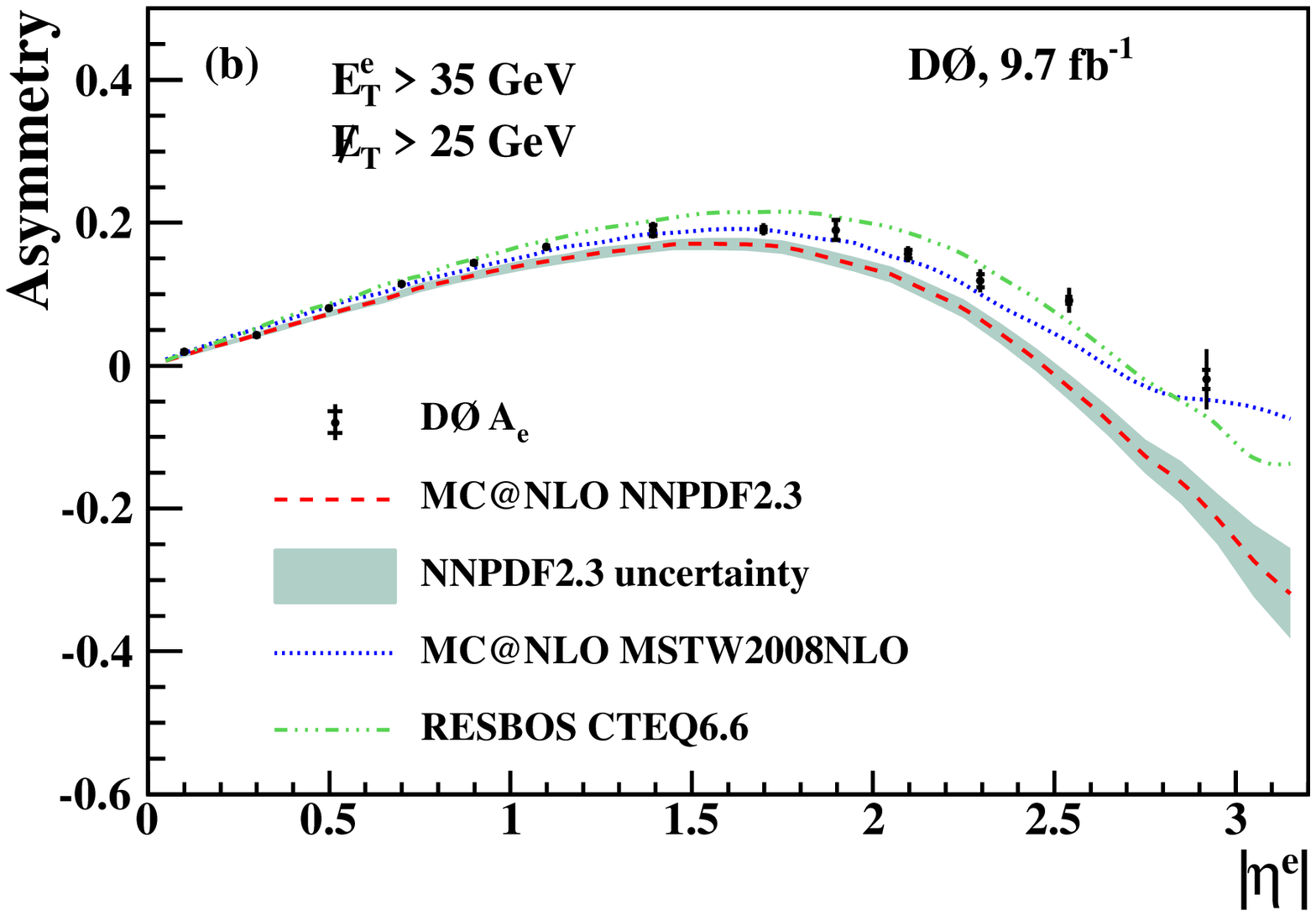, scale=0.43}
\epsfig{file=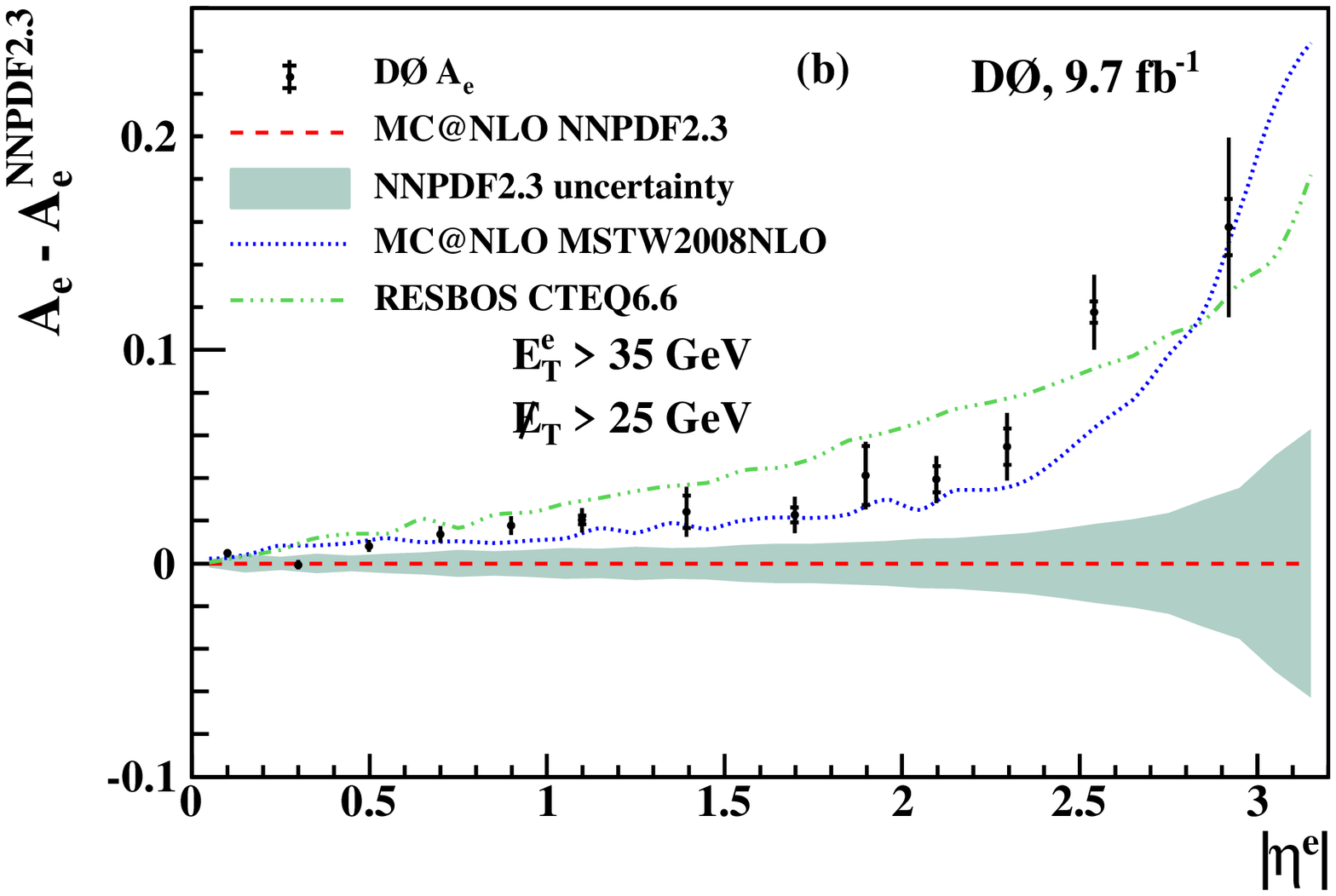, scale=0.43}
\caption{(color online). The electron charge asymmetry distribution after CP folding 
with asymmetric kinematic cuts $E_T^e > 35$~GeV, $\met > 25$~GeV. (a) Comparison between
the measured asymmetry and predictions and (b) the differences between the data and MC predictions and the predicted central value from 
{\sc mc@nlo} using the NNPDF2.3 PDF set.
The black dots show the measured electron charge asymmetry, with the
horizontal bars showing statistical uncertainty and the vertical lines showing the total uncertainty.
The red dashed lines and cyan bands are the central value and uncertainty band from
{\sc mc@nlo} using the NNPDF2.3 PDF sets. 
The blue dotted lines show the prediction from {\sc mc@nlo} using the MSTW2008NLO central PDF set, 
and the green dot-dashed lines show the prediction from {\sc resbos} using the CTEQ6.6 central PDF set.}
\label{fig:compare_asym5}
\end{figure*}
\end{center}

The asymmetry results for $\eta^e>0$ are found to be consistent with those for $\eta^e<0$, 
so we assume CP invariance with $A(\eta^e)$ being equivalent to $-A(-\eta^e)$.
The data for $\eta^e<0$ are folded appropriately with those for $\eta^e>0$ to 
increase the statistics, and results are presented for $|\eta^e|$.
We perform the electron charge asymmetry measurement in five kinematic bins.
Results from the different kinematic bins probe different ranges of $y_W$, and thus
different ranges of the fraction of proton momentum carried by the parton. 
The measured electron asymmetries with symmetric kinematic cuts on $E_T^e$ and $\met$
($E_T^e>25$~GeV, $\met > 25$~GeV; $25<E_T^e<35$~GeV, $25<\met < 35$~GeV;
and $E_T^e>35$~GeV, $\met > 35$~GeV)
and the differences between measured values and 
{\sc mc@nlo}~\cite{mcatnlo} with the NNPDF2.3~\cite{nnpdf} PDF set predictions 
are shown in Figs.~\ref{fig:compare_asym1}--\ref{fig:compare_asym3}.
For the measured electron asymmetries with asymmetric kinematic cuts
($25<E_T^e<35$~GeV, $\met > 25$~GeV; $E_T^e>35$~GeV, $\met > 25$~GeV),
the differences between measured values and predictions are shown in Figs.~\ref{fig:compare_asym4} and
~\ref{fig:compare_asym5}. 
The PDF bands are obtained from {\sc mc@nlo} using the NNPDF2.3 
NLO PDF uncertainty sets.
The central value of predictions from {\sc mc@nlo} with {\sc herwig}~\cite{herwig} using the MSTW2008NLO~\cite{mstw} central
PDF set and from {\sc resbos} with {\sc photos}~\cite{photos} 
(for QED final state radiation) using the CTEQ6.6 central PDF set are also included. 
The theory curves are generated with selection criteria applied to the electron
and neutrino generator-level transverse momenta, 
with all the radiated photons merged into 
the electron if they fall within a cone of radius $\Delta R<0.3$.
Generator-level $W$ bosons are further required to have a transverse mass in the range 
between 50 and 200 GeV and to have a transverse momentum less than 120 GeV.

The measured electron charge asymmetries are consistent with predictions for the inclusive kinematic bin
$E_T^e>25$~GeV, $\met > 25$~GeV. In the kinematic bins with asymmetric cuts 
($25<E_T^e<35$~GeV, $\met > 25$~GeV; $E_T^e>35$~GeV, $\met > 25$~GeV),
the measured electron charge asymmetries are consistent with predictions from
{\sc resbos} using the CTEQ6.6 central PDF set, but in the kinematic bins with symmetric cuts
($25<E_T^e<35$~GeV, $25<\met < 35$~GeV; $E_T^e>35$~GeV, $\met > 35$),
the measured electron charge asymmetries are not consistent with any of the considered predictions,
with the $\chi^2/d.o.f.$ between measured asymmetry and 
the {\sc mc@nlo} with NNPDF2.3 predictions equal to 47.1/13 and 95.5/13, respectively. 
The results presented here are in good agreement with those of
Ref.~\cite{d0_muon} for the muon charge asymmetry for both $E_T^{\ell}>25$~GeV, $\met > 25$~GeV
and $E_T^{\ell}>35$~GeV, $\met > 35$~GeV. This agreement is noteworthy since
the analysis techniques and dominant systematic uncertainties in the two measurements are quite different. 
The results are consistent with the previously published results~\cite{d0_results_em_old}
in the $|\eta^e| < 2$ region, and disagree in the high $|\eta^e|$ region.
In this paper, compared to the previous results~\cite{d0_results_em_old},
there are several improvements for the modeling of electrons in high $|\eta^e|$ region, 
including the $\eta$-dependent energy scale corrections, recoil system modeling, and 
the positron/electron identification efficiency corrections. 
This measurement thus supersedes the results of Ref.~\cite{d0_results_em_old}.

The electron charge asymmetry measurements for various bins in $\eta^e$ 
for the five kinematic regions, and 
their uncertainties together with {\sc mc@nlo} predictions using the NNPDF2.3 
PDF sets are listed in Tables~\ref{Tab:Data_lept_fold_asym1}--\ref{Tab:Data_lept_fold_asym3}. In most $\eta^e$ bins and kinematic bins, the experimental uncertainties
are smaller than the uncertainties from the predictions, especially in the high $\eta^e$ region, 
demonstrating the importance of this analysis for improving the accuracy of future PDF fits.

\begin{table} 
\begin{center} 
\caption{CP-folded electron charge asymmetry for data and predictions from {\sc mc@nlo} 
using NNPDF2.3 PDFs multiplied by 100. 
$\langle|\eta^e|\rangle$ is the cross section weighted average of electron pseudorapidity 
in each bin from {\sc resbos} with {\sc photos}. For data, the first uncertainty is statistical 
and the second is systematic. The uncertainties on the prediction are due to uncertainties on the PDFs.}
\begin{tabular}{crr} 
\hline 
\hline 
 \multirow{3}{*}{$\langle|\eta^e|\rangle$} & \multicolumn{2}{c}{$E_T^{e} > 25$ GeV}        \\ 
   & \multicolumn{2}{c}{$\met > 25$ GeV}     \\ \cline{2-3} 
   & \multicolumn{1}{c}{Data} & Prediction     \\ \hline 
$0.10$ & $2.10 \pm 0.12 \pm 0.11$ &$1.90 \pm 0.16$    \\
$0.30$ & $5.23 \pm 0.11 \pm 0.14$ &$5.55 \pm 0.31$    \\
$0.50$ & $9.16 \pm 0.11 \pm 0.18$ &$8.93 \pm 0.44$    \\
$0.70$ & $11.97 \pm 0.11 \pm 0.25$ &$12.04 \pm 0.54$    \\
$0.90$ & $14.52 \pm 0.12 \pm 0.32$ &$14.50 \pm 0.60$    \\
$1.10$ & $15.59 \pm 0.18 \pm 0.41$ &$15.74 \pm 0.66$    \\
$1.39$ & $15.37 \pm 0.67 \pm 0.61$ &$15.41 \pm 0.70$    \\
$1.70$ & $11.05 \pm 0.31 \pm 0.49$ &$11.50 \pm 0.83$    \\
$1.90$ & $6.66 \pm 1.19 \pm 0.53$ &$5.84 \pm 0.92$    \\
$2.10$ & $-1.55 \pm 0.53 \pm 0.61$ &$-1.68 \pm 1.03$    \\
$2.30$ & $-9.97 \pm 0.71 \pm 0.88$ &$-11.00 \pm 1.17$    \\
$2.54$ & $-19.10 \pm 0.41 \pm 1.16$ &$-24.05 \pm 1.38$    \\
$2.92$ & $-39.97 \pm 0.93 \pm 2.23$ &$-43.73 \pm 1.94$    \\
\hline 
\hline 
\end{tabular} 
\label{Tab:Data_lept_fold_asym1} 
\end{center} 
\end{table} 

\begin{table*} 
\begin{center} 
\caption{CP-folded electron charge asymmetry for data and predictions from {\sc mc@nlo} 
using NNPDF2.3 PDFs multiplied by 100.
$\langle|\eta^e|\rangle$ is the cross section weighted average of electron pseudorapidity 
in each bin from {\sc resbos} with {\sc photos}. For data, the first uncertainty is statistical 
and the second is systematic. The uncertainties on the prediction are due to uncertainties on the PDFs.}
\begin{tabular}{crrrr} 
\hline 
\hline 
 \multirow{3}{*}{$\langle|\eta^e|\rangle$} &   \multicolumn{2}{c}{$25 < E_T^{e} < 35$ GeV}  & \multicolumn{2}{c}{$25 < E_T^{e} < 35$ GeV}        \\ 
  & \multicolumn{2}{c}{$\met > 25$ GeV}  & \multicolumn{2}{c}{$25< \met < 35$ GeV}      \\ \cline{2-5} 
  & \multicolumn{1}{c}{Data} & Prediction  & \multicolumn{1}{c}{Data} & Prediction  \\ \hline 
$0.10$ & $2.32 \pm 0.16 \pm 0.24$ &$2.47 \pm 0.21$    &$2.30 \pm 0.19 \pm 0.27$ &$2.07 \pm 0.24$    \\
$0.30$ & $6.36 \pm 0.15 \pm 0.24$ &$7.18 \pm 0.38$    &$6.93 \pm 0.18 \pm 0.28$ &$6.04 \pm 0.51$    \\
$0.50$ & $10.53 \pm 0.15 \pm 0.27$ &$11.26 \pm 0.60$    &$11.31 \pm 0.17 \pm 0.31$ &$9.00 \pm 0.78$    \\
$0.70$ & $12.60 \pm 0.14 \pm 0.28$ &$14.73 \pm 0.73$    &$12.97 \pm 0.17 \pm 0.31$ &$11.55 \pm 0.98$    \\
$0.90$ & $14.58 \pm 0.16 \pm 0.30$ &$17.10 \pm 0.80$    &$14.92 \pm 0.18 \pm 0.32$ &$12.44 \pm 1.02$    \\
$1.10$ & $14.11 \pm 0.23 \pm 0.39$ &$17.36 \pm 0.87$    &$13.85 \pm 0.27 \pm 0.44$ &$10.98 \pm 1.14$    \\
$1.39$ & $9.95 \pm 0.74 \pm 0.38$ &$13.74 \pm 0.87$    &$6.63 \pm 0.88 \pm 0.45$ &$3.78 \pm 1.17$    \\
$1.70$ & $-1.40 \pm 0.44 \pm 0.34$ &$3.24 \pm 0.94$    &$-7.99 \pm 0.51 \pm 0.40$ &$-12.19 \pm 1.24$    \\
$1.90$ & $-12.70 \pm 1.72 \pm 0.54$ &$-8.31 \pm 0.98$    &$-21.85 \pm 1.70 \pm 0.65$ &$-27.66 \pm 1.23$    \\
$2.10$ & $-28.36 \pm 0.76 \pm 0.78$ &$-21.63 \pm 1.09$    &$-40.05 \pm 0.85 \pm 0.95$ &$-42.94 \pm 1.28$    \\
$2.30$ & $-41.27 \pm 0.93 \pm 1.04$ &$-33.54 \pm 1.15$    &$-52.93 \pm 1.00 \pm 1.17$ &$-53.65 \pm 1.27$    \\
$2.54$ & $-50.86 \pm 0.48 \pm 1.26$ &$-44.33 \pm 1.32$    &$-59.43 \pm 0.49 \pm 1.51$ &$-61.49 \pm 1.38$    \\
$2.92$ & $-60.00 \pm 1.04 \pm 2.75$ &$-55.99 \pm 2.05$    &$-64.68 \pm 1.07 \pm 2.94$ &$-69.79 \pm 2.13$    \\

\hline 
\hline 
\end{tabular} 
\label{Tab:Data_lept_fold_asym2} 
\end{center} 
\end{table*} 

\begin{table*} 
\begin{center} 
\caption{CP-folded electron charge asymmetry for data and predictions from {\sc mc@nlo} 
using NNPDF2.3 PDFs multiplied by 100.
$\langle|\eta^e|\rangle$ is the cross section weighted average of electron pseudorapidity 
in each bin from {\sc resbos} with {\sc photos}. For data, the first uncertainty is statistical 
and the second is systematic. The uncertainties on the prediction are due to uncertainties on the PDFs.}
\begin{tabular}{crrrr} 
\hline 
\hline 
 \multirow{3}{*}{$\langle|\eta^e|\rangle$} &   \multicolumn{2}{c}{$E_T^{e} > 35$ GeV}    &   \multicolumn{2}{c}{$ E_T^{e} > 35$ GeV}       \\ 
 & \multicolumn{2}{c}{$\met > 25$ GeV}  & \multicolumn{2}{c}{ $ \met > 35$ GeV}      \\ \cline{2-5} 
 & \multicolumn{1}{c}{Data} & Prediction  & \multicolumn{1}{c}{Data} & Prediction  \\ \hline 

$0.10$ & $1.94 \pm 0.14 \pm 0.13$ &$1.47 \pm 0.28$ & $1.65 \pm 0.16 \pm 0.14$ &$1.70\pm 0.32$ \\
$0.30$ & $4.26 \pm 0.14 \pm 0.18$ &$4.33 \pm 0.35$ & $3.78 \pm 0.15 \pm 0.15$ &$5.25\pm 0.42$ \\
$0.50$ & $8.04 \pm 0.13 \pm 0.25$ &$7.22 \pm 0.39$ & $6.89 \pm 0.15 \pm 0.18$ &$8.67\pm 0.39$ \\
$0.70$ & $11.42 \pm 0.13 \pm 0.35$ &$10.06 \pm 0.55$ & $9.94 \pm 0.15 \pm 0.22$ &$12.15\pm 0.56$ \\
$0.90$ & $14.40 \pm 0.14 \pm 0.42$ &$12.62 \pm 0.60$ & $12.61 \pm 0.16 \pm 0.28$ &$15.47\pm 0.59$ \\
$1.10$ & $16.63 \pm 0.21 \pm 0.54$ &$14.60 \pm 0.68$ & $15.02 \pm 0.23 \pm 0.35$ &$18.05\pm 0.69$ \\
$1.39$ & $18.95 \pm 0.76 \pm 0.88$ &$16.53 \pm 0.75$ & $18.25 \pm 0.69 \pm 0.54$ &$21.34\pm 0.77$ \\
$1.70$ & $19.07 \pm 0.36 \pm 0.76$ &$16.80 \pm 0.91$ & $19.66 \pm 0.40 \pm 0.49$ &$23.33\pm 0.94$ \\
$1.90$ & $18.98 \pm 1.38 \pm 0.77$ &$14.86 \pm 1.00$ & $21.06 \pm 1.33 \pm 0.51$ &$23.10\pm 1.00$ \\
$2.10$ & $15.61 \pm 0.61 \pm 0.92$ &$11.68 \pm 1.16$ & $19.50 \pm 0.68 \pm 0.73$ &$22.15\pm 1.20$ \\
$2.30$ & $11.89 \pm 0.85 \pm 1.33$ &$6.43 \pm 1.34$ & $18.08 \pm 0.93 \pm 1.21$ &$19.65\pm 1.35$ \\
$2.54$ & $9.14 \pm 0.51 \pm 1.69$ &$-2.63 \pm 1.76$ & $17.58 \pm 0.58 \pm 1.63$ &$14.16\pm 1.77$ \\
$2.92$ & $-1.93 \pm 1.32 \pm 4.00$ &$-17.68 \pm 3.04$ & $11.07 \pm 1.56 \pm 4.03$ &$4.13\pm 3.51$ \\
\hline 
\hline 
\end{tabular} 
\label{Tab:Data_lept_fold_asym3} 
\end{center} 
\end{table*} 

To estimate the correlation between the measured asymmetry within 
different kinematic bins as a function of $\eta^e$ bin, we use the numbers 
of selected electrons and positrons in data and the migration matrices and acceptances
obtained from $W\rightarrow e\nu$ MC to study the statistical correlations between 
kinematic bins. The correlation matrix is defined as $c_{ij}/\sqrt{c_{ii}c_{jj}}$, 
where $c_{ij}$ represents the element of the statistical covariance matrix between
$\eta^e$ bins $i$ and $j$ calculated by summing partial derivatives of the asymmetry,
\begin{eqnarray}
c_{ij}=\sum_f\sum_k \frac{\partial A_i}{\partial f_k}\cdot\frac{\partial A_j}{\partial f_k}\cdot(\Delta f_k)^2
\end{eqnarray}
where $k$ represents the number of $\eta^e$ bins, 
$f$ represents the sources of uncertainty from each non-overlapping
kinematic bin, and $A_i$ is the measured asymmetry in $\eta^e$ bin $i$. 
The correlation matrices between central
values in each $\eta^e$ bin for the five kinematic bins after CP folding are given 
in Tables~\ref{Tab:Correlation_All_Fold_Type_1}--\ref{Tab:Correlation_All_Fold_Type_5}. From
these tables, we can see that the off-diagonal elements of the statistical correlation 
matrices are small, which
indicates the migration effects are small between different $\eta^e$ bins. 
The statistical uncertainties in Tables~\ref{Tab:Data_lept_fold_asym1}--\ref{Tab:Data_lept_fold_asym3}
are calculated using the covariance matrix, with
$\sigma_i = \sqrt{c_{ii}}$.

Besides small migration between $\eta^e$ bins, there are significant migration effects between kinematic bins,
due to detector resolution effects. 
In Table~\ref{Tab:Kine_fraction} we show the fraction of MC signal events originating in a different 
generator-level kinematic bin that are found in a given reconstruction-level bin.
The categories in Table~\ref{Tab:Kine_fraction} are not independent. 
In Table~\ref{Tab:Kine_fraction}, $20<p_T^e <25~\text{\sc or}~20<p_T^{\nu} <25$~GeV 
denotes $W$ boson events in which 
either the electron $p_T^e$ at the generator level ($E_T^e$ at the reconstruction level) 
is in the range 20 to 25~GeV or the neutrino 
$p_T^{\nu}$ ($\met$) is in the range 20 to 25~GeV, while the other lepton $p_T$ ($E_T$) is above $25$~GeV.
Also $p_T^e >25~\text{\sc and}~p_T^{\nu} >25$~GeV denotes $W$ boson events in which the 
electron $p_T^e$ ($E_T^e$) is above $25$~GeV 
and the neutrino $p_T^{\nu}$ ($\met$) is above $25$~GeV.

%%%%%%%%%%%%%%%%%%%%%%%%%%%%%%%%%%%%%%%%%%%%%%%

\section{Conclusions}
\indent In summary, we have measured the electron charge asymmetry 
in $p\bar{p}\rightarrow W^{\pm} +X\rightarrow e^{\pm}\nu+X$
events using 9.7~fb$^{-1}$ of integrated luminosity collected by 
the D0 experiment in $p\bar{p}$ collisions at $\sqrt{s}=1.96$~TeV. 
In this analysis, the electron pseudorapidity coverage is extended to $|\eta^e| = 3.2$ and is thus 
sensitive to $W$ bosons created by small- and large-$x$ partons. 
Our measurement is the most precise lepton charge asymmetry measurement to date. 
The uncertainty on the measured asymmetry is smaller than the PDF uncertainty for most of the bins.
We provide the correlation coefficients for different $|\eta^e|$ bins and the correlation 
coefficients between different kinematic bins, to be used for future PDF determinations.

This measurement supersedes the results of Ref.~\cite{d0_results_em_old}.
It also complements and provides more details on the results of Ref.~\cite{d0_w} which 
measured the $W^{\pm}$ boson charge asymmetry using the same data set. 
These asymmetries are in good agreement with those measured in the muon decay channel~\cite{d0_muon}.
The electron asymmetries presented here include the effects of the $W$ boson decay asymmetry, 
whereas the Ref.~\cite{d0_w} analysis solely addresses the production asymmetry. 
Both measurements should be useful in future analyses of the PDFs.
%%%%%%%%%%%%%%%%%%%%%%%%%%%%%%%%%%%%%%%%%%%%%%%%

% acknowledgement_APS_full_names.tex             18 October 2014
%
% Acknowledgement paragraph in English of Oct. 15, 2014 for APS journals

We thank the staffs at Fermilab and collaborating institutions,
and acknowledge support from the
Department of Energy and National Science Foundation (United States of America);
Alternative Energies and Atomic Energy Commission and
National Center for Scientific Research/National Institute of Nuclear and Particle Physics  (France);
Ministry of Education and Science of the Russian Federation, 
National Research Center ``Kurchatov Institute" of the Russian Federation, and 
Russian Foundation for Basic Research  (Russia);
National Council for the Development of Science and Technology and
Carlos Chagas Filho Foundation for the Support of Research in the State of Rio de Janeiro (Brazil);
Department of Atomic Energy and Department of Science and Technology (India);
Administrative Department of Science, Technology and Innovation (Colombia);
National Council of Science and Technology (Mexico);
National Research Foundation of Korea (Korea);
Foundation for Fundamental Research on Matter (The Netherlands);
Science and Technology Facilities Council and The Royal Society (United Kingdom);
Ministry of Education, Youth and Sports (Czech Republic);
Bundesministerium f\"{u}r Bildung und Forschung (Federal Ministry of Education and Research) and 
Deutsche Forschungsgemeinschaft (German Research Foundation) (Germany);
Science Foundation Ireland (Ireland);
Swedish Research Council (Sweden);
China Academy of Sciences and National Natural Science Foundation of China (China);
and
Ministry of Education and Science of Ukraine (Ukraine).
%

%%%%%%%%%%%%%%%%%%%%%%%%%%%%%%%%%%%%%%%%%%%%%%%%%%%%%%%%%

%%%%%%%%%%%%%%%%%%%%%%%%%%%%%%%%%%%%%%%%%%%%%%%%%%%%%%%%%
\begin{table*} 
\begin{center} 
\caption{Correlation matrix of the statistical uncertainties between different $|\eta^e|$ bins for the kinematic bin
$E_T^{e} >25$~GeV, $\met>25$~GeV. The matrix elements are multiplied by 100.} 
\begin{tabular}{c|ccccccccccccc} 
\hline 
\hline 
$|\eta^e|$ bins & 1    & 2    & 3    & 4    & 5    & 6    & 7    & 8    & 9    & 10   & 11   & 12   & 13  \\ \hline 
1 & 100 & 0.76 & 0.00 & 0.00 & 0.00 & 0.00 & 0.00 & 0.00 & 0.00 & 0.00 & 0.00 & 0.00 & 0.00 \\
2 &  & 100 & 0.69 & 0.00 & 0.00 & 0.00 & 0.00 & 0.00 & 0.00 & 0.00 & 0.00 & 0.00 & 0.00 \\
3 &  &  & 100 & 0.66 & 0.00 & 0.00 & 0.00 & 0.00 & 0.00 & 0.00 & 0.00 & 0.00 & 0.00 \\
4 &  &  &  & 100 & 0.66 & 0.00 & 0.00 & 0.00 & 0.00 & 0.00 & 0.00 & 0.00 & 0.00 \\
5 &  &  &  &  & 100 & 0.66 & 0.00 & 0.00 & 0.00 & 0.00 & 0.00 & 0.00 & 0.00 \\
6 &  &  &  &  &  & 100 & 0.32 & 0.00 & 0.00 & 0.00 & 0.00 & 0.00 & 0.00 \\
7 &  &  &  &  &  &  & 100 & 0.53 & 0.00 & 0.00 & 0.00 & 0.00 & 0.00 \\
8 &  &  &  &  &  &  &  & 100 & 0.40 & 0.00 & 0.00 & 0.00 & 0.00 \\
9 &  &  &  &  &  &  &  &  & 100 & 0.07 & 0.00 & 0.00 & 0.00 \\
10 &  &  &  &  &  &  &  &  &  & 100 & 0.41 & 0.00 & 0.00 \\
11 &  &  &  &  &  &  &  &  &  &  & 100 & 0.15 & 0.00 \\
12 &  &  &  &  &  &  &  &  &  &  &  & 100 & 0.31 \\
13 &  &  &  &  &  &  &  &  &  &  &  &  & 100 \\
\hline 
\hline 
\end{tabular} 
\label{Tab:Correlation_All_Fold_Type_1} 
\end{center} 
\end{table*} 

\begin{table*} 
\begin{center} 
\caption{Correlation matrix of the statistical uncertainties between different $|\eta^e|$ bins for the kinematic bin 
$25<E_T^{e}<35$~GeV, $\met > 25$~GeV.
The matrix elements are multiplied by 100.} 
\begin{tabular}{c|ccccccccccccc} 
\hline 
\hline 
$|\eta^e|$ bins & 1    & 2    & 3    & 4    & 5    & 6    & 7    & 8    & 9    & 10   & 11   & 12   & 13  \\ \hline 
1 & 100 & 0.75 & 0.00 & 0.00 & 0.00 & 0.00 & 0.00 & 0.00 & 0.00 & 0.00 & 0.00 & 0.00 & 0.00 \\
2 &  & 100 & 0.69 & 0.00 & 0.00 & 0.00 & 0.00 & 0.00 & 0.00 & 0.00 & 0.00 & 0.00 & 0.00 \\
3 &  &  & 100 & 0.65 & 0.00 & 0.00 & 0.00 & 0.00 & 0.00 & 0.00 & 0.00 & 0.00 & 0.00 \\
4 &  &  &  & 100 & 0.67 & 0.00 & 0.00 & 0.00 & 0.00 & 0.00 & 0.00 & 0.00 & 0.00 \\
5 &  &  &  &  & 100 & 0.68 & 0.00 & 0.00 & 0.00 & 0.00 & 0.00 & 0.00 & 0.00 \\
6 &  &  &  &  &  & 100 & 0.00 & 0.00 & 0.00 & 0.00 & 0.00 & 0.00 & 0.00 \\
7 &  &  &  &  &  &  & 100 & 0.83 & 0.00 & 0.00 & 0.00 & 0.00 & 0.00 \\
8 &  &  &  &  &  &  &  & 100 & 0.38 & 0.00 & 0.00 & 0.00 & 0.00 \\
9 &  &  &  &  &  &  &  &  & 100 & 0.28 & 0.00 & 0.00 & 0.00 \\
10 &  &  &  &  &  &  &  &  &  & 100 & 0.43 & 0.00 & 0.00 \\
11 &  &  &  &  &  &  &  &  &  &  & 100 & 0.14 & 0.00 \\
12 &  &  &  &  &  &  &  &  &  &  &  & 100 & 0.33 \\
13 &  &  &  &  &  &  &  &  &  &  &  &  & 100 \\
\hline 
\hline 
\end{tabular} 
\label{Tab:Correlation_All_Fold_Type_2} 
\end{center} 
\end{table*} 

\begin{table*} 
\begin{center} 
\caption{Correlation matrix of the statistical uncertainties between different $|\eta^e|$ bins for the kinematic bin 
$25 <E_T^{e}<35$~GeV, $25<\met <35$~GeV. 
The matrix elements
are multiplied by 100.} 
\begin{tabular}{c|ccccccccccccc} 
\hline 
\hline 
$|\eta^e|$ bins  & 1    & 2    & 3    & 4    & 5    & 6    & 7    & 8    & 9    & 10   & 11   & 12   & 13  \\ \hline 
1 & 100 & 0.75 & 0.00 & 0.00 & 0.00 & 0.00 & 0.00 & 0.00 & 0.00 & 0.00 & 0.00 & 0.00 & 0.00 \\
2 &  & 100 & 0.67 & 0.00 & 0.00 & 0.00 & 0.00 & 0.00 & 0.00 & 0.00 & 0.00 & 0.00 & 0.00 \\
3 &  &  & 100 & 0.65 & 0.00 & 0.00 & 0.00 & 0.00 & 0.00 & 0.00 & 0.00 & 0.00 & 0.00 \\
4 &  &  &  & 100 & 0.64 & 0.00 & 0.00 & 0.00 & 0.00 & 0.00 & 0.00 & 0.00 & 0.00 \\
5 &  &  &  &  & 100 & 0.66 & 0.00 & 0.00 & 0.00 & 0.00 & 0.00 & 0.00 & 0.00 \\
6 &  &  &  &  &  & 100 & 0.00 & 0.00 & 0.00 & 0.00 & 0.00 & 0.00 & 0.00 \\
7 &  &  &  &  &  &  & 100 & 0.81 & 0.00 & 0.00 & 0.00 & 0.00 & 0.00 \\
8 &  &  &  &  &  &  &  & 100 & 0.38 & 0.00 & 0.00 & 0.00 & 0.00 \\
9 &  &  &  &  &  &  &  &  & 100 & 0.30 & 0.00 & 0.00 & 0.00 \\
10 &  &  &  &  &  &  &  &  &  & 100 & 0.39 & 0.00 & 0.00 \\
11 &  &  &  &  &  &  &  &  &  &  & 100 & 0.14 & 0.00 \\
12 &  &  &  &  &  &  &  &  &  &  &  & 100 & 0.35 \\
13 &  &  &  &  &  &  &  &  &  &  &  &  & 100 \\
\hline 
\hline 
\end{tabular} 
\label{Tab:Correlation_All_Fold_Type_3} 
\end{center} 
\end{table*} 

\begin{table*} 
\begin{center} 
\caption{Correlation matrix of the statistical uncertainties between different $|\eta^e|$ bins for the kinematic bin
$E_T^{e} >35$~GeV, $\met> 25$~GeV. The matrix elements 
are multiplied by 100.} 
\begin{tabular}{c|ccccccccccccc} 
\hline 
\hline 
$|\eta^e|$ bins  & 1    & 2    & 3    & 4    & 5    & 6    & 7    & 8    & 9    & 10   & 11   & 12   & 13  \\ \hline 
1 & 100 & 0.75 & 0.00 & 0.00 & 0.00 & 0.00 & 0.00 & 0.00 & 0.00 & 0.00 & 0.00 & 0.00 & 0.00 \\
2 &  & 100 & 0.68 & 0.00 & 0.00 & 0.00 & 0.00 & 0.00 & 0.00 & 0.00 & 0.00 & 0.00 & 0.00 \\
3 &  &  & 100 & 0.65 & 0.00 & 0.00 & 0.00 & 0.00 & 0.00 & 0.00 & 0.00 & 0.00 & 0.00 \\
4 &  &  &  & 100 & 0.65 & 0.00 & 0.00 & 0.00 & 0.00 & 0.00 & 0.00 & 0.00 & 0.00 \\
5 &  &  &  &  & 100 & 0.63 & 0.00 & 0.00 & 0.00 & 0.00 & 0.00 & 0.00 & 0.00 \\
6 &  &  &  &  &  & 100 & 0.31 & 0.00 & 0.00 & 0.00 & 0.00 & 0.00 & 0.00 \\
7 &  &  &  &  &  &  & 100 & 0.52 & 0.00 & 0.00 & 0.00 & 0.00 & 0.00 \\
8 &  &  &  &  &  &  &  & 100 & 0.51 & 0.00 & 0.00 & 0.00 & 0.00 \\
9 &  &  &  &  &  &  &  &  & 100 & 0.05 & 0.00 & 0.00 & 0.00 \\
10 &  &  &  &  &  &  &  &  &  & 100 & 0.39 & 0.00 & 0.00 \\
11 &  &  &  &  &  &  &  &  &  &  & 100 & 0.16 & 0.00 \\
12 &  &  &  &  &  &  &  &  &  &  &  & 100 & 0.32 \\
13 &  &  &  &  &  &  &  &  &  &  &  &  & 100 \\
\hline 
\hline 
\end{tabular} 
\label{Tab:Correlation_All_Fold_Type_4} 
\end{center} 
\end{table*} 

\begin{table*} 
\begin{center} 
\caption{Correlation matrix of the statistical uncertainties between different $|\eta^e|$ bins for the kinematic bin
$E_T^{e} >35$~GeV, $\met > 35$~GeV. 
The matrix elements are multiplied by 100.} 
\begin{tabular}{c|ccccccccccccc} 
\hline 
\hline 
$|\eta^e|$ bins  & 1    & 2    & 3    & 4    & 5    & 6    & 7    & 8    & 9    & 10   & 11   & 12   & 13  \\ \hline 
1 & 100 & 0.75 & 0.00 & 0.00 & 0.00 & 0.00 & 0.00 & 0.00 & 0.00 & 0.00 & 0.00 & 0.00 & 0.00 \\
2 &  & 100 & 0.67 & 0.00 & 0.00 & 0.00 & 0.00 & 0.00 & 0.00 & 0.00 & 0.00 & 0.00 & 0.00 \\
3 &  &  & 100 & 0.65 & 0.00 & 0.00 & 0.00 & 0.00 & 0.00 & 0.00 & 0.00 & 0.00 & 0.00 \\
4 &  &  &  & 100 & 0.64 & 0.00 & 0.00 & 0.00 & 0.00 & 0.00 & 0.00 & 0.00 & 0.00 \\
5 &  &  &  &  & 100 & 0.62 & 0.00 & 0.00 & 0.00 & 0.00 & 0.00 & 0.00 & 0.00 \\
6 &  &  &  &  &  & 100 & 0.31 & 0.00 & 0.00 & 0.00 & 0.00 & 0.00 & 0.00 \\
7 &  &  &  &  &  &  & 100 & 0.53 & 0.00 & 0.00 & 0.00 & 0.00 & 0.00 \\
8 &  &  &  &  &  &  &  & 100 & 0.38 & 0.00 & 0.00 & 0.00 & 0.00 \\
9 &  &  &  &  &  &  &  &  & 100 & 0.07 & 0.00 & 0.00 & 0.00 \\
10 &  &  &  &  &  &  &  &  &  & 100 & 0.41 & 0.00 & 0.00 \\
11 &  &  &  &  &  &  &  &  &  &  & 100 & 0.16 & 0.00 \\
12 &  &  &  &  &  &  &  &  &  &  &  & 100 & 0.35 \\
13 &  &  &  &  &  &  &  &  &  &  &  &  & 100 \\
\hline 
\hline 
\end{tabular} 
\label{Tab:Correlation_All_Fold_Type_5} 
\end{center} 
\end{table*} 

%%%%%%%%%%%%%%%%%%%%%%%%%%%%%%%%%%%%%%%%%%%%%%%%%%%%%%%%%
\begin{table*}
\begin{center}
\caption{The fraction of events in each reconstruction-level kinematic bin that come from different
generator-level kinematic bins. These bins are not all independent. }
\begin{tabular}{rcl|c|c|c|c|c|c}
\hline
\hline
 &&& $20<p_T^e<25$ & $p_T^e>25$ &  $25<p_T^e<35$ & $25<p_T^e<35$ & $p_T^e > 35$ & $p_T^e > 35$ \\
 &&& $\text{\sc or}$ & $\text{\sc and}$ & $\text{\sc and}$ & $\text{\sc and}$ & $\text{\sc and}$ & $\text{\sc and}$ \\
 &&& $20<p_T^{\nu}<25$ & $p_T^{\nu}>25$ & $p_T^{\nu}>25$ &  $25<p_T^{\nu}<35$ & $p_T^{\nu}>25$ & $p_T^{\nu}>35$ \\ \hline
 $20<E_T^e<25$ & $\text{\sc or}$ & $20<\met<25$ & 0.64 & 0.36 & 0.32 & 0.27 & 0.04 & 0.02 \\ 
 $E_T^e>25$ & $\text{\sc and}$ & $\met>25$ & 0.08 & 0.92 & 0.35 & 0.22 & 0.57 & 0.42 \\ 
$25<E_T^e<35$ & $\text{\sc and}$ & $\met>25$  & 0.12 & 0.88 & 0.80 & 0.51 & 0.08 & 0.06 \\ 
$25<E_T^e<35$ & $\text{\sc and}$ & $25<\met<35$  & 0.15 & 0.85 & 0.79 & 0.60 & 0.06 & 0.03 \\ 
$E_T^e>35$ & $\text{\sc and}$ & $\met>25$ & 0.05 & 0.95 & 0.06 & 0.03 & 0.89 & 0.66 \\ 
$E_T^e>35$ & $\text{\sc and}$ & $\met>35$ & 0.03 & 0.97 & 0.05 & 0.02 & 0.92 & 0.72 \\ 
\hline
\hline
\end{tabular}
\label{Tab:Kine_fraction}
\end{center}
\end{table*}
%%%%%%%%%%%%%%%%%%%%%%%%%%%%%%%%%%%%%%%%%%%%%%%%%%%%%%%%%

\begin{thebibliography}{99}
\bibitem{d0_coordinate}
 D0 uses a cylindrical coordinate system with the $z$ axis 
along the beam axis in the proton direction. Angles $\theta$ and $\phi$ are the polar and
azimuthal angles, respectively. Pseudorapidity is defined as
$\eta=-\ln [\tan(\theta/2)]$, where $\theta$ is measured with respect
to the $p\bar{p}$ interaction vertex. In the massless limit, $\eta$ is equivalent to
the rapidity $y=(1/2) \ln[(E+p_z)/(E-p_z)]$.
$\eta_{\text{det}}$ is the pseudorapidity measured with respect to the
center of the detector.
\end{thebibliography}
\end{document}